\documentclass[12pt, a4paper, oneside]{Thesis} % Paper size, default font size and one-sided paper
\usepackage[greek,english]{babel}
\usepackage[square, numbers, comma, sort&compress]{natbib} % Use the natbib reference package - read up on this to edit the reference style; if you want text (e.g. Smith et al., 2012) for the in-text references (instead of numbers), remove 'numbers' 
\usepackage{longtable}
\usepackage{lipsum}
\usepackage[strict]{changepage}
\usepackage{graphicx}
\usepackage{array}
\usepackage[export]{adjustbox}[2011/08/13]
\usepackage{amsmath}
\usepackage{amsfonts}
\usepackage{amssymb}
\usepackage{lscape}
\usepackage{pdflscape}
\graphicspath{{Pictures/}} % Specifies the directory where pictures are stored
\hypersetup{urlcolor=blue, colorlinks=true} % Colors hyperlinks in blue - change to black if annoying
\title{\ttitle} % Defines the thesis title - don't touch this

\begin{document}

\frontmatter % Use roman page numbering style (i, ii, iii, iv...) for the pre-content pages

\setstretch{1.3} % Line spacing of 1.3
\setlength{\parindent}{1cm}
\setlength{\parskip}{0cm}

% Define the page headers using the FancyHdr package and set up for one-sided printing
\fancyhead{}  % Clears all page headers and footers
\rhead{\thepage}  % Sets the right side header to show the page number
\lhead{}  % Clears the left side page header

\pagestyle{fancy}  % Finally, use the "fancy" page style to implement the FancyHdr headers

\newcommand{\HRule}{\rule{\linewidth}{0.5mm}}  % New command to make the lines in the title page

% PDF meta-data
\hypersetup{pdftitle={\ttitle}}
\hypersetup{pdfsubject=\subjectname}
\hypersetup{pdfauthor=\authornames}
\hypersetup{pdfkeywords=\keywordnames}

%----------------------------------------------------------------------
%	TITLE PAGE - ENGLISH
%----------------------------------------------------------------------

\begin{titlepage}
\begin{center}

\begin{figure}[htbp]
  \centering
    \includegraphics[width=35mm]{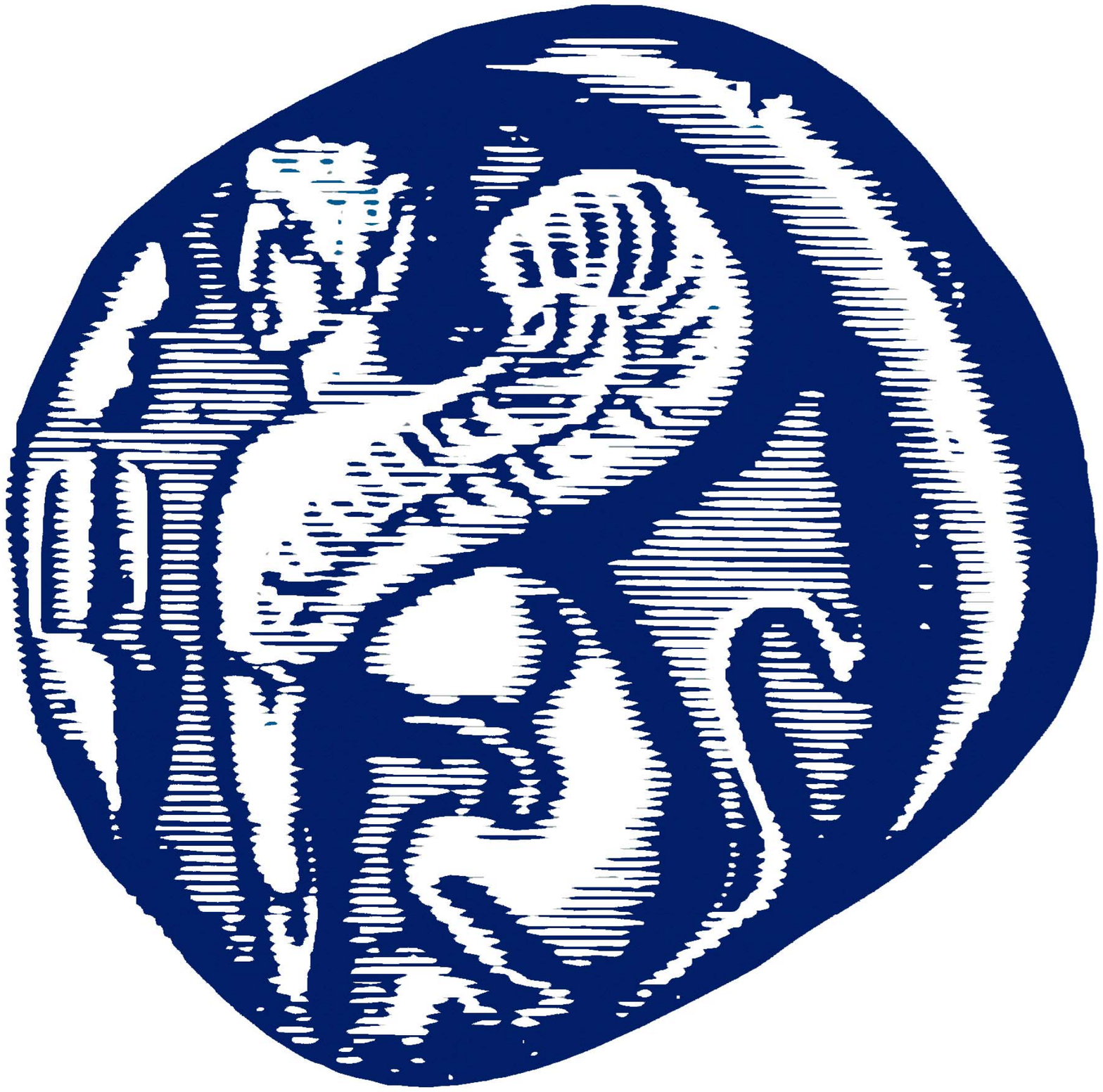}    %University/department logo - uncomment to place it                                 
\end{figure}

\textsc{\LARGE \univname}\\[1.5cm] % University name
\textsc{\Large Doctoral Thesis}\\[0.5cm] % Thesis type

\HRule \\[0.4cm] % Horizontal line
{\huge \bfseries \ttitle}\\[0.4cm] % Thesis title
\HRule \\[1.5cm] % Horizontal line
 
%\begin{minipage}{0.4\textwidth}
%\begin{flushleft} 
\Large
\begin{center}
\href{http://www.icsd.aegean.gr/group/members-data.php?group=L5&member=1177}{\authornames}\\[2cm] % Author name- remove the \href bracket to remove the link
\end{center}
 
%\end{flushleft}
%\end{minipage}
%\begin{minipage}{0.4\textwidth}
%\begin{flushright} \large
%\emph{Supervisor:} \\
%\href{http://www.icsd.aegean.gr/group/members-data.php?group=L5&member=44}{\supname} % Supervisor name - remove the \href bracket to remove the link  
%\end{flushright}
%\end{minipage}\\[2cm]
 
\large \textit{A thesis submitted in fulfilment of the requirements\\ for the degree of \degreename}\\[0.3cm] % University requirement text
\textit{in the}\\[0.4cm]
\groupname\\\deptname\\[2cm] % Research group name and department name
 
{\large Samos, April 2016}\\[4cm] % Date

\vfill
\end{center}

\end{titlepage}
%----------------------------------------------------------------------
%	TITLE PAGE - GREEK
%----------------------------------------------------------------------

\begin{center}

\thispagestyle{empty}   % Do not put a number in this page
\begin{figure}[htbp]
  \centering
    \includegraphics[width=35mm]{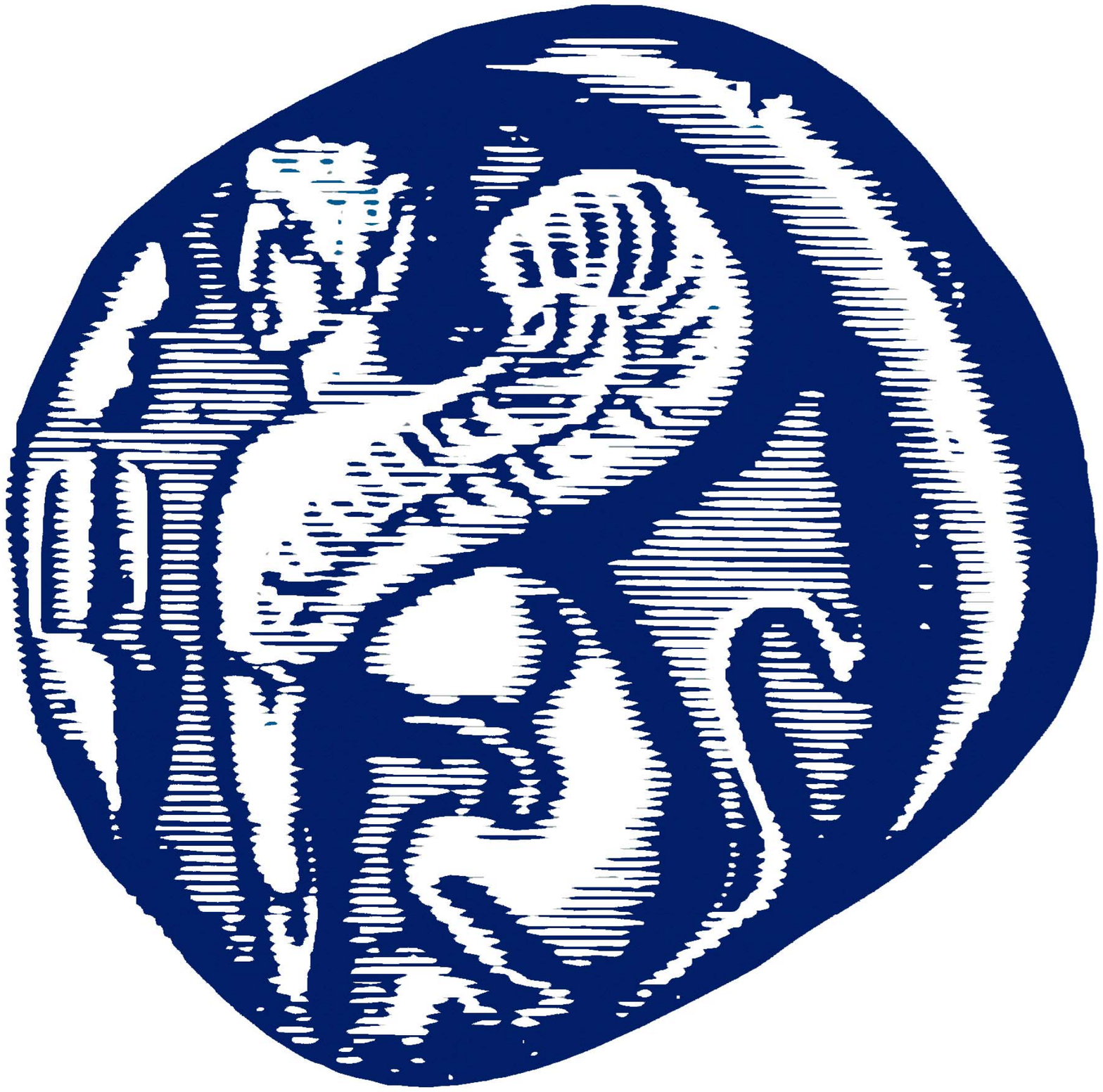}    %University/department logo - uncomment to place it                                 
\end{figure}

\textsc{\LARGE \textgreek{PANEPISTHMIO AIGAIOU}}\\[1.5cm] % University name
\textsc{\Large \textgreek{DIDAKTORIKH DIATRIBH}}\\[0.5cm] % Thesis type

\HRule \\[0.4cm] % Horizontal line
{\huge \bfseries \textgreek{Asumptwtik'ec Idi'othtec Kosmologi'wn se Jewr'iec Bar'uthtac Uyhl'oterhc T'axhc}}\\[0.4cm] % Thesis title
\HRule \\[1.5cm] % Horizontal line

%\begin{minipage}{0.4\textwidth}
%\begin{flushleft} 
\large

\begin{center}
\href{http://www.icsd.aegean.gr/group/members-data.php?group=L5&member=1177}{\Large \bfseries \textgreek{Ge'wrgioc Koli'wnhc}}\\[2.5cm] % Author name- remove the \href bracket to remove the link
\end{center}
 
%\end{flushleft}
%\end{minipage}
%\begin{minipage}{0.4\textwidth}
%\begin{flushright} \large
%\emph{Supervisor:} \\
%\href{http://www.icsd.aegean.gr/group/members-data.php?group=L5&member=44}{\supname} % Supervisor name - remove the \href bracket to remove the link  
%\end{flushright}
%\end{minipage}\\[2cm]
 
\large \textgreek{Didaktorik'h diatrib'h upoblhje'isa sto}\\[0.3cm] % University requirement text
\textgreek{Tm'hma Mhqanik'wn Plhroforiak'wn kai Epikoinwniak'wn Susthm'atwn}\\
\textgreek{tou Panepisthm'iou Aiga'iou}\\[0.3cm]
\textgreek{gia thn ap'okthsh tou t'itlou tou}\\
\textgreek{Did'aktora tou Panepisthm'iou Aiga'iou}\\[3cm]

\large \textgreek{S'amoc}, \textgreek{Apr'ilioc 2016}\\ % Date

\vfill
\end{center}

\clearpage
%---------------------------------------------------------------------
%	    COMMITTEES - ENGLISH
%---------------------------------------------------------------------

% CODE FOR VERTICAL ALIGNMENT
%
%\vspace*{\fill} 
%\begin{center} 
%\begin{minipage}{\textwidth} 
%\centering{This is some text to be centred vertically.} 
%\end{minipage} 
%\end{center} 
%\vfill % equivalent to \vspace{\fill} \clearpage 

\addtotoc{Advisory Committee} % Add the "Committees" page entry to the Contents
\thispagestyle{empty}  % Do not put a number in this page

\vspace*{\fill} 
\begin{center} 
\begin{minipage}{\textwidth} 
\centering{
\textbf{The advisory committee}
\vspace{0.5cm}\\
Georgios Kofinas, Assistant Professor, University of the Aegean\\Supervisor\\
John Miritzis, Associate Professor, University of the Aegean\\Member\\
Nicolas Hadjisavvas, Professor Emeritus, University of the Aegean\\Member\\
\vspace{3cm}
%\textbf{The examining committee}
%\vspace{0.5cm}\\
%Stefanos Gritzalis, Professor\\
%University of the Aegean \\
%John Miritzis, Associate Professor\\
%University of the Aegean \\
%Nicolas Hadjisavvas, Professor Emeritus\\
%University of the Aegean\\
%Georgios Kofinas, Assistant Professor\\
%University of the Aegean\\
%Kyriakos Papadopoulos, Assistant Professor\\
%American University of the Middle East, Kuwait\\
%Ifigeneia Klaoudatou, Assistant Professor\\
%American University of the Middle East, Kuwait\\
%Georgia Kittou, Assistant Professor\\
%American University of the Middle East, Kuwait
} 
\end{minipage} 
\end{center} 
\vfill

\clearpage

%---------------------------------------------------------------------
%	    COMMITTEES - GREEK
%---------------------------------------------------------------------
%\addtotoc{Committees} % Add the "Committees" page entry to the Contents

\thispagestyle{empty}  % Do not put a number in this page
\vspace*{\fill} 
\begin{center} 
\begin{minipage}{\textwidth} 
\centering{
\textbf{\textgreek{H sumbouleutik'h epitrop'h}}
\vspace{0.5cm}\\
\textgreek{Ge'wrgioc Kofin'ac}, \textgreek{Ep'ikouroc Kajhght'hc}, \textgreek{Panepist'hmio Aiga'iou}\\ \textgreek{Epibl'epwn}\\
\textgreek{Gi'annhc Muritz'hc}, \textgreek{Anaplhrwt'hc Kajhght'hc}, \textgreek{Panepist'hmio Aiga'iou}\\ \textgreek{M'elos}\\
\textgreek{Nik'olac Qatzhs'abbac}, \textgreek{Om'otimoc Kajhght'hc}, \textgreek{Panepist'hmio Aiga'iou}\\ \textgreek{M'elos}\\
\vspace{3cm}
%\textbf{\textgreek{H exetastik'h epitrop'h}}
%\vspace{0.5cm}\\
%\textgreek{St'efanoc Gkr'itzalhs}, \textgreek{Kajhght'hc}\\
%\textgreek{Panepist'hmio Aiga'iou} \\\
%\textgreek{Gi'annhc Muritz'hc}, \textgreek{Anaplhrwt'hc Kajhght'hc},\\
%\textgreek{Panepist'hmio Aiga'iou} \\
%\textgreek{Nik'olac Qatzhs'abbac}, \textgreek{Om'otimoc Kajhght'hc}\\
%\textgreek{Panepist'hmio Aiga'iou} \\
%\textgreek{Ge'wrgioc Kofin'ac}, \textgreek{Ep'ikouroc Kajhght'hc}\\
%\textgreek{Panepist'hmio Aiga'iou} \\
%\textgreek{Kuri'akoc Papad'opouloc}, \textgreek{Ep'ikouroc Kajhght'hc}\\
%American University of the Middle East, Kuwait\\
%\textgreek{Ifig'eneia Klaoud'atou}, \textgreek{Ep'ikourh Kajhg'htria}\\
%American University of the Middle East, Kuwait\\
%\textgreek{Gewrg'ia K'ittou}, \textgreek{Ep'ikourh Kajhg'htria}\\
%American University of the Middle East, Kuwait
} 
\end{minipage} 
\end{center} 
\vfill

\clearpage

%----------------------------------------------------------------------
%	DECLARATION PAGE
%	Your institution may give you a different text to place here
%----------------------------------------------------------------------

\Declaration{

%\addtocontents{toc}{\vspace{1em}} % Add a gap in the Contents, for aesthetics

I, \authornames, declare that this thesis entitled, '\ttitle' and the work presented in it are my own. I confirm that:

\begin{itemize} 
\item[\tiny{$\blacksquare$}] This work was done wholly or mainly while in candidature for a research degree at this University.
\item[\tiny{$\blacksquare$}] Where any part of this thesis has previously been submitted for a degree or any other qualification at this University or any other institution, this has been clearly stated.
\item[\tiny{$\blacksquare$}] Where I have consulted the published work of others, this is always clearly attributed.
\item[\tiny{$\blacksquare$}] Where I have quoted from the work of others, the source is always given. With the exception of such quotations, this thesis is entirely my own work.
\item[\tiny{$\blacksquare$}] I have acknowledged all main sources of help.
\end{itemize}

\vspace{10mm}

Signed:\\
\rule[1em]{25em}{0.5pt} % This prints a line for the signature

Date:\\
\rule[1em]{25em}{0.5pt} % This prints a line to write the date
}

\clearpage % Start a new page

%--------------------------------------------------------------------
%	 PUBLISHED WORK
%--------------------------------------------------------------------

%\addtotoc{Published work} % Add the "Published work" page entry to the Contents

\thispagestyle{empty}  % Do not put a number in this page
\vspace*{\fill} 
\begin{center} 
\begin{minipage}{\textwidth} 
\centering{
This thesis is partly based on the following publications:
\begin{itemize}

\item S. Cotsakis, G. Kolionis and A. Tsokaros. \textit{The initial state of generalized radiation universes}. Physics Letters B\textbf{721}: 1-6, 2013. ISSN 0370-2693. doi: http://dx.doi.org/10.1016/j.physletb.2013.02.048. URL \href{http://www.sciencedirect.com/science/article/pii/S0370269313001913}{http://www.sciencedirect.com/science/article/pii/S0370269313001913}.
\href{http://arxiv.org/abs/1211.5255}{arXiv:1211.5255}.
\item S. Cotsakis, S. Kadry, G. Kolionis and A. Tsokaros. \textit{Asymptotic vacua with higher derivatives}. Physics Letters B\textbf{755}: 387-392, 2016. ISSN 0370-2693. doi: http://dx.doi.org/10.1016/j.physletb.2016.02.036. URL \href{http://www.sciencedirect.com/science/article/pii/S0370269316001325}{http://www.sciencedirect.com/science/article/pii/S0370269316001325}.
\href{http://arxiv.org/abs/1303.2234}{arXiv:1303.2234}.
\item S. Cotsakis, G. Kolionis and A. Tsokaros. \textit{Asymptotic states of generalized universes with higher derivatives}. In Proceedings of the Thirteenth Marcel Grossman Meeting on General Relativity, K.  Rosquist, R. T Jantzen, R. Ruffini (eds.), World Scientific,  2015, p. 1868-1870. URL \href{http://www.worldscientific.com/doi/abs/10.1142/9789814623995_0302}{http://www.worldscientific.com/doi/abs/10.1142/9789814623995\texttt{\_}0302}. \href{http://arxiv.org/abs/1302.6089}{arXiv:1302.6089}.
\end{itemize}
} 
\end{minipage} 
\end{center} 
\vfill

\clearpage % Start a new page
%---------------------------------------------------------------------
%	QUOTATION PAGE
%---------------------------------------------------------------------

\pagestyle{empty} %No headers or footers for the following pages

\null\vfill %  Add some space to move the quote down the page a bit

\begin{flushright}
\textit{$\Delta\tilde{\omega}\varsigma \hspace{1mm} \mu o\iota \hspace{1mm} \pi\tilde{\alpha} \hspace{1mm} \sigma\tau\tilde{\omega} \hspace{1mm}  \kappa\alpha\grave{\iota} \hspace{1mm} \tau\grave{\alpha}\nu \hspace{1mm} \gamma\tilde{\alpha}\nu \hspace{1mm} \kappa\iota\nu\acute{\alpha}\sigma\omega$.}
\end{flushright}

\begin{flushright}
Archimedes
\end{flushright}

\vfill\vfill\vfill\vfill\vfill\vfill\null % Add some space at the bottom to position the quote just right

\clearpage % Start a new page

%--------------------------------------------------------------------
%	ABSTRACT PAGE  - ENGLISH 
%--------------------------------------------------------------------

\addtotoc{Abstract} % Add the "Abstract" page entry to the Contents

\abstract{
We study the early-time behavior of isotropic and homogeneous solutions in vacuum as well as radiation-filled cosmological models in the full, effective, four-dimensional gravity theory with higher derivatives. We use asymptotic methods to analyze all possible ways of approach to the initial singularity of such universes. In order to do so, we construct autonomous dynamical systems that describe the evolution of these models, and decompose the associated vector fields. We prove that, at early times, all flat vacua as well as general curved ones are globally attracted by the `universal' square root scaling solution. Open vacua, on the other hand show in both, future and past directions a dominant asymptotic approach to horizon-free, Milne states that emerge from initial data sets of smaller dimension. Closed universes exhibit more complex logarithmic singularities. Our results on asymptotic stability show a possible relation to cyclic and ekpyrotic cosmologies at the passage through the singularity. In the case of radiation-filled universes of the same class we show the essential uniqueness and stability of the resulting asymptotic scheme, once more dominated by $t^{1/2}$, in all cases except perhaps that of the conformally invariant Bach-Weyl gravity. In all cases, we construct a formal series representation valid near the initial singularity of the general solution of these models and prove that curvature as well as radiation play a subdominant role in the dominating form. A discussion is also made on the implications of these results for the generic initial state of the theory. 
}

\clearpage % Start a new page

%----------------------------------------------------------------------
%	ABSTRACT PAGE  - GREEK
%----------------------------------------------------------------------

\vspace*{\fill} 
\begin{center} 

\begin{minipage}{\textwidth} 
\centering{ 
\normalsize
\textsc{\textgreek{PANEPISTHMIO AIGAIOU}}\\ % University name
%\normalsize \normalfont \textgreek{Sqol'h Jetik'wn Episthm'wn}\\ % University name
\textgreek{Tm'hma Mhqanik'wn Plhroforiak'wn kai Epikoinwniak'wn Susthm'atwn }\\[1cm]

\bfseries {\huge \textit{\textgreek{Per'ilhyh}}}\\[0.3cm] % University name
\normalfont\textgreek{Didaktorik'hc diatrib'hc}\\[0.1cm]
\large\bfseries {\textgreek{Asumptwtik'ec Idi'othtec Kosmologi'wn se Jewr'iec Bar'uthtac Uyhl'oterhc T'axhc}}\\[0.1cm] % Thesis title
\normalfont\textsc{\textgreek{GEWRGIOU KOLIWNH}}\\[1cm]
}
\end{minipage} 
\end{center} 
\vspace{1.5cm}

\normalsize
\textgreek{Melet'ame thn pr'wimh sumperifor'a, is'otropwn kai omogen'wn l'usewn t'oso sto ken'o 'oso kai se kosmologik'a mont'ela me aktinobol'ia stis pl'hreic, lusitele'ic} (effective), \textgreek{tetradi'astatec barutik'ec jewr'iec uyhl'oterhc t'axhc. Qrhsimopoio'ume asumptwtik'ec mej'odouc gia na anal'usoume 'olous touc pijano'uc tr'opouc pros'eggishc thc arqik'hc idiomorf'iac se t'etoiou e'idouc s'umpanta. Efarm'ozoume aut'a ta majhmatik'a ergale'ia kataskeu'azontac aut'onoma dunamik'a sust'hmata pou perigr'afoun thn exelixh aut'wn twn mont'elwn kai diasp'ame ta ant'istoiqa dianusmatik'a ped'ia. Apodeikn'uoume 'oti se pr'wimo qr'ono 'ola ta ep'ipeda kaj'wc ep'ishc kai ta kampulwm'ena k'ena s'umpanta 'elkontai sunolik'a ap'o thn `kajolik'h' l'ush tou par'agonta kl'imakac wc tetragwnik'hc r'izac thc qronik'hc sunist'wsac. Ta anoiqt'a ken'a s'umpanta epideikn'uoun pio pol'uplokec logarijmik'ec idiomorf'iec. Ta apotel'esmat'a mac gia thn asumptwtik'h sumperifor'a, anadeikn'uoun mia pijan'h susq'etish me tic kuklik'ec kai ekpurwtik'ec kosmolog'iec kat'a th met'abash diam'esou thc idiomorf'iac. Sthn 'idia kathgoria sump'antwn me aktinobol'ia, de'iqnoume thn ousi'wdh monadik'othta kai eust'ajeia twn prokupt'ontwn asumptwtik'wn sqhm'atwn, sta opo'ia kai p'ali epikrate'i h $t^{1/2}$, se 'olec tic peript'wseic ekt'oc 'iswc ap'o th s'ummorfh anallo'iwth bar'uthta} Bach-Weyl. \textgreek{Se k'aje per'iptwsh kataske'uazoume mia anapar'astash twn genik'wn l'usewn aut'wn twn mont'elwn se morf'h tupik'hc} (formal), \textgreek{ seir'ac g'urw ap'o thn arqik'h idiomorf'ia kai apodeikn'uoume 'oti ta qarakthristik'a t'oso thc kampul'othtac 'oso kai thc aktinobol'iac diadramat'izoun upole'iponta r'olo sthn epikrato'usa sumperifor'a. Anafer'omaste ep'ishc stic genik'oterec sunepeiec twn apotelesm'atwn mac gia thn tupik'h arqik'h kat'astash thc en l'ogw jewr'iac.}

\vfill      % equivalent to \vspace{\fill} 
\clearpage 

%----------------------------------------------------------------------------------------
%	ACKNOWLEDGEMENTS
%----------------------------------------------------------------------------------------

\setstretch{1.3} % Reset the line-spacing to 1.3 for body text (if it has changed)

\acknowledgements{\addtocontents{toc}{\vspace{1em}} % Add a gap in the Contents, for aesthetics
This thesis would have never been realized without the unwavering support of my supervisor, Professor Spiros Cotsakis. From the beginning until the end he has been a mentor, a guide and a friend by showing the limitless understanding of a true teacher and sharing his passion for the precision and uncompromising integrity of scientific research. I am truly grateful.

I would also like to thank the members of my advisory committee Assistant Professor Georgios Kofinas, Associate Professor Johh Miritzis and Professor Emeritus Nicolas Hadjisavvas for their kind support and valuable advice.

The Department of Information and Communication Systems Engineering provided the ideal conditions for the completion of this project including an administrative part-time job. My sincere thanks to all the staff of the Department with whom I have cooperated during my work there and especially to Mrs. Eleni Papagrigoriou, Director of the University's Administrative Unit of Samos, for our kind and inspiring discussions and her sincere interest. 

I am indebted to Dr. Antonios Tsokaros whose help was always available and his time for fruitful discussions endless.

Many thanks to Assistant Professor Ifigeneia Klaoudatou for useful discussions and guidance especially in the first years of my research.

My deep gratitude to Dimitris Trachilis and Assistant Professor Georgia Kittou for all those years of working together, all the time we shared and all the fun we had between hard work.

I am grateful to my wife Katja Alexiadou who has been shedding the light of her love and understanding even in the darkest times of this journey.

I also take this opportunity to express my sincere sense of gratitude to my parents who paved the path before me with their sacrifices and their efforts.

My friends Foteini Malagari, Kostas Karastathis, Katerina Gioroglou, Ioannis Stamatiou, Kalia Anastasopoulou, Iro Papachristou, Giannis Mamoutzis, Iro Papadopoulou, Lida Mamoutzi, Roula Kalozoumi, Pigi Kladi, Stelios Mamoutzis, Foivos Gkourogiannis and Spartakos Papadakis shared with me longer or shorter parts of this journey. They were all there and I thank them deeply. 
}  

\clearpage % Start a new page

%----------------------------------------------------------------------
%	LIST OF CONTENTS/FIGURES/TABLES PAGES
%----------------------------------------------------------------------

\pagestyle{fancy} % The page style headers have been "empty" all this time, now use the "fancy" headers as defined before to bring them back

\lhead{\emph{Contents}} % Set the left side page header to "Contents"
\tableofcontents % Write out the Table of Contents

\mainmatter % Begin numeric (1,2,3...) page numbering

\pagestyle{fancy} % Return the page headers back to the "fancy" style

% Include the chapters of the thesis as separate files from the Chapters folder
% Uncomment the lines as you write the chapters

% Chapter Template

\chapter{Introduction} % Main chapter title

\label{Chapter1} % Change X to a consecutive number; for referencing this chapter elsewhere, use \ref{ChapterX}

\lhead{Chapter 1. \emph{Introduction}} % Change X to a consecutive number; this is for the header on each page - perhaps a shortened title

It is generally accepted that modern cosmology was born with the discovery of the theory of general relativity (GR)\cite{ein-15,ein-16} about 100 years ago. GR was developed as a theory about the nature of gravity. However, since gravity holds a special place among the fundamental forces of nature in the sense that it is the only one that seems to play an essential role at the macroscopic level, gravitational theories are those that form the basis upon which the cosmological models are built. Therefore, naturally the discovery of such a radical and universal gravitational theory gave rise to a dizzying development of cosmology. In just 100 years and in combination with the rapid technological development that boosted both our observational as well as our experimental capability, our knowledge about the universe rocketed to such an extent that the creation of a universal unified theory of `everything' escaped the realm of fiction and began to be considered a feasible scientific objective.

The original mathematical background of GR is Riemannian geometry which had been already developed from the mid-19th century. In the mathematical basis of GR we consider a four-dimensional Lorentzian manifold, which is the unified expression of space and time in a single entity called spacetime. Gravitational phenomena are implemented through the metric defined on this manifold, a rank-2 tensor $g_{\mu\nu}$ for which we employ the \emph{space-like convention}, such that it has the signature $(\:-\:+\:+\:+\:)$ when diagonalized and thus, the line element has the form
\be
ds^2=-dt^2+dx^2+dy^2+dz^2.
\ee
Following the sign conventions of \cite{mis-tho-whe-73}, the Riemann and Einstein tensors are given by
\be
R^{\mu}_{\:\:\:\nu\alpha\beta}=\partial_{\alpha}\Gamma^{\mu}_{\:\:\:\nu\beta}-\partial_{\beta}\Gamma^{\mu}_{\:\:\:\nu\alpha}+\Gamma^{\mu}_{\:\:\:\sigma\alpha}\Gamma^{\sigma}_{\:\:\:\nu\beta}-\Gamma^{\mu}_{\:\:\:\sigma\beta}\Gamma^{\sigma}_{\:\:\:\nu\alpha}
\ee
\be
G_{\mu\nu}=R_{\mu\nu}-\frac{1}{2}g_{\mu\nu}R
\ee
where
\be 
R_{\mu\nu}=R^{\sigma}_{\:\:\:\mu\sigma\nu} \:\:\:\text{and}\:\:\: R=R^{\sigma}_{\:\:\:\sigma}
\ee
are the Ricci tensor and the scalar curvature respectively. Under these conventions we write the field equation of GR, also known as the Einstein equation, as
\be\label{einsteineqs}
R_{\mu\nu}-\frac{1}{2}g_{\mu\nu}R=8\pi G T_{\mu\nu} - g_{\mu\nu}\Lambda
\ee
Here $T_{\mu\nu}$ is the energy-momentum tensor and $\Lambda$ is the cosmological constant. This equation forms a set of 10 partial differential equations (PDEs) - one for each of the 10 independent components of the metric tensor - which contain up to second-order derivatives of the metric $g_{\mu\nu}$ in 4 variables.

Within this general mathematical background and simultaneously with the formulation of GR, David Hilbert showed \cite{hil-15} that the derivation of the Einstein equation under the Hamiltonian formulation of a classical field theory was possible for a Lagrangian density of the form
\be \label{GRlagrangian}
\mathcal{L}=\frac{1}{16\pi G} (R-2\Lambda) + \mathcal{L}_{M}(g_{\mu\nu},\psi).
\ee
The assumption that the action
\be \label{GRaction}
\mathcal{S}=\frac{1}{16\pi G}\int (R-2\Lambda) d\mu_{g} +\int \mathcal{L}_{M}(g_{\mu\nu},\psi) d\mu_{g}
\ee
where $d\mu_{g}=\sqrt{-g}d^4x$, is an extremum under arbitrary variations of the metric $g_{\mu\nu}$ leads to Eq. (\ref{einsteineqs}). As we will see below very soon it became apparent that a different Lagrangian density could lead to other more exotic gravity theories.

Since a detailed presentation of GR deviates from the purpose of this thesis we refer the interested reader to the classical textbooks \cite{mis-tho-whe-73,wal-84,wei-72,one-83,lan-lif-75,haw-ell-73,cho-09}.

In order to built a cosmological model based on GR one must basically determine a metric $g_{\mu\nu}$ that will satisfy the Einstein equations (\ref{einsteineqs}) under certain constraints based on specific assumptions about the physical interpretation of the specific model. The assumption of homogeneity and isotropy of the universe leads to the well known Friedmann-Robertson-Walker (FRW) metric $g_{\mu\nu}$ given, in spherical coordinates, by the line element \cite{fri-22,fri-24,lem-31,rob-35,rob-36,wal-37},
\be \label{frwmetric}
ds^2=-dt^{2}+a^{2}(t)\,(\frac{dr^{2}}{1-kr^2}+r^2(d\theta^{2}+\sin^{2}\theta d\phi^{2})) \equiv g_{\mu\nu}dx^{\mu}dx^{\nu},
\ee
where $a(t)$ is the scale factor and $k$ is the constant curvature of the spacetime, normalized to take the three values $0, +1$ or $-1$ for the complete, simply connected, flat, closed or open space sections respectively \cite{mis-tho-whe-73, wal-84, wei-08}. We note that by assuming this form of the metric one imposes the assumption of homogeneity and isotropy in the cosmological model under consideration regardless of the chosen gravity theory which is generaly expressed through the specific form of the gravitational field equations. During the course of this thesis we will assume homogeneity and isotropy over the universes we will be studying, hence it is exactly the form (\ref{frwmetric}) that will be imposed on the field equations of the modified theory of gravity we will be studying and will be explained in the next section.

An important FRW cosmological solution of the Einstein equations, (\ref{einsteineqs}), which plays a key role in the interpretation of our results is the \emph{Milne universe}. In this cosmological model the constant curvature $k$ takes the value $-1$ and the universe is considered to be completely empty. In this case $a(t)=t$ and the metric takes the form \cite{mil-35}
\be \label{milnemetric}
ds^2=-dt^{2}+t^2\,(\frac{dr^{2}}{1+r^2}+r^2(d\theta^{2}+\sin^{2}\theta d\phi^{2})) \equiv g_{\mu\nu}dx^{\mu}dx^{\nu},
\ee
As expected, since the Milne universe is an exact solution of GR for an isotropic universe without matter, it is in fact a piece of the Minkowski spacetime in expanding coordinates \cite{muk-05, mis-tho-whe-73}.

%----------------------------------------------------------------------------------------
%	SECTION   -  Modified theories of gravity
%----------------------------------------------------------------------------------------
\section{Modified theories of gravity}

Despite the enormous influence and impact of GR on the scientific community, it did not take long at all for the first proposals to appear either for its modification or its expansion. The task for its unification with the quantum theory of fields which constitutes the other pillar of modern science in order to be led to a unified theory, is definitely one of the strongest motives for every scientist in this field ever since. Already before 1920, the ideas of Eddington \cite{edd-23}, Weyl \cite{wey-19a} and Kaluza \cite{kal-21} set out the basic guidelines for some of the most influential and productive modifications of GR.

The investigation of the possibility of the dependence of Newton's constant of the time, led to the creation in 1961 of the theory of Brans-Dicke \cite{bra-dic-61,dic-62,ber-68}, which in turn became the foundation of the so called \emph{scalar-tensor theories}. In this great class of gravity theories which in general assumes the existence of extra fields in the Lagrangian of the theory, one can also include the Einstein-Aether theories \cite{jac-mat-01}, the bimetric theories of gravity \cite{ros-73,ros-78,dru-01} and the tensor-vector-scalar theories \cite{bek-04} among others. While in GR gravity is mediated through a 2nd order tensor field, nothing precludes the existence in the field equations of other fields, scalar, vector, tensor or even of higher order. The existence of such fields is usually implemented through weak couplings in order to enable their study in scales which are comparable to the GR. This does not mean of course that there have not been efforts in other directions. 
%[Clifton-Ferreira Padilla-Skordis-2012]
%In General Relativity the gravitational force is mediated by a single rank-2 tensor field, or a massless spin-2 particle in the
%quantum field theory picture. While there is good reason to couple matter fields to gravity in this way, there is less reason
%to think that the field equations of gravity should not contain other fields, and one is in general free to speculate on the
%existence of such additional fields in the gravitational sector. The simplest scenario that one could consider in this context is
%the addition of an extra scalar field, but one might also choose to consider extra vectors, tensors, or even higher rank fields
%[508,1242]. Of course, the effect of such additional fields needs to be suppressed at scales where General Relativity has been
%well tested, such as in the lab or solar system. This is usually achieved making couplings very weak, although novel screening
%mechanisms such as the chameleon mechanism [684,683] and Vainshtein mechanism [1236] have also been explored.
%This section represents an overview of four-dimensional gravity theories with extra fields, focusing on additional scalars,
%vectors and tensors. Wenote that some theories in other sections of this review can also be considered as theories with extra
%fields (e.g. f (R) gravity, galileons, and ghost condensates). The reader is referred to later sections for details of this.

Another feature of GR which turned to a great incentive for the creation of a large class of modified gravity theories, is that GR is the most general theory whose field equations contain at  most second order derivatives of a single metric \cite{lov-71,lov-72}. The development of generalized gravitational theories with higher-order derivatives of the metric was based largely on the hypothesis that the Einstein-Hilbert action is a simplification of a more general action which due to quantum fluctuations of spacetime contains higher power corrections. These corrections may take many different forms, and the exploration of this class of theories has given rise to a large discussion on its phenomenological implications. In the next section we will discuss more specifically this class of gravity theories since the cosmological models that will be explored during the course of this thesis will be built on a specific subset of such higher-order gravity (HOG) theories.

The systematic studies for the construction of a quantum field theory of gravity has been the third major motivation for the creation of modified gravity theories. Given that developement of Riemannian geometry is not restricted to four dimensions, the mathematical tools were already available for evolving any kind of proposal in higher dimensional spaces. The work of Kaluza and Klein already introduced in 1919 the extra dimensions in the study of gravity, and was the ideal substrate for the subsequent development of superstring and supergravity theories that followed the appearance of the notion of supersymmetry. The discovery of D-branes \cite{pol-95} made a great impact towards that direction as well. Since there exists a fundamental contradiction between the experimental behavior of gravity in more than four dimensions and the lack of ability of superstring theory to be formulated consistently in less than ten, there has been a considerable effort for the solution of that problem and various proposals have been presented to that end. The most prominent of them along with some references for more detailed descriptions being the Kaluza-Klein theories \cite{kal-21}, Randal-Sundrum gravity \cite{ran-sun-99a,ran-sun-99b}, brane-world gravity \cite{maa-04}, Dvali-Gabadadze-Porrati gravity \cite{dva-gab-por-00} and Lovelock gravity \cite{lov-71,lov-72,cha-09} among many others.

Modified theories of gravity have been developed in a multitude of directions that are impossible to meet the narrow scope of this Introduction. For a more complete description of modified gravity theories and their implications in cosmology see \cite{cli-fer-pad-sko-12,sot-07}. In the next section, we take a closer look to the higher-order gravity theories which constitute the main core of this thesis.

%=================================================
\subsection{Higher-order gravity theories}

As mentioned earlier, allowing the action to include higher-than-second derivatives of the metric, is the main feature of the large and extensively studied class of HOG theories. One of the most important advantages of such theories is the fact that they show improved renormalization properties, by allowing the graviton propagator to drop off faster in the violet spectrum. There are quite a few ways that can lead to that result, and have been followed by different researchers of this class of gravity theories. The simplest of them include either adding specific higher order scalar curvature invariants to the Einstein-Hilbert action or, a more straightforward approach, considering the Lagrangian as a general function of the scalar curvature. This latter kind of HOG theories are generally known as $f(R)$ theories and as we will see right after they show some very interesting phenomenological behavior addressing several problems of modern cosmology. $f(R)$ theories manage to avoid certain instabilities that emerge from the higher order derivatives of the metric by causing them to act in a way that turns some commonly non-dynamical sectors into dynamical ones \cite{cli-fer-pad-sko-12}. Similar approaches involve adding to the Einstein-Hilbert action general combinations of the Ricci and the Riemann curvature invariants or, as it happens in the case of Ho\v{r}ava-Lifschitz gravity \cite{hor-09a,hor-09b,hor-09c,kir-kof-09}, allowing only higher-order spatial derivatives and excluding the higher-order time derivatives in order to prevent the existence of ghost instabilities. Since the cosmological models that this thesis describes fall, as shown in Chapter two in the category of the f(R) HOG theories, we will now see these theories in greater detail.

%=================================================
\subsection{$f(R)$ Theories}

As discussed previously, $f(R)$ gravity theories belong to the general class of HOG theories that admit higher-order derivatives of the metric in the field equations. More specificaly, due to Lovelock's theorem \cite{lov-71,lov-72} f(R) theories contain up to fourth-order derivatives of the metric in their field equations. One of the arguably most important reasons which acted as a strong motive for the $f(R)$ generalizations of the Einstein-Hilbert action, is the fact, that while GR is not renormalizable and as a consequence cannot be part of a quantum gravity field theory since it cannot by quantized in a conventional way, in 1962 it was shown by Utiyama and DeWitt \cite{uti-dew-62} that by adding higher-order curvature invariants in the Einstein-Hilbert action, the theory could admit renormalization properties at one loop. A result which was confirmed in 1977 by the results of Stelle \cite{ste-77}.

In addition to that, the key role that quadratic corrections to the GR Lagrangian might play near spacetime singularities has been stressed both by the Starobinski's work \cite{sta-80} concerning the inflationary scenario for an $R+aR^2$ cosmological model, as well as by the work  Branderberger, Mukhanov and Sornborger \cite{muk-bra-92,bra-92,bra-93,bra-muk-sor-93} concerning non-singular universes.

%=================================================
\subsection{Action and field equations of $f(R)$ gravity theories in metric formalism}

In order to write the field equations in the general case of f(R) theories one has to consider an action of the form
\be \label{fRactionvac}
\mathcal{S}=\frac{1}{16\pi G}\int f(R) d\mu_{g} 
\ee
which by adding a matter term becomes
\be \label{fRactionmat}
\mathcal{S}=\frac{1}{16\pi G}\int f(R) d\mu_{g} + \mathcal{S}_{M}\left(g_{\mu\nu},\psi \right)
\ee
where $\psi$ denotes the matter fields. Variation with respect to the metric $g_{\mu\nu}$ up to the surface terms leads to \cite{buc-70}
\be \label{fRfe}
f'(R)R_{\mu\nu}-\frac{1}{2}f(R)g_{\mu\nu}- \nabla_{\mu}\nabla_{\nu}f'(R) +g_{\mu\nu}\Box f'(R) = 8\pi G T_{\mu\nu}
\ee
where 
\be
T_{\mu\nu}=\frac{-2}{\sqrt{-g}}\frac{\delta \mathcal{S}_{M}}{\delta g^{\mu\nu}}. 
\ee
A prime denotes differentiation with respect to the scalar curvature R, $\:\Box \equiv \nabla^{\mu}\nabla_{\mu}$ and $\nabla_{\mu}$ is the covariant derivative associated with the Levi-Civita connection.

In the derivation of the field equations (\ref{fRfe}) there is a number of surface area terms that occur in the same way as in the respective case of GR. Nevertheless, in GR these terms can be grouped into a total divergence which in turn can be `eliminated' by the addition of the Gibbons-Hawking-York surface term \cite{yor-72,gib-haw-77}. On the contrary in the general case of an $f(R)$ Lagrangian the surface terms cannot be obtained from a total divergence because of the $f'(R)$ term present in them \cite{sot-07, sot-far-10}. In order to avoid this, it is assumed that the higher-order derivatives included in the action, permit the fixing of more degrees of freedom on the boundary than those of the metric. Following this hypothesis we accept that the boundary terms vanish and thus, Eqs. (\ref{fRfe}) are obtained. We note also that the same reasons that apply to the field equations of GR for the general covariance, also apply for Eq. (\ref{fRfe}). 

Subsequently, there is another important issue concerning the derivation of the field equations (\ref{fRfe}) that has to be addressed. It is well known \cite{mis-tho-whe-73,wal-84} that the field equations of GR can be obtained by applying two different variational principles to the Einstein-Hilbert action. The first of the them is the \emph{metric variation} where it is accepted that the connection is the Levi-Civita one and the variation takes place with respect to the metric. This was also the way that Eqs. (\ref{fRfe}) where obtained by the action (\ref{fRactionmat}). The second way is the one termed the \emph{Palatini variation} which is performed under the hypothesis that the metric and the connection are independent and as such, the variation of the action should be performed with respect to both variables. While the application of the Palatini variation in the case of GR leads to the demand that the connection is the Levi-Civita one and hence both variational principles lead to the Einstein equations, this is not the case in the $f(R)$ theories of gravity, where the two different variational principle lead to two completely different field equations. 
Although the variational principle which will be used through out this thesis is the standard metric one, we give here the forms of the field equations in the Palatini version \cite{sot-far-10},
\begin{eqnarray}
f'(\mathcal{R})\mathcal{R}_{(\mu\nu)}-\frac{1}{2}f(\mathcal{R})g_{\mu\nu}&=&8\pi G T_{\mu\nu},\nonumber\\[10pt]
\bar{\nabla}_{\lambda}\left[\sqrt{-g}f'(\mathcal{R})g^{\mu\nu}\right]&=&0
\end{eqnarray}
where $(\mu\nu)$ shows symmetrization over the indices $\mu$ and $\nu$, $\bar{\nabla}_{\mu}$ denotes the covariant derivative defined by the independent connection $\Gamma^{\lambda}_{\:\mu\nu}$ and $\mathcal{R}$ is the scalar curvature constructed with that same connection.
In addition to that there is even a third variational method, the \emph{metric-affine variation}, that occurs in the case of the Palatini variation with the additional assumption that the matter action is also depended on the connection \cite{sot-lib-07,sot-far-10}. For an overview of the $f(R)$ gravity theories see \cite{noj-odi-07,fel-tsu-10,sot-far-10,sch-07,noj-odi-11}.

%

%----------------------------------------------------------------------------------------
%	SECTION  -  Stability
%----------------------------------------------------------------------------------------  
\section{Asymptotic stability}

Cosmology in HOG theories has advanced into a field of special importance for a number of interesting questions concerning the structure of the early universe. More specifically, an already formed subfield of HOG theories involves the stability of cosmological solutions in generalized gravity theories. This area combines the search for fundamental physical phenomena that might have taken place during the earliest moments of the universe with intriguing mathematical problems. There exist two main facets of the cosmological stability problem. The first is the perturbation theory aspect which plays an important role in studies of structure formation. The second is the asymptotic stability aspect which is used mainly for issues involving geometric and nonlinear dynamics. The latter problem naturally comes to surface when we investigate whether or not there is a possibly more general significance in a given exact solution of the field equations, whether there are any common properties in a certain \emph{set} of solutions, or whether we need to understand what will be the fate, if we wait long enough, of a particular universe in this context.

The asymptotic stability of homogeneous and isotropic solutions of cosmological models in HOG theories is a problem that has two major aspects itself. On one hand, there is the question about the \emph{late-time stability}  -that is, deciding the behavior of these universes in the distant future. On the other hand, there is the \emph{early-time stability} which is about examining the evolution towards the past, at early times, in the neighborhood of a possible initial singularity. The seminal work of Barrow and Ottewill \cite{bar-ott-83} examined the issue of existence and stability of various cosmological solutions emphasizing on the de Sitter and FRW ones and renewed interest in late-time evolution cosmological problems in higher order gravity (for related general late-time stability results for FRW universes in the same context see also \cite{cot-fle-93a}). 

The \emph{cosmic no-hair conjecture} is one such interesting late-time stability problem that has received a lot of attention. The original cosmic no-hair theorems in HOG theories, have been evolved in \cite{mae-89,mij-ste-88,cou-mad-89}, while one can see \cite{bar-87a,bar-87b,bar-her-06a} for limitations of this property. For generalizations in higher-order gravity of the respective situation that emerges in general relativistic cosmology, cf. \cite{pag-87,mul-sch-sta-88,sch-88,ber-90,ber-91,cot-fle-93b,mir-cot-96}.

Another important late-time stability issue is the \emph{recollapse problem}. In \cite{cot-mir-96,cot-mir-98} various recollapse theorems in generalized cosmological theory are examined, while \cite{mir-03,mir-05,mir-09} include more elaborate and complete  approaches and results on this subject. It is a probably true that cosmic no-hair and recollapse of closed models are not unrelated. A ``premature'' recollapse problem in closed universes that inflate has been formulated and studied in \cite{bar-88} in an interesting approach of those two issues.

The problem of the early-time evolution for homogeneous and isotropic universes in HOG theories is concerned with clarifying the different possible behaviors that could exist on approach to the initial singularity. In the first papers on this issue, cf. \cite{ruz-ruz-70,ker-82} there was already present a duality between bouncing and singular early time solutions. While in \cite{fre-bre-82} these first solutions were shown to have the impressive characteristic of being horizon-breaking, it was later shown that they could be unstable \cite{rot-ann-91}. These first results were the ones that also led to the better understanding of the fact that the problem of the possible \emph{early time asymptotes} of the admissible cosmological solutions of the higher order gravity equations was more involved \cite{mul-sch-85,mul-86}. In fact, it became quite clear that even from the simplest `radiation fluid' solution $t^{1/2}$, a completely different set of properties than the corresponding situation in GR could be obtained, since it is a solution in both the radiation filled case \emph{and} in vacuum in these theories. 

Conditions concerning instability as well as general properties of the early-time stability of the flat and the curved, radiation-filled, isotropic solutions were studied in \cite{cot-fle-95}. In that paper, there are various possible \emph{stability} results that one could derive from the general conditions and equations by the appropriate selection of various constants in order of specific forms to be taken, although the interest of the study was in finding certain instability properties of these systems as they advance towards the initial singularity. We recall that, with respect to any kind of perturbation, the corresponding radiation solutions in GR are unstable and also non generic, cf. \cite{lan-lif-75}.

In the interesting as well as important works \cite{bar-mid-07,bar-mid-08}, stable solutions in vacuum were found in the neighborhood of the initial singularity in the case of a flat isotropic cosmology with a term of the form $\textrm{Ric}^{2n}, n\in\mathbb{Q}$, added in the basic quadratic Lagrangian $R+\a R^2$. In these two papers, a linear perturbation analysis of the $t^{1/2}$ solution in vacuum, \emph{flat} FRW universes is used as a method to show that the various perturbations vanish asymptotically at early times. 

It has been shown that this vacuum solution is stable under anisotropic, spatially homogeneous  perturbations, cf. \cite{cot-dem-der-que-93,bar-her-06b}. Therefore, the precise extent that a generic perturbation of the flat, vacuum, $t^{1/2}$ solution occupies in the whole space of solutions of the higher order gravity equations is an interesting open question.

As a consequence, we can distinguish two separate asymptotic problems concerning the early-time stability of the flat FRW $t^{1/2}$ solution in HOG. One one hand, we need to examine its stability as a solution of the vacuum field equations, and, on the other hand, that as solution of the HOG field equations filled with a radiation fluid. Both problems need to be in all different levels of stability. We know \cite{cot-tso-07} that, in four spacetime dimensions, the flat, radiation solution is asymptotically stable at early times in the space of all flat solutions of the theory. What remains to be examined is the precise behavior of this solution consecutively with respect to curved FRW perturbations, anisotropic perturbations and generic inhomogeneous perturbations. As far as it regards the vacuum early-time problem, as noted above, there are clear indications that the flat, vacuum solution is stable with respect to various FRW and anisotropic perturbations, the strongest known results being its stability with respect to anisotropic perturbations \cite{bar-her-06b}, and with respect to perturbations in the $R+\b R^2$ action by adding a term of the form $\textrm{Ric}^{2n}$, cf. \cite{bar-mid-07,bar-mid-08}.

In GR, one cannot trivially obtain vacuum states for simple isotropic universes. We have to go beyond them to anisotropic, or more general inhomogeneous cosmologies for a vacuum to start making sense \cite{cho-09}. Nevertheless, in effective theories with higher derivatives, isotropic vacua are very common, see e.g.,  \cite{sta-80,bar-mid-08}. Such classical vacua are usually thought of as acquiring a physical significance when viewed as possible low-energy manifestations of a more fundamental superstring theory, although their treatment shows an intrinsic interest quite independently of the various quantum considerations.

In the first part of this thesis, we consider the possible asymptotic limits towards singularities of vacuum universes coming from effective theories with higher derivatives. Such a study is  related to the existence and stability of an inflationary stage at early times in these contexts, and also to the intriguing possibilities of solutions with no particle horizons. For flat vacua, we find the general asymptotic solution with an early-time singularity. This result is then extended to cover general curved vacuum isotropic  solutions and we give the precise form of the attractor of all such universes with a past singularity. We also obtain special asymptotic states valid specifically for open or closed vacua starting from lower-dimensional initial data. These results have a potential importance for the ekpyrotic and cyclic scenarios as they strongly point to the dynamical stability of the reversal phase under higher derivative corrections in these universes.

Subsequently, we treat the problem of the early-time behavior of the flat radiation $t^{1/2}$ solution of higher order gravity with respect to curved FRW perturbations. That is, considered as a solution of the curved FRW equations for the  $R+\b R^2$ action, what is the behavior as we approach the initial singularity, i.e., as $t\rightarrow 0$ of all solutions  which are initially (that is, for some $t^*>0$) near this radiation solution? For this purpose, as we will see in the next section, we approach the problem via the use of the \emph{method of asymptotic splittings} developed in \cite{cot-bar-07,gor-01}, and trace all possible asymptotic behaviors that solutions to the higher-order curved FRW equations may develop at early times. Following this geometric approach, we are able to show that the exact radiation solution is stable asymptotically at early times, meaning that the initial state of these universes proves to be a very simple one indeed. Given that this theory is known to admit an inflationary stage \cite{sta-80}, this also means that any pre-inflationary period in such universes is necessarily  isotropic and flat.

%--------------------------------------------------------------------------------------------------------------------
%	SECTION   - Definition of Asymptotic Solution
%--------------------------------------------------------------------------------------------------------------------

\section{Asymptotic solutions}

In this thesis we are particularly interested in the behavior of quadratic, vacuum or radiation-filled universes in the neighborhood of the initial singularity, taken at $t=0$. The position of the initial singularity is really arbitrary. We could have placed it at any $t_0$ and used the variable $\tau=t-t_0$ instead of $t$. We will fully describe any such initial state by giving the possible modes of approach of the various solutions to it. These modes are, subsequently, identified by the behavior of the corresponding vector field near the initial singularity. In order to find this behavior, we shall use the method of \emph{asymptotic splittings}, cf. \cite{cot-bar-07,gor-01}. According to this method, the associated vector fields are asymptotically decomposed in such a way as to reveal their most important dominant features on approach to the singularity. This leads to a detailed construction of all possible local asymptotic solutions valid near the finite-time singularity. These, in turn, provide  a most accurate picture of all possible dominant features that the fields possess as they are driven to a blow up (for previous applications of this asymptotic technique to cosmological singularities, apart from \cite{cot-tso-07}, we refer to  \cite{ant-cot-kla-08,ant-cot-kla-10,cot-kit-12,ant-cot-kla-12}).

It is expected that the vector fields describing the evolution of this class of universes will show some dominant features as on approach to the finite-time singularity at $t=0$, and these will correspond to the different, inequivalent ways that it splits in the neighborhood of the blow up. We need two definitions to describe the situation precisely. Firstly, we say
that a solution $b(t)$ of the dynamical system describing the evolution of cosmological model is \emph{asymptotic} to another solution
$a(t)$ provided that the following two conditions hold (the first is
subdivided):
\begin{enumerate}
\item[(i)] Either $(1)$ $a(t)$ is an exact solution of the system, or
 $(2)$ $a(t)$ is a solution of the system (substitution gives $0=0$) as $t\rightarrow\infty$,
\item [(ii)] $b(t)=a(t)[1+g(t)],\, g(t)\rightarrow 0$, as $t\rightarrow\infty$.
\end{enumerate}
If either of these two conditions is not satisfied, then $b(t)$ cannot be asymptotic to $a(t)$. Additionaly,  a solution of the dynamical system is called \emph{dominant} near the singularity if, for constants $\mathbf{a}=(\theta, \eta, \rho)\in\mathbb{C}^3$, and $\mathbf{p}=(p, q, r)\in\mathbb{Q}^3$, it is asymptotic to the form
\be \label{domsol3}
\mathbf{x}(t)=\mathbf{a}t^{\mathbf{p}}=(\theta t^{p}, \eta t^{q}, \rho t^{r}). 
\ee
For any given dominant solution of a dynamical system describing our universe near the singularity, we call the pair $(\mathbf{a},\mathbf{p})$ a \emph{dominant balance} of the associated vector field.

Near their blow up singularities, vector fields are characterized by dominant balances and the corresponding asymptotic integral curves. By a solution with a finite-time singularity we mean one where there is a time at which at least one of its components diverges. It must be noted that the usual dynamical systems analysis through linearization is not relevant here, for in that one does not deal with singularities but with equilibria. Each vector field $\mathbf{f}$ itself is decomposed asymptotically into a dominant part and another, subdominant part:
\be\label{general split3}
\mathbf{f}=\mathbf{f}^{(0)} + \mathbf{f}^{\,(\textrm{sub})},
\ee
and such a candidate asymptotic splitting (or decomposition) needs to be checked for consistency in various different ways before it is to be admitted as such.
By direct substitution of the dominant balance forms in our system, we look for the possible scale invariant solutions of the system. A vector field $\mathbf{f}$ is called \emph{scale invariant} if $\mathbf{f}(\mathbf{a}\t^{\mathbf{p}})=\t^{\mathbf{p-1}} \mathbf{f}(\mathbf{a})$, for a more detailed treatment, cf. \cite{cot-bar-07}.

%=================================================
\section{Structure of this Thesis}

The structure of this Thesis is as follows.

In the first chapter we introduce the reader to the broader field of generalized gravitational theories and describe more specifically the $f(R)$ theories of gravity and more general theories that include various combinations of the Ricci and the Riemann curvatures. We write the field equations in the general case and we close with a discussion for the asymptotic stability of the solutions of these theories. Additionally, we analyze some specific concepts that will help in understanding the specific conclusions of this thesis.

In the second chapter we present the general field equations of vacuum $f(R)$ gravity theories focusing in the case of homogeneous and isotropic cosmological models. We consider the equivalent autonomous dynamical system and the corresponding vector field and give an outline of the mathematical method we will use for the asymptotic analysis of the behavior of these cosmological models in the neighborhood of the initial singularity.

The third chapter contains our analysis of the asymptotic behavior of the solutions of the theory as we approach the singularity. In particular we analyze all possible cases in which the vector field or the equivalent autonomous dynamical system may decompose asymptotically. We present qualitative as well as analytical arguments in order to decide which cases show a dominant asymptotic behavior of the dynamical system and lead to the construction of asymptotic solutions of the field equations. Then we proceed to the construction of these asymptotic solutions in the form of Fuchsian series for the case of vacuum flat universes.

In the fourth chapter we proceed to the study of the case of vacuum curved cosmological models for homogeneous and isotropic universes. Following again the method of asymptotic splittings we construct asymptotic solutions of these cosmologies in the neighborhood of the initial singularity after having analyzed extensively all the possible ways of approaching the singularity. Additionally we make specific comments on the stability of these solutions in the context of the gradual modification of the specific characteristic of curvature.

In the fifth chapter the quadratic, curved, radiation-filled isotropic and homogeneous cosmologies are studied. We investigate thoroughly the asymptotic form of specific solutions in the neighborhood of the singularity emphasizing in the way that the characteristics of curvature and radiation interfere with the asymptotic behavior of these cosmologies. We compare these new results with those of the previous chapters. In any case we draw conclusions about the stability of the solutions found based on their final form as Fuchsian series.

In the final chapter we summarize the conclusions of this work and we further make various general remarks on the asymptotic behavior of the cosmologies we studied in previous chapters. Our approach allows us to examine how this behavior changes with the gradual addition of certain features, such as the curvature and the radiation. We also correlate our results with other gravitational theories and examine cosmological models related to those we have seen here. We conclude this work with a discussion about possible open problems emerging from our current results and overall approach.

% Chapter Template

\chapter{The basic vacuum vector fields} % Main chapter title

\label{Chapter2} % Change X to a consecutive number; for referencing this chapter elsewhere, use \ref{ChapterX}

\lhead{Chapter 2. \emph{The basic vacuum vector fields}} % Change X to a consecutive number; this is for the header on each page - perhaps a shortened title

%--------------------------------------------------------------------------------------------------------------------
%   Introduction
%--------------------------------------------------------------------------------------------------------------------

In this chapter we derive the basic dynamical systems and the equivalent vector fields which describe the dynamical evolution of any vacuum FRW universe in higher order gravity.

%--------------------------------------------------------------------------------------------------------------------
%	SECTION  - Field equations
%--------------------------------------------------------------------------------------------------------------------

\section{Field equations}

This section is devoted to the derivation of the field equations coming from the most general quadratic action in four dimensions.
As a first step, we start by giving some basic properties of the general form of Lagrangian density which includes all possible curvature invariants as quadratic corrections to terms linear in the scalar curvature $R$,
\be
\label{eq:action}
\mathcal{S}=\int_{\mathcal{M}}\mathcal{L}(R)d\mu_{g},
\ee
where
\be
\label{eq:lagra}
\mathcal{L}(R)=\mathcal{L}(0)+ aR + bR^2 + cR^{\mu\nu}R_{\mu\nu} + dR^{\mu\nu\kappa\lambda}R_{\mu\nu\kappa\lambda},
\ee
where $\mathcal{L}(0), a, b, c, d$ are constants and $\mathcal{L}(0)$ plays the role of the cosmological constant. By a (Riemannian) curvature invariant we mean a smooth function of the metric $g_{\mu\nu}$ and its derivatives which is a local invariant under smooth coordinate transformations (diffeomorphisms)\footnote{In distinction, a smooth function of the metric which is invariant under conformal transformations of the metric is called a local \emph{conformal} invariant.}.
The fact that the Lagrangian (\ref{eq:lagra}) contains, in addition to the last three quadratic curvature invariant terms, the two terms $\mathcal{L}(0)$ and $aR$, implies the basic fact that this theory cannot be scale invariant. The reason behind this is the fact that because of the presence of the first two terms (\ref{eq:lagra}) cannot be a homogeneous polynomial in the derivatives of the metric.

Another property of the gravity theory defined by (\ref{eq:lagra}) is that not all quadratic curvature invariants appearing in it are algebraically independent \cite{ste-58,hig-59}.

To see this, we can use the following simple variational argument. We consider, as usual, a family of metrics $\{g_s:s\in\mathbb{R}\}$, and denote its compact variation by $\dot{g}_{\mu\nu}=(\partial g/\partial s)_{s=0}$ (cf. e.g., \cite{haw-ell-73} page 65). Since in four dimensions we have the Gauss-Bonnet identity,
\be\label{eq:gentity}
\dot{\mathcal{S}}_{GB}= \int_{\mathcal{M}}(R^2_{GB}d\mu_{g})^{\cdot}=0,\quad
R^2_{GB}=R^2 - 4\textrm{Ric}^2 + \textrm{Riem}^2,
\ee
it follows that in the derivation of the field equations through a $g$-variation  of the action (\ref{eq:lagra}),
only terms up to $\textrm{Ric}^2$ will matter. In particular, because of Eq. (\ref{eq:gentity}) there is no necessity to include the $\textrm{Riem}^{2}$ term.

Hence, we may replace (\ref{eq:lagra}) by the following gravitational action in four dimensions in which the curvature invariants are algebraically independent (we set $8\pi G=c=1$, and the sign conventions are those of \cite{mis-tho-whe-73}),
\be\label{eq:genlagrareduced}
\mathcal{S}=\int_{\mathcal{M}}\mathcal{L}(R)d\mu_{g}, \quad\mathcal{L}(R)=\quad\mathcal{L}(0)+\alpha R +\beta R^2 +\gamma R^{\mu\nu}R_{\mu\nu},
\ee
where $\alpha=a$, $\beta=b-1$ and $\gamma=c+4d$. Each one of the terms in the action (\ref{eq:genlagrareduced}) leads to the following variations\footnote{We use $\delta$ to mean a compact variation of the fields, `$\cdot$', as above.}:
\be\label{eq:Term1variation}
\delta\int\quad\mathcal{L}(0)d\mu_{g} =\int\mathcal{L}(0)g^{\mu\nu}\delta g_{\mu\nu}d\mu_{g},
\ee

\be\label{eq:Term2variation}
\delta\int\alpha R d\mu_{g}=\alpha\int\left(R_{\mu\nu}-\frac{1}{2}g_{\mu\nu}\delta g^{\mu\nu}\right)d\mu_{g}
+\alpha\int_{\partial_{\mathcal{M}}}g^{\mu\kappa}\delta R_{\mu\kappa}dS_{\mu},
\ee

\begin{eqnarray}\label{eq:Term3variation}
\delta\int\beta R^2d\mu_{g}&=&\beta\int [-2RR^{\mu\nu}+\frac{1}{2}R^2g^{\mu\nu} \nonumber \\[15pt] 
&+& 2\left(g^{\mu\kappa}g^{\nu\lambda}-g^{\mu\nu}g^{\kappa\lambda}\right)\nabla_{\kappa}\nabla_{\lambda}R] \delta g^{\mu\nu}\delta g_{\mu\nu}d\mu_{g}\nonumber \\[15pt] 
&+& \beta \int_{\partial_{\mathcal{M}}} 2[R\nabla^{\nu}\delta g_{\mu\nu}-\nabla^{\nu}R \delta g_{\mu\nu} \nonumber \\[15pt] 
&-&R\nabla_{\mu}\left(g^{\nu\rho}\delta g_{\nu\rho} \right)+\nabla_{\mu}R g^{\nu\rho}\delta g_{\nu\rho}]dS_{\mu},
\end{eqnarray}

\begin{eqnarray}\label{eq:Term4variation}
\delta\int\gamma R^{\mu\nu}R_{\mu\nu}d\mu_{g}&=&\gamma\int(-2R^{\mu\lambda}R^{\nu}_{\lambda}\delta g_{\mu\nu}+\frac{1}{2}g^{\mu\nu}R^{\kappa\lambda}R_{\kappa\lambda}\nonumber\\[15pt]
&-&\nabla_{\kappa}\nabla^{\kappa}R^{\mu\nu}-g^{\kappa\lambda}\nabla_{\lambda}\nabla_{\kappa}R^{\kappa\lambda}+2\nabla_{\kappa}\nabla^{\nu}R^{\kappa\mu})\delta g_{\mu\nu}d\mu_{g} \nonumber\\[15pt]
&+&\gamma \int_{\partial_{\mathcal{M}}} ( R^{\kappa\lambda}g^{\alpha\beta}\nabla_{\kappa}\delta g_{\lambda\beta}+R^{\kappa\lambda}g^{\alpha\beta}\nabla_{\lambda}\delta g_{\kappa\beta}-R^{\kappa\lambda}\nabla^{\alpha}\delta g_{\kappa\lambda}\nonumber\\[15pt]
&+&\nabla^{\alpha}R^{\kappa\lambda}\delta g_{\kappa\lambda}+g^{\mu\nu}\nabla_{\kappa}R^{\kappa\alpha}\delta g_{\mu\nu}-R^{\alpha\lambda}g^{\mu\nu}\nabla_{\lambda}\delta g_{\mu\nu}\nonumber\\[15pt]
&-&2\nabla^{\beta}R^{\alpha\lambda}\delta g_{\lambda\beta})dS_{\mu},
\end{eqnarray}

The boundary integrals in Eqs. (\ref{eq:Term2variation}), (\ref{eq:Term3variation}) and (\ref{eq:Term4variation}) can be set equal to $\int_{\partial_{\mathcal{M}}}\Phi^{\mu}dS_{\mu}$, $\int_{\partial_{\mathcal{M}}}X^{\mu}dS_{\mu}, \int_{\partial_{\mathcal{M}}}\Psi^{\mu}dS_{\mu}$ respectively with the vector fields $\Phi^{\mu}$, $X^{\mu}$ and $\Psi^{\mu}$ given by

\begin{eqnarray}
\Phi^{\mu}&=&g^{\mu\kappa}\delta R_{\mu\kappa},\label{eq:boundary2}\\[20pt]
X^{\mu}&=&2[R\nabla^{\nu}\delta g_{\mu\nu}-\nabla^{\nu}R \delta g_{\mu\nu} \nonumber \\[15pt] 
&-&R\nabla_{\mu}\left(g^{\nu\rho}\delta g_{\nu\rho} \right)+\nabla_{\mu}R g^{\nu\rho}\delta g_{\nu\rho}],\label{eq:boundary3}\\[20pt]
\Psi^{\mu}&=&R^{\kappa\lambda}g^{\alpha\beta}\nabla_{\kappa}\delta g_{\lambda\beta}+R^{\kappa\lambda}g^{\alpha\beta}\nabla_{\lambda}\delta g_{\kappa\beta}-R^{\kappa\lambda}\nabla^{\alpha}\delta g_{\kappa\lambda}\nonumber\\[15pt] 
&+&\nabla^{\alpha}R^{\kappa\lambda}\delta g_{\kappa\lambda}+g^{\mu\nu}\nabla_{\kappa}R^{\kappa\alpha}\delta g_{\mu\nu}-R^{\alpha\lambda}g^{\mu\nu}\nabla_{\lambda}\delta g_{\mu\nu}\nonumber\\[15pt] 
&-&2\nabla^{\beta}R^{\alpha\lambda}\delta g_{\lambda\beta}.\label{eq:boundary4}
\end{eqnarray}
These integrals are zero since the compact variation $\dot{g}_{\mu\nu}=(\partial g/\partial s)_{s=0}$ vanishes at the boundary $\partial_{\mathcal{M}}$. So, we have
\be
\int_{\partial_{\mathcal{M}}}(\alpha\Phi^{\mu}+\beta X^{\mu}+\gamma\Psi^{\mu})dS_{\mu}=0
\ee
Accordingly, the field equations that stem from the variation of the gravitational action (\ref{eq:genlagrareduced}) read as follows:
\begin{eqnarray}\label{eq:basicfe1}
\frac{1}{2}\mathcal{L}(0)g^{\mu\nu}&-&\alpha(R^{\mu\nu}-\frac{1}{2}g^{\mu\nu}R)\nonumber\\[15pt]
&+&\beta[-2RR^{\mu\nu}+\frac{1}{2}R^2g^{\mu\nu}+2(g^{\mu\kappa}g^{\nu\lambda}-g^{\mu\nu}g^{\kappa\lambda})\nabla_{\kappa}\nabla_{\lambda}R]\nonumber\\[15pt]
&+&\gamma(-2R^{\mu\lambda}R^{\nu}_{\lambda}+\frac{1}{2}g^{\mu\nu}R^{\kappa\lambda}R_{\kappa\lambda}-\nabla_{\kappa}\nabla^{\kappa}R^{\mu\nu}\nonumber\\[15pt]
&-&g^{\kappa\lambda}\nabla_{\lambda}\nabla_{\kappa}R^{\kappa\lambda}+2\nabla_{\kappa}\nabla^{\nu}R^{\kappa\mu})=0
\end{eqnarray}
By taking $\alpha=1$ (since $8\pi G=c=1$) and $\mathcal{L}(0)=0$, in order to focus in the case where the cosmological constant vanishes, we get:
\begin{eqnarray}\label{eq:basicfe2}
R^{\mu\nu}&-&\frac{1}{2}g^{\mu\nu}R-\beta[-2RR^{\mu\nu}+\frac{1}{2}R^2g^{\mu\nu}+2(g^{\mu\kappa}g^{\nu\lambda}-g^{\mu\nu}g^{\kappa\lambda})\nabla_{\kappa}\nabla_{\lambda}R]\nonumber\\[15pt]
&-&\gamma(-2R^{\mu\lambda}R^{\nu}_{\lambda}+\frac{1}{2}g^{\mu\nu}R^{\kappa\lambda}R_{\kappa\lambda}-\nabla_{\kappa}\nabla^{\kappa}R^{\mu\nu}\nonumber\\[15pt]
&-&g^{\kappa\lambda}\nabla_{\lambda}\nabla_{\kappa}R^{\kappa\lambda}+2\nabla_{\kappa}\nabla^{\nu}R^{\kappa\mu})=0
\end{eqnarray}

%--------------------------------------------------------------------------------------------------------------------
%	SECTION   - The vacuum vector fields
%--------------------------------------------------------------------------------------------------------------------
\section{The vacuum vector fields}

We consider a vacuum, FRW universe with scale factor $a(t)$ determined by the Friedmann-Robertson-Walker metric of the form
\be \label{rwmetrics}
g_{4}=-dt^{2}+a^{2}\, g_{3}.
\ee
Each slice is given the 3-metric
\be
g_{3}=\frac{1}{1-kr^2}dr^{2}+r^2g_{2},
\ee
$k$ being the (constant) curvature normalized to take the three values $0, +1$ or $-1$ for the complete, simply connected, flat, closed or open space sections respectively, and the 2-dimensional sections are such that
\be
g_{2}=d\theta^{2}+\sin^{2}\theta d\phi^{2}.
\ee
Below we focus on the case where $\mathcal{M}$ is a homogeneous and isotropic universe with the FRW metric (\ref{rwmetrics}). 
In this case, it is well known \cite{bar-ott-83} that the following identity holds: 
\be
\int_{\mathcal{M}} ((R^2 - 3\textrm{Ric}^2)d\mu_{g})^{\cdot}=0 . \:
\label{eq:isontity}
\ee
This further enables us to combine the contributions of the $\textrm{Ric}^2$ and the $R^2$ terms into  (\ref{eq:basicfe1}), altering only the arbitrary constants. Consequently, the field equations (\ref{eq:basicfe1}) will become:
\begin{eqnarray}\label{eq:fe1}
\frac{1}{2}\mathcal{L}(0)g^{\mu\nu}&-&\alpha(R^{\mu\nu}-\frac{1}{2}g^{\mu\nu}R)+(\beta+\frac{1}{3}\gamma)[-2RR^{\mu\nu}+\frac{1}{2}R^2g^{\mu\nu}+\nonumber \\[15pt]
&+&2(g^{\mu\kappa}g^{\nu\lambda}-g^{\mu\nu}g^{\kappa\lambda})\nabla_{\kappa}\nabla_{\lambda} R]=0
\end{eqnarray}
Finally, the field equations derived from the variation of the gravitational action (\ref{eq:genlagrareduced}) have the following form:
\be
R^{\mu\nu}-\frac{1}{2}g^{\mu\nu}R+
      \frac{\xi}{6} \left[2RR^{\mu\nu}-\frac{1}{2}R^2g^{\mu\nu}-2(g^{\mu\rho}g^{\nu\s}-g^{\mu\nu}g^{\rho\s})\nabla_{\rho}\nabla_{\s}R \right]=0,
\label{eq:fe}
\ee
where we have set  
\be
\xi=2(3\beta+\gamma).
\label{eq:xi}
\ee
We note that (\ref{eq:fe}) is identical to the field equations that result from the variation of the purely quadratic action
\be
\label{eq:Rsquareaction}
\mathcal{S}=\int_{\mathcal{M}}\mathcal{L}(R)d\mu_{g},
\ee
where
\be
\label{eq:Rsquare}
\mathcal{L}(R)=R + \zeta R^2,
\ee
with $\zeta$ arbitrary. However, this is not quite so true because the parameter $\xi$ in (\ref{eq:xi}) depends not only on the coefficient $b=\beta+1$ multiplying $R^2$ in (\ref{eq:lagra}) but also, through $\gamma=c+4d$, on the coefficients $c$ and $d$ of $\textrm{Ricci}^2$ and $\textrm{Riem}^2$. Therefore we conclude that because of the form of the coefficient $\xi$, some `memory' of the original fully quadratic theory (\ref{eq:lagra}) remains, and the final effective action leading to the field equations (\ref{eq:fe}) is not equivalent to a `standard' $R+\zeta R^2$ action with $\zeta$ \emph{arbitrary}, but here $\zeta$ is a function depending on $\beta$ and $\gamma$, given by (\ref{eq:xi}). A use of the former action, in the present context, would imply taking into account only the \emph{algebraic} dependence of the action on the quadratic curvature invariants with $\zeta$ being a free parameter of the theory instead of a function of $\beta=b-1$ and $\gamma=c+4d$ as it actually is.

We now proceed to the derivation of the field equations for the class of universes in question. Eq. (\ref{eq:fe}) naturally splits into $00$- and $ij$-components ($i,j=1,2,3$). Using  the metric (\ref{rwmetrics}), the field equation (\ref{eq:fe}) takes the following form (from now on an overdot denotes differentiation with respect to the proper time, $t$) for the $00$- and $ij$-components respectively,
\be
\frac{k+\dot{a}^2}{a^2}+\xi\left[2\: \frac{\dddot{a}\:\dot{a}}{a^2} + 2\:\frac{\ddot{a}\dot{a}^2}{a^3}-\frac{\ddot{a}^2}{a^2} -3\:
\frac{\dot{a}^4}{a^4} -2k\frac{\dot{a}^2}{a^4} + \frac{k^2}{a^4}\right] = 0,
\label{eq:fe2}
\ee
\begin{eqnarray}
-2\frac{\ddot{a}}{a}-\frac{\dot{a}^2}{a^2}-\frac{k}{a^2}+\xi[2\frac{{a}^{(4)}}{a}&+&12\frac{\dot{a}^2\ddot{a}}{a^3}-4\frac{\dot{a}\dddot{a}}{a^2}-3\frac{\ddot{a}^2}{a^2}-3\frac{\dot{a}^4}{a^4}+ \nonumber\\[15pt]
&+&\frac{k^2}{a^4}+4k\frac{\ddot{a}}{a^3}-2k\frac{\dot{a}^2}{a^4}] = 0.
\label{eq:ijcomponent}
\end{eqnarray}

Due to symmetry reasons, it is sufficient to use the 00-component, (\ref{eq:fe2}), as the only field equation \cite{ker-82,bar-ott-83}.
 
In what follows, we are interested in tracing all possible vacuum asymptotics, especially those solutions for which curvature and  vacuum enter in the dominant part of the vector field asymptotically. In order to do that, we will introduce new variables and write Eq.(\ref{eq:beq}) below as an autonomous dynamical using the method of asymptotic splittings presented in \cite{cot-bar-07}. 

We expect that in terms of suitable variables, the eventual dynamical system which will emerge during this process of reduction will show novel asymptotes for the vacuum problem that are not obtainable from the radiation problem (cf. Chapter 4) when letting the radiation terms tend to zero. These new asymptotes will only be possible in decompositions allowing the curvature as well as other terms characterizing the vacuum state be present in the \emph{dominant} part of the field asymptotically, something impossible in the radiation problem. 

We also expect to find other decompositions in the new variables which will indeed lead to vacuum solutions obtained from radiation ones by letting suitable radiation terms tend to zero and these solutions will exactly correspond to, and stem from, decompositions having the curvature and vacuum terms only in the subdominant part asymptotically.

For the differential equation (\ref{eq:fe2}) we can obtain these new variables as follows. First, we rewrite (\ref{eq:fe2}) using the Hubble expansion rate $H=\dot{a}/a$, in the form
\be \ddot{H}=\frac{1}{2}\frac{\dot{H}^2}{H}-3H\dot{H}+\frac{k}{a^2}H-\frac{1}{2}\frac{k^2}{a^4}\frac{1}{H}
-\frac{1}{12\epsilon}H-\frac{k}{12\epsilon a^2}\frac{1}{H}
\label{eq:beq}
\ee
where now we have put $\epsilon =\xi/6$. We then introduce new variables for the present problem by setting 

\be\label{vars1}
x=H,\hspace{0.2cm} y=\dot{H},\hspace{0.2cm} z=a^{-2}.
\ee
Then Eq. (\ref{eq:beq})  can be written as an autonomous dynamical system in the  form
\begin{eqnarray}
\label{eq:ds}
\dot{x} &=& y \:\:\:\:\:\nonumber \\[15pt]
\dot{y} &=& \frac{y^{2}}{2x}-3xy+kxz-\frac{k^2 z^2}{2x}- \frac{x}{12\epsilon}-\frac{kz}{12\epsilon x}\:\:\:\:\:\label{basic dynamical system1}  \\[15pt]
\dot{z} &=& -2xz.\:\:\:\:\:\nonumber
\end{eqnarray}
This can be expressed equivalently  as a \emph{vacuum}, 3-dimensional  vector field $\mathbf{f}_{\textsc{VAC}}:\mathbb{R}^3\rightarrow\mathbb{R}^3$ with
\be\label{basic dynamical system}
\mathbf{\dot{x}}=\mathbf{f}_{\textsc{VAC}}(\mathbf{x}),\quad \mathbf{x}=(x,y,z),
\ee
and
\be\label{vf}
\mathbf{f}_{\textsc{VAC}}(x,y,z)=\left( y,\frac{y^{2}}{2x}-3xy+kxz-\frac{k^2 z^2}{2x}- \frac{x}{12\epsilon}-\frac{kz}{12\epsilon x},-2xz \right).
\ee
This  vector field completely describes the dynamical evolution of a vacuum, flat or curved, FRW universe in the gravity theory defined by the full quadratic action (\ref{eq:lagra}). We shall assume that $x\neq 0$, that is $\dot{a}/a \neq 0$, that is we consider only non-static universes in sharp contrast to the situation in GR (cf. \cite{one-83}).

%--------------------------------------------------------------------------------------------------------------------
%	SECTION  - The solution space of higher order gravity
%--------------------------------------------------------------------------------------------------------------------

\section{The solution space of higher-order gravity}

There are various instances that indicate that the solution spaces of GR and HOG theories derived from the action (\ref{eq:action}) are not identical. In particular, the solution space of HOG includes that of GR. In the present case, for example, we can see that by setting $\rho=p=0$ in the standard FRW cosmological equations  \cite{one-83} which govern an FRW universe with a perfect fluid in GR,
\be
\frac{\rho}{3}=H^2+\frac{k}{a^2}
\ee
and
\be
\dot{\rho}=-3H(\rho+p)
\ee
where $p$ is the pressure and $\rho$ is the mass-energy density of the fluid, one cannot obtain expanding solutions for all $k$, that is solutions with $x = \dot{a}/a > 0$. On the contrary, in equation (\ref{eq:beq}) there is room for assuming the existence of non-static, vacuum solutions whereas this is impossible in GR. So the ability to obtain non-static solutions leads to the fact that the space of cosmological solutions of HOG is larger than the space of such solutions in the framework of GR.

Not only do the Einstein equations have solutions which are included in the solution space of HOG, but there are such solutions that do not belong to the set of solutions of GR, making clear that the later is a subset of the former space. By looking at the action (\ref{eq:Rsquareaction})-(\ref{eq:Rsquare}) and taking $\zeta$ tending to zero, solutions of GR will be obtained as limiting cases.

There is another way that we may view the solution space of the two theories, GR and HOG. This is to regard HOG as a way of perturbing the solutions of GR not `inside' GR but `outside' it, in the framework of HOG. The solutions of the action (\ref{eq:Rsquareaction})-(\ref{eq:Rsquare}) can be considered as $\zeta$-perturbations of the Einstein equations in the sense that the term $\zeta R^2$ resembles the second term of a certain kind of a Taylor expansion around zero. HOG Lagrangians viewed in such a way can be considered as perturbations of the GR Lagrangian around zero. This is due to the fact that being sufficiently close to zero scalar curvature, the linear term in an expansion may be considered as a sufficiently reliable approximation but as we approach higher values of the curvature we need to consider more terms in the action to achieve the necessary precision.

As we will see in Chapter 5, the radiation solutions of GR can be obtained from HOG as limiting cases asymptotically towards the singularity thus confirming that such solutions maybe stable not with respect to perturbations inside GR but outside it in the sense discussed about.

%--------------------------------------------------------------------------------------------------------------------
%	SECTION   - Conclusion
%--------------------------------------------------------------------------------------------------------------------

\section{Conclusion}

In this chapter we started with the most general action of a quadratic gravity theory of the form (\ref{eq:lagra}) and reduced it to the new action (\ref{eq:genlagrareduced}) stating explicitly its dependence on the various quadratic curvature invariants. We underlined the dependence of the parameter $\zeta$ appearing in the reduced action $R+\zeta R^2$ on the coefficients of the curvature invariants in the original quadratic Lagrangian. This is in contrast to $\zeta$ being an arbitrary parameter appearing the theory. This is important for, as we will see later, the coefficients tending to various limits will indicate the relative importance of the associated quadratic invariants ($a$ of $R^2$, $b$ of $Ric^2$ and $d$ of $Riem^2$).

Following that, we derived the basic field equations for the general quadratic action, Eq. (\ref{eq:genlagrareduced}). In the second Section of this chapter, we specialized our calculations to a vacuum FRW universe, and wrote down the basic field equations corresponding to these solutions. 

Finally, we introduced new variables in the field equations and derived the autonomous dynamical system (\ref{eq:ds}). This will provide a useful basis for further study in later chapters of (the possible asymptotic regimes) a systematic asymptotic approach of such universes to the initial singularity. 

In the following chapter, we will consider the case where the curvature terms as well as terms describing vacuum terms enter the asymptotic decompositions subdominantly.

% Chapter Template

\chapter{Asymptotic analysis of flat vacua} % Main chapter title

\label{Chapter3} % Change X to a consecutive number; for referencing this chapter elsewhere, use \ref{ChapterX}

\lhead{Chapter 3. \emph{Asymptotic analysis of the flat vacua}} % Change X to a consecutive number; this is for the header on each page - perhaps a shortened title

%--------------------------------------------------------------------
%Introduction
%--------------------------------------------------------------------

In this chapter we treat the flat cases of vacuum FRW universes in HOG. We use asymptotic arguments to describe the behavior of the associated vector field and its integral curves as we approach the spacetime singularity. These integral curves - solutions of the relevant dynamical system - are obtained by asymptotically decomposing the vector field in such a way as to reveal all possible dominant characteristics emerging near the blow up singularity.

%-----------------------------------------------------------------------
%	SECTION 3.1 - Asymptotic splittings of flat-vacuum field
%-----------------------------------------------------------------------
\section{Asymptotic splittings of the flat-vacuum field}

%=====================================================================
\subsection{Flat vector field and the associated dynamical system}

In order to describe the asymptotics of the flat, quadratic, vacuum, FRW universe, we set $k=0$ in the general field equations (\ref{eq:beq}) and obtain the simpler form,
\be \ddot{H}=\frac{1}{2}\frac{\dot{H}^2}{H}-3H\dot{H}-\frac{1}{12\epsilon}H,
\label{eq:bfeflat}
\ee
in terms of the expansion rate, H.
Consequently, using the variables introduced in (\ref{vars1}), namely
\be\label{vars1flat}
x=H,\hspace{0.2cm} y=\dot{H},
\ee
and since the use of  the variable $z$ is not necessary currently, the autonomous dynamical system (\ref{eq:ds}) becomes
\begin{eqnarray}
\label{dsflat}
\dot{x} &=& y, \:\:\:\:\:\nonumber \\
\dot{y} &=& \frac{y^{2}}{2x}-3xy- \frac{x}{12\epsilon}.
\end{eqnarray}
The equivalent expressions of Eqs. (\ref{basic dynamical system})-(\ref{vf}) are given by the following forms which describe the \emph{flat}, vacuum, 2-dimensional  vector field $\mathbf{f}_{\,0,\textsc{VAC}}:\mathbb{R}^2\rightarrow\mathbb{R}^2$.
\be\label{dsflat2}
\mathbf{\dot{x}}=\mathbf{f}_{\,0,\textsc{VAC}}(\mathbf{x}),\quad \mathbf{x}=(x,z),
\ee
\be\label{fvf}
\mathbf{f}_{\,0,\textsc{VAC}}(x,y)=\left( y,\frac{y^{2}}{2x}-3xy-\frac{x}{12\epsilon} \right).
\ee
We note here that in the notation $\mathbf{f}_{\,0,\textsc{VAC}}$, the first index refers to the normalized curvature $k$ (zero in this case), while the second index, \footnotesize VAC \normalsize, denotes the vacuum state, in analogy to the curved, radiation-filled universes that we shall study in later chapters, $\mathbf{f}_{\,k,\textsc{RAD}}$.

%============================================================================================================
\subsection{Definition of a finite-time singularity}

Our main interest in the rest of this chapter is to study the behavior of the universe according to Eqs.(\ref{dsflat})-(\ref{fvf}), assuming that the sought-for solutions admit a finite-time, blow-up singularity appearing at some parameter value $t_{*}$ of the proper time $t$. That is, we assume that there exists a $t_{*}\in\mathbb{R}$  and an $x_{0} \in \mathcal{M}$, such that for all $M \in \mathbb{R}$ there exists a $\delta > 0$ with
\be\label{blowup}
\left\|\textbf{x}(t;\textbf{x}_{0})\right\|_{L^{p}} > M,
\ee
for all $t$ satisfying $\left|t-t_{*}\right| < \delta$. Here $\textbf{x}: (0,b)\rightarrow \mathcal{M}$ is a solution $\textbf{x}(t;c_1,\ldots,c_k)$, $k\leq 2$ ($c_k$ being the arbitrary integration constants), $\textbf{x}_{0}=\textbf{x}(t_{0})$ is a set of initial conditions for some $t_{0} \in (0,b)$, and $\left\|\hspace{0.1cm}\cdot\hspace{0.1cm}\right\|$ is any ${L^{p}}$-norm defined on the differentiable manifold $\mathcal{M}$. Without any loss of generality, we set $t_{*}=0$, stressing the fact that this specific value is really arbitrary since we could have placed it at any finite $t_*$ and used the variable $\tau=t-t_*$ instead of $t$.

Alternatively, the above precise definition of a finite-time singularity in the solutions of our dynamical system can be translated, using the vector field (\ref{fvf}), to a condition about the existence of an integral curve passing through the point $\textbf{x}_0$ of $\mathcal{M}$, such that at least one of its ${L^{p}}$-norms diverges at $t=t_*$, that is
\be
\lim_{t\rightarrow t_{*}} \left\|\textbf{x}(t;\textbf{x}_0)\right\|_{L^{p}}=\infty.
\ee
Additionally, we note that we can assign the meaning `now' to $t_0$ since it is an arbitrary point in the domain $(0,b)$. In the following, a finite-time singularity can be characterized for that matter to be a \emph{past} singularity when $t_{*}<t_0$, or a \emph{future} singularity when $t_{*}>t_0$.

%============================================================================================================
\subsection{Possible behavior of the vector field in the neighborhood of the singularity}

In order to describe the behavior of the vector field in the neighborhood of the initial singularity, we will follow the approach of the method of asymptotic splittings of Refs.\cite{cot-bar-07,gor-01,cot-13}. The basic notion is the fact that there are two different behaviors a vector field can adopt sufficiently close to the singularity. The first is to show some dominant feature meaning that the most nonlinear terms of the vector field approaching the singularity will determine a distinctly dominant behavior of the solutions. The second possible behavior is for the solutions of the vector field to `spiral' around the singularity forever in a way that condition (\ref{blowup}) is satisfied and the dynamics of the system are controlled by the subdominant terms.

%============================================================================================================
\subsection{Definition of weight-homogeneous decompositions}
 
As a first step in the study of possible dominant behavior of solutions of the vector field (\ref{fvf}) in the neighborhood of the initial singularity, we need to find suitable asymptotic decompositions of the vector field. That is to know all possible ways it can be split in dominant and subdominant \emph{components}. 

We say that a nonlinear vector field \textbf{f} on $\mathcal{M}^n$ admits a weight-homogeneous decomposition with respect to a given vector \textbf{p}, if it splits as a combination of the form \cite{cot-bar-07},
\be\label{whd}
\textbf{f}=\textbf{f}^{(0)}+\textbf{f}^{(1)}+\cdots+\textbf{f}^{(k)},
\ee
where the components $\textbf{f}^{(j)}, j=0,\cdots,k$, are weight-homogeneous vector fields. That is
\be\label{whc}
\textbf{f}^{(j)}(\textbf{a}\tau^{\textbf{p}})= \tau^{\textbf{p}+\textbf{1}(q^{(j)}-1)}\textbf{f}^{(j)}(\textbf{a}), \hspace{0.3cm} j=0,\cdots,k,
\ee
for some non-negative numbers $q^{(j)}$ and all \textbf{a} in some domain $\mathcal{E}$ of $\mathbb{R}^n$. This last condition (\ref{whc}) can be written for each individual component in the form,
\be\label{whcperterm}
f_{i}^{(j)}(\textbf{a}\tau^{\textbf{p}})= \tau^{p_i+q^{(j)}-1}f_i^{(j)}(\textbf{a}), \hspace{0.3cm} i=0,\cdots,n, \hspace{0.3cm} j=0,\cdots,k.
\ee
From the previous definition, Eq. (\ref{whd}), it follows that a weight-homogeneous decomposition splits the field \textbf{f} into $k+1$ weight-homogeneous components each with degree $\textbf{p}+\textbf{1}(q^{(j)}-1), \hspace{0.3cm} j=0,\cdots,k,$. The first of these vector fields, namely $f^{(0)}$, is the lowest order component and is scale invariant, since the non-negative numbers $q^{(j)}$, also called \textit{subdominant exponents} can be ordered so that,
\be
0=q^{(0)}<q^{(j_1)}<q^{(j_2)}, \hspace{0.4cm} when \hspace{0.3cm} j_1<j_2.
\ee

%============================================================================================================
\subsection{Vector field decompositions}

Accordingly, we find that the vector field (\ref{fvf}) possesses the following $2^3-1=7$ possible asymptotic decompositions of the form (\ref{whd}), or more specifically, 
\be\label{flatdecmode}
\mathbf{f}_{\,0,\textsc{VAC}}=\mathbf{f}^{(0)}_{\,0,\textsc{VAC}} + \mathbf{f}^{\,(\textrm{sub})}_{\,0,\textsc{VAC}},
\ee
where $\mathbf{f}^{(0)}_{\,0,\textsc{VAC}}$ is the \emph{dominant part}, and $\mathbf{f}^{\,(\textrm{sub})}_{\,0,\textsc{VAC}}\equiv \sum_{j=1}^{k}\textbf{f}^{(j)}$ the \emph{subdominant part} in each asymptotic decomposition valid in the neighborhood of the initial singularity. We have:
\begin{align}\label{flatvacdec1}
&\mathbf{f^1}_{\,0,\textsc{VAC}}=\mathbf{f^1}^{(0)}_{\,0,\textsc{VAC}} + \mathbf{f^1}^{\,(\textrm{sub})}_{\,0,\textsc{VAC}},\nonumber \\ \nonumber \\ 
&\mathbf{f^1}^{(0)}_{\,0,\textsc{VAC}}(\mathbf{x})= \left(y, \frac{y^2}{2x}, -2xz \right), \hspace{0,5cm}
\mathbf{f^1}^{\,(\textrm{sub})}_{\,0,\textsc{VAC}}(\mathbf{x})= \left(0, -\frac{x}{12\epsilon}-3xy, 0 \right),
\end{align}

\begin{align}\label{flatvacdec2}
&\mathbf{f^2}_{\,0,\textsc{VAC}}=\mathbf{f^2}^{(0)}_{\,0,\textsc{VAC}} + \mathbf{f^2}^{\,(\textrm{sub})}_{\,0,\textsc{VAC}},\nonumber \\ \nonumber \\ 
&\mathbf{f^2}^{(0)}_{\,0,\textsc{VAC}}(\mathbf{x})=\left(y, -3xy, -2xz \right), \hspace{0,5cm}
\mathbf{f^2}^{\,(\textrm{sub})}_{\,0,\textsc{VAC}}(\mathbf{x})=\left(0, \frac{y^2}{2x}-\frac{x}{12\epsilon}, 0 \right),
\end{align}

\begin{align}\label{flatvacdec3}
&\mathbf{f^3}_{\,0,\textsc{VAC}}=\mathbf{f^3}^{(0)}_{\,0,\textsc{VAC}} + \mathbf{f^3}^{\,(\textrm{sub})}_{\,0,\textsc{VAC}},\nonumber \\ \nonumber \\ 
&\mathbf{f^3}^{(0)}_{\,0,\textsc{VAC}}(\mathbf{x})=\left(y, -\frac{x}{12\epsilon}, -2xz \right),\hspace{0,5cm}
\mathbf{f^3}^{\,(\textrm{sub})}_{\,0,\textsc{VAC}}(\mathbf{x})=\left(0, \frac{y^2}{2x}-3xy, 0 \right),
\end{align}

\begin{align}\label{flatvacdec4}
&\mathbf{f^4}_{\,0,\textsc{VAC}}=\mathbf{f^4}^{(0)}_{\,0,\textsc{VAC}} + \mathbf{f^4}^{\,(\textrm{sub})}_{\,0,\textsc{VAC}},\nonumber \\ \nonumber \\ 
&\mathbf{f^4}^{(0)}_{\,0,\textsc{VAC}}(\mathbf{x})=\left(y, \frac{y^2}{2x}-3xy, -2xz \right), \hspace{0,5cm}
\mathbf{f^4}^{\,(\textrm{sub})}_{\,0,\textsc{VAC}}(\mathbf{x})=\left(0, -\frac{x}{12\epsilon}, 0 \right),
\end{align}

\begin{align}\label{flatvacdec5}
&\mathbf{f^5}_{\,0,\textsc{VAC}}=\mathbf{f^5}^{(0)}_{\,0,\textsc{VAC}} + \mathbf{f^5}^{\,(\textrm{sub})}_{\,0,\textsc{VAC}},\nonumber \\ \nonumber \\ 
&\mathbf{f^5}^{(0)}_{\,0,\textsc{VAC}}(\mathbf{x})=\left(y, \frac{y^2}{2x}- \frac{x}{12\epsilon}, -2xz \right),\hspace{0,5cm}
\mathbf{f^5}^{\,(\textrm{sub})}_{\,0,\textsc{VAC}}(\mathbf{x})=\left(0, -3xy, 0 \right),
\end{align}

\begin{align}\label{flatvacdec6}
&\mathbf{f^6}_{\,0,\textsc{VAC}}=\mathbf{f^6}^{(0)}_{\,0,\textsc{VAC}} + \mathbf{f^6}^{\,(\textrm{sub})}_{\,0,\textsc{VAC}},\nonumber \\ \nonumber \\ 
&\mathbf{f^6}^{(0)}_{\,0,\textsc{VAC}}(\mathbf{x})=\left(y, -3xy-\frac{x}{12\epsilon}, -2xz \right),\hspace{0,5cm}
\mathbf{f^6}^{\,(\textrm{sub})}_{\,0,\textsc{VAC}}(\mathbf{x})=\left(0, \frac{y^2}{2x}, 0 \right),
\end{align}

\begin{align}\label{flatvacdec7}
&\mathbf{f^7}_{\,0,\textsc{VAC}}=\mathbf{f^7}^{(0)}_{\,0,\textsc{VAC}} + \mathbf{f^7}^{\,(\textrm{sub})}_{\,0,\textsc{VAC}},\nonumber \\ \nonumber \\ 
&\mathbf{f^7}^{(0)}_{\,0,\textsc{VAC}}(\mathbf{x})=\left(y, \frac{y^2}{2x}-3xy- \frac{x}{12\epsilon}, -2xz \right), \hspace{0,5cm}
\mathbf{f^7}^{\,(\textrm{sub})}_{\,0,\textsc{VAC}}(\mathbf{x})=\left(0, 0, 0 \right).
\end{align}
We will eventually construct convergent, asymptotic series solutions that encode information about the leading order behavior of all solutions, as well as their generality (number of arbitrary constants) near the spacetime singularity at $t=0$. 

%============================================================================================================
\subsection{Dominant balances}

For any given dominant asymptotic decomposition (\ref{flatvacdec1})-(\ref{flatvacdec7}) of the system (\ref{dsflat}), we call the pair $(\mathbf{a},\mathbf{p})$ a \emph{dominant balance} of the vector field $\mathbf{f}_{\,0,\textsc{VAC}}$, where $\mathbf{a}=(\theta, \eta)\in\mathbb{C}^2$ are constants and $\mathbf{p}=(p, q)\in\mathbb{Q}^2$, and look for a leading-order behavior of the form,
\be
\mathbf{x}(t)=\mathbf{a}t^{\mathbf{p}}=(\theta t^{p}, \eta
t^{q}). \label{eq:domisol}
\ee
Such behaviors geometrically correspond to the possible asymptotic forms of the integral curves  of the vacuum field $\mathbf{f}_{\,0,\textsc{VAC}}$, as we take it to a neighborhood of the singularity. 

Substituting the forms (\ref{eq:domisol}) into the dominant system 
\be \label{vfdomsys}
\mathbf{\dot{x}}(t)=\mathbf{f}^{(0)}_{\,0,\textsc{VAC}}(\mathbf{x}(t))
\ee 
and solving the resulting nonlinear algebraic system to determine  the dominant balance $(\mathbf{a},\mathbf{p})$ as an exact, scale invariant solution, we find that only three of the seven possible decompositions (\ref{flatvacdec1})-(\ref{flatvacdec7}) lead to acceptable dominant balances. 
Namely, asymptotic decompositions (\ref{flatvacdec1}),(\ref{flatvacdec2}) and (\ref{flatvacdec4}) lead to the following dominant balances $\mathcal{B}_{\,0,\textsc{VAC}}\in\mathbb{C}^2\times\mathbb{Q}^2$,  which need to be further tested in order to be fully accepted for the construction of the asymptotic solutions of the initial dynamical system.
\begin{eqnarray}
\mathcal{B}^{1.1}_{\,0,\textsc{VAC}}&=&(\mathbf{a_{1.1}},\mathbf{p_{1.1}})= \left(\left(\theta, 0 \right),\:
\left(0, -1\right)\right),\label{eq:flat1-1balance} \\[20pt]
\mathcal{B}^{1.2}_{\,0,\textsc{VAC}}&=&(\mathbf{a_{1.2}},\mathbf{p_{1.2}})= \left(\left(\theta, 2\theta \right),\:
\left(2, 1\right)\right),\label{eq:flat1-2balance} \\[20pt]
\mathcal{B}^{2}_{\,0,\textsc{VAC}}&=&(\mathbf{a_2},\mathbf{p_2})= \left(\left(\frac{2}{3},-\frac{2}{3} \right),\:
\left(-1, -2\right)\right),\label{eq:flat2balance} \\[20pt]
\mathcal{B}^{4}_{\,0,\textsc{VAC}}&=&(\mathbf{a_4},\mathbf{p_4})= \left(\left(\frac{1}{2},-\frac{1}{2} \right),\:
\left(-1,-2\right)\right),\label{eq:flat4balance}
\end{eqnarray}
where we note that we use the notation $\mathcal{B}^{i.j}_{\,0,\textsc{VAC}}$ for the $j$-th dominant balance corresponding to the $\mathbf{f^i}_{\,0,\textsc{VAC}}$ asymptotic decomposition of the flat, vacuum vector field $\mathbf{f}_{\,0,\textsc{VAC}}$. In particular, this means that the vector fields $\mathbf{f^1}^{(0)}_{\,0,\textsc{VAC}}$, $\mathbf{f^2}^{(0)}_{\,0,\textsc{VAC}}$ and $\mathbf{f^4}^{(0)}_{\,0,\textsc{VAC}}$ are \emph{scale-invariant systems}, cf. \cite{cot-bar-07,gor-01,cot-13}.

%============================================================================================================
\subsection{Subdominant condition for each possible asymptotic balance}

Further, we need to show that the terms $\mathbf{f^1}^{\,(\textrm{sub})}_{\,0,\textsc{VAC}}(\mathbf{x})$, $\mathbf{f^2}^{\,(\textrm{sub})}_{\,0,\textsc{VAC}}(\mathbf{x})$ and $\mathbf{f^4}^{\,(\textrm{sub})}_{\,0,\textsc{VAC}}(\mathbf{x})$ in the basic decompositions (\ref{flatvacdec1}), (\ref{flatvacdec2}) and (\ref{flatvacdec4}) of the flat-vacuum field (\ref{fvf}) are themselves weight-homogeneous with respect to the corresponding flat-vacuum balances (\ref{eq:flat1-1balance})-(\ref{eq:flat4balance}) for this splittings to be finally acceptable. For this we need to check that these candidate subdominant parts are indeed subdominant by calculating the expression,
\be\label{vfsubcon}
\lim_{t\rightarrow 0} \frac{\mathbf{f}^{\,(\textrm{sub})}_{\,0,\textsc{VAC}}(\mathbf{a}t^{\mathbf{p}})}{t^{\mathbf{p}-1}}.
\ee
Using the balances $\mathcal{B}^{1.2}_{\,0,\textsc{VAC}}, \mathcal{B}^{1.1}_{\,0,\textsc{VAC}}, \mathcal{B}^{2}_{\,0,\textsc{VAC}}$ and $\mathcal{B}^{4}_{\,0,\textsc{VAC}}$ defined by Eqs. (\ref{eq:flat1-1balance})-(\ref{eq:flat4balance}), we find that,
\begin{eqnarray} 
\frac{\mathbf{f^1}^{\,(\textrm{sub})}_{\,0,\textsc{VAC}}(\mathbf{a_1}t^{\mathbf{p_1}})}{t^{\mathbf{p_1}-1}}
&=&\mathbf{f^1}^{\,(\textrm{sub})}_{\,0,\textsc{VAC}}(\mathbf{a_1})\,t^2 =\left(0, -\frac{\theta}{12\epsilon}\right)t^2, \label{eq:fvsubcon1-1}\\[20pt] 
\frac{\mathbf{f^1}^{\,(\textrm{sub})}_{\,0,\textsc{VAC}}(\mathbf{a_2}t^{\mathbf{p_2}})}{t^{\mathbf{p_2}-1}}
&=&\left(0, - \frac{\theta}{12\epsilon} \right)t^2+\left(0, \theta \right)t^3,\label{eq:fvsubcon1-2} \\[20pt]
\frac{\mathbf{f^2}^{\,(\textrm{sub})}_{\,0,\textsc{VAC}}(\mathbf{a}t^{\mathbf{p}})}{t^{\mathbf{p}-1}} 
&=&\left(0, \frac{1}{3}\right)t^0+\left(0,  \frac{1}{18\epsilon}\right)t^2,\label{eq:fvsubcon2} \\[20pt] 
\frac{\mathbf{f^4}^{\,(\textrm{sub})}_{\,0,\textsc{VAC}}(\mathbf{a}t^{\mathbf{p}})}{t^{\mathbf{p}-1}}
&=&\mathbf{f^4}^{\,(\textrm{sub})}_{\,0,\textsc{VAC}}(\mathbf{a})\,t^2 =\left(0, - \frac{1}{24\epsilon} \right)t^2.\label{eq:fvsubcon4}
\end{eqnarray}
Taking the limit of these expressions as $t\rightarrow 0$, we can see that all of them, except (\ref{eq:fvsubcon2}), go to zero asymptotically. This, on one hand, means that asymptotic decomposition (\ref{flatvacdec2}) is not an acceptable decomposition of the vector field (\ref{fvf}), however, on the other hand that dominant balances (\ref{eq:flat1-1balance}), (\ref{eq:flat1-2balance}) and (\ref{eq:flat4balance}) are indeed candidates for the construction of an asymptotic solution of (\ref{fvf}) around the initial singularity, provided that the forms $\mathbf{f^{1}}^{(sub)}_{\,0,\textsc{VAC}}(\mathbf{a_1})$ and $\mathbf{f^{4}}^{(sub)}_{\,0,\textsc{VAC}}(\mathbf{a})$ are different from zero. 

This happens only when $\epsilon \neq 0$, that is for all cases except when $3\beta +\gamma =0$. We conclude that when this constraint holds true the basic decompositions (\ref{flatvacdec1}) and (\ref{flatvacdec4}) are acceptable asymptotically in any higher order gravity theory. The so-called conformally invariant Bach-Weyl gravity cf. \cite{wey-19a} is a physical example that is excluded from this analysis and consequently needs a separate treatment. We note that the same constraint appears in the stability analysis of purely radiation universes in these theories, cf. Chapter 5.

%============================================================================================================
\subsection{Section conclusion}

We have completed in this Section the first part of our asymptotic analysis via the method of asymptotic splittings, that is we found all possible asymptotic systems on approach to the initial singularity. This process amounts to dropping all terms that are \emph{small}, and replace exact by asymptotic relations (by this we mean using (\ref{vfdomsys}) in conjunction with (\ref{flatdecmode}) and (\ref{vfsubcon}) instead of (\ref{dsflat}) and (\ref{fvf})). 

This first part of the application of the method of asymptotic splittings allows us to conclude that there are essentially two such systems, and we were able to extract preliminary qualitative results about the behavior of our basic vector field, without actually solving the systems. In the next Section, we shall proceed to study solutions of our asymptotic systems through the processes of \emph{balance}, \emph{subdominance} and \emph{consistency}.

%--------------------------------------------------------------------------------------------------------------------
%	SECTION 3.2  - Dominant solutions
%--------------------------------------------------------------------------------------------------------------------
\section{Dominant solutions}

%============================================================================================================
\subsection{Construction of the K-matrices}

We now proceed to test our asymptotic solutions in terms of their internal consistency with respect to the general framework of our asymptotic analysis (cf. \cite{cot-bar-07,gor-01,cot-13} for more details and proofs). We will eventually construct series representations of these asymptotic solutions valid locally around the initial singularity, so that it is dominated by the dominant balance solutions we have built so far. 

The degree of generality of these formal series solutions depends on the number of arbitrary constants in them. As explained in \cite{cot-bar-07}, the arbitrary constants of any (particular or general) solution first appear in those terms in the asymptotic series solution whose coefficients $\mathbf{c}_{i}$ have indices $i=\varrho s$, where $\varrho$ is a non-negative $\mathcal{K}$-exponent, and $s$ denotes the least common multiple of the denominators of the set of all subdominant exponents and those of all the $\mathcal{K}$-exponents with positive real parts. These exponents are complex numbers belonging to the spectrum of the Kovalevskaya matrix given by
\be
\mathcal{K}=D\,\mathbf{f}^{(0)}_{0,\textsc{VAC}}(\mathbf{a})-\textrm{diag}(\mathbf{p}),
\ee
for which the following expression also stands,
\be
\mathcal{K}\mathbf{a}\mathbf{p}=-\mathbf{a}\mathbf{p},
\ee
Hence, the $\mathcal{K}$-matrix always has $\varrho_{1}=-1$ as an eigenvalue with $\mathbf{a}\mathbf{p}=\mathbf{f}^{(0)}_{0,\textsc{VAC}}(\mathbf{a})$ in this case, as the corresponding eigenvector. In this way the $\mathcal{K}$-exponents depend on the dominant part of the vector field as well as the dominant balance. 

In the present case, the Kovalevskaya matrices for each of the possible dominant balances are,

\be
\mathcal{K}^{1.1}_{\,0,\textsc{VAC}}=\left(
                     \begin{array}{cc}
                       0 & 1  \\
                       0 & 1  \\
                     \end{array}
                   \right),
\ee
with spectrum,

\be
\textrm{spec}(\mathcal{K}^{1.1}_{\,0,\textsc{VAC}})=\{1,0\}.
\ee

\be
\mathcal{K}^{1.2}_{\,0,\textsc{VAC}}=\left(
                     \begin{array}{cc}
                       -2 & 1  \\
                       -2 & 1  \\
                     \end{array}
                   \right),
\ee
with spectrum

\be
\textrm{spec}(\mathcal{K}^{1.2}_{\,0,\textsc{VAC}})=\{-1,0\}.
\ee

\be
\mathcal{K}^{4}_{\,0,\textsc{VAC}}=\left(
                     \begin{array}{cc}
                       1 & 1     \\
                       1 & -1/2  \\
                     \end{array}
                   \right),
\ee
with spectrum
\be
\textrm{spec}(\mathcal{K}^{4}_{\,0,\textsc{VAC}})=\{-1,3/2\}.
\ee
We conclude that $\mathcal{K}^{1.1}_{\,0,\textsc{VAC}}$ does not correspond to valid asymptotic balance, since it does not have $-1$ as an eigenvalue, while $\mathcal{K}^{1.2}_{\,0,\textsc{VAC}}$ which does not have any eigenvalues with positive real parts, leads to an acceptable balance but it is one that may be valid at infinity.

%============================================================================================================
\subsection{Final asymptotic balance}

Thus, it is only the asymptotic balance (\ref{eq:flat4balance}) that is fully consistent with overall approximation scheme we are using and can lead to a series representation of the asymptotic solutions valid locally around the initial singularity. The least common multiple of the denominators of the set of all subdominant exponents and those of all the $\mathcal{K}$-exponents with positive real parts, $s=2$ in this case. 

As we discussed, the number of non-negative $\mathcal{K}$-exponents equals the number of arbitrary constants that appear in the series expansions. The $-1$ exponent corresponds to the position of the singularity, and because the $\textrm{spec}(\mathcal{K}^{4}_{\,0,\textsc{VAC}})$ in our case possesses one non-negative eigenvalue, the balance $\mathcal{B}^{4}_{\,0,\textsc{VAC}}$ indeed corresponds to the dominant behavior of a \emph{general} solution having the form of a formal series and valid locally around the initial singularity. 

%============================================================================================================
\subsection{Construction of the formal series}

In order to find that solution, we substitute the \emph{Fuchsian series expansions} \footnote{A series expansion with no constant first term and rational exponents.} and their derivatives

\begin{equation} \label{eq:flatseries}
x(t) = \sum_{i=0}^{\infty} c_{1i}\: t^{\frac{i}{2}-1}, \:\:\:\:\:
y(t) = \sum_{i=0}^{\infty} c_{2i}\:  t^{\frac{i}{2}-2},\:\:\:\:\:
\end{equation}
\begin{equation} \label{eq:flatseriesdev}
\dot{x}(t) = \sum_{i=0}^{\infty} c_{1i} \left(\frac{i}{2}-1\right)\: t^{\frac{i}{2}-2},\:\:\:\:\:
\dot{y}(t) = \sum_{i=0}^{\infty} c_{2i} \left(\frac{i}{2}-2\right)\: t^{\frac{i}{2}-3},\:\:\:\:\:
\end{equation}
where because of the form of the balance, Eq. (\ref{eq:flat4balance}), we have $c_{10}=1/2$ and $\:\:c_{20}=-1/2$, 
in the following equivalent form of the original system (\ref{dsflat}), assuming $x\neq0$,
\begin{eqnarray}
\label{dsf}
\dot{x} &=& y, \:\:\:\:\:\nonumber \\
2x\dot{y} &=& y^{2}-6x^{2}y-\frac{1}{6\epsilon}x^{2},
\end{eqnarray}
from which we will be led to various recursion relations that determine the unknowns $c_{1i}, c_{2i}$ term by term.

More specifically, from the first equation of (\ref{dsf}), after substitution we have,

\be
\sum_{i=0}^{\infty} c_{1i} \left(\frac{i}{2}-1\right) t^{\frac{i}{2}-1}=\sum_{i=0}^{\infty} c_{2i}  t^{\frac{i}{2}-2},
\ee
which leads to,
\be
c_{1i} \left(\frac{i}{2}-2\right) = c_{2i}\:.\label{eq:substi1}
\ee
From the second equation of (\ref{dsf}), we calculate separately each term after substitution:

\begin{eqnarray}
2x\dot{y}&=& 2\left(\sum_{i=0}^{\infty}c_{1i}t^{\frac{i}{2}-1}\right)\left(\sum_{i=0}^{\infty}c_{2i} \left(\frac{i}{2}-2\right) t^{\frac{i}{2}-3} \right)\nonumber\\[15pt]
&=& 2t^{-4} \sum_{i=0}^{\infty} \sum_{k=0}^{i} c_{2(i-k)} \left(\frac{i-k}{2}-2\right)c_{1k} t^{\frac{i}{2}},\\[25pt]
y^{2}&=&  \left( \sum_{i=0}^{\infty} c_{2i}  t^{\frac{i}{2}-2} \right)\left( \sum_{i=0}^{\infty} c_{2i}  t^{\frac{i}{2}-2} \right) \nonumber\\
&=& t^{-4} \sum_{i=0}^{\infty} \sum_{k=0}^{i} c_{2(i-k)} c_{2k} t^{\frac{i}{2}},
\end{eqnarray}

\begin{eqnarray}
-6x^{2}y &=& -6  \left(\sum_{i=0}^{\infty} c_{1i} t^{\frac{i}{2}-1}\right) \left(\sum_{i=0}^{\infty} c_{1i} t^{\frac{i}{2}-1}\right) \left( \sum_{i=0}^{\infty} c_{2i}  t^{\frac{i}{2}-2} \right) \nonumber\\[15pt]
&=& -6t^{-4} \sum_{i=0}^{\infty} \sum_{k=0}^{i} \sum_{l=0}^{k} c_{1(i-k)} c_{1(k-l)} c_{2l} t^{\frac{i}{2}}, \\[25pt]
-\frac{1}{6\epsilon}x^{2}&=& -\frac{1}{6\epsilon}\left(\sum_{i=0}^{\infty} c_{1i} t^{\frac{i}{2}-1}\right) \left(\sum_{i=0}^{\infty} c_{1i} t^{\frac{i}{2}-1}\right)  \nonumber \\[15pt]
&=& -\frac{1}{6\epsilon}t^{-2} \sum_{i=0}^{\infty} \sum_{k=0}^{i} c_{1(i-k)} c_{1k} t^{\frac{i}{2}}.
\end{eqnarray}
Consequently, we are led to the following form of the second equation of (\ref{dsf}),
\begin{eqnarray}
2t^{-4} \sum_{i=0}^{\infty} \sum_{k=0}^{i} c_{2(i-k)} \left(\frac{i-k}{2}-2\right)c_{1k} t^{\frac{i}{2}}=t^{-4} \sum_{i=0}^{\infty} \sum_{k=0}^{i} c_{2(i-k)} c_{2k} t^{\frac{i}{2}}-\nonumber\\[15pt]
-6t^{-4} \sum_{i=0}^{\infty} \sum_{k=0}^{i} \sum_{l=0}^{k} c_{1(i-k)} c_{1(k-l)} c_{2l} t^{\frac{i}{2}} -\frac{1}{6\epsilon}t^{-2} \sum_{i=0}^{\infty} \sum_{k=0}^{i} c_{1(i-k)} c_{1k} t^{\frac{i}{2}}\label{eq:substi2}
\end{eqnarray}

%============================================================================================================
\subsection{Calculation of the final series coefficients}

Eqs. (\ref{eq:substi1}) and (\ref{eq:substi2}) constitute the system from which we may solve for the coefficients $c_{1i}$ and $c_{2i}$ in the expansions,

\begin{eqnarray}
x(t)&=&\theta\:\:t^{p} +c_{11}\:\:t^{-1/2} +c_{12}\:\:t^{0}+c_{13}\:\:t^{1/2} +c_{14}\:\:t^{1} + \cdots,\nonumber\\[10pt]
y(t)&=&\eta\:\:t^{q} +c_{21}\:\:t^{-3/2} +c_{22}\:\:t^{-1}+c_{23}\:\:t^{-1/2} +c_{24}\:\:t^{0} + \cdots.
\end{eqnarray}

Below, we will construct a set of equations from each of the Eqs. (\ref{eq:substi1}) and (\ref{eq:substi2}) by comparing the coefficients of the various powers of $t$.

%============================================================================================================
\subsubsection*{1st set of equations}

For the coefficients of the different powers of $t$, Eq. (\ref{eq:substi1}) will lead to the following equations,
\bq
\text{for the coefficients of the term}\:\: &t^{-3/2},\:\:\:\:c_{11} \left(\frac{1}{2}-1\right) = c_{21}\:,\label{eq:cvgencoe1-1}\\[12pt]
\text{for the coefficients of the term}\:\: &t^{-1},\:\:\:\:c_{12} \left(\frac{2}{2}-1\right) = c_{22}\:,\label{eq:cvgencoe1-2}\\[12pt]
\text{for the coefficients of the term}\:\: &t^{-1/2},\:\:\:\:c_{13} \left(\frac{3}{2}-1\right) = c_{23}\:,\label{eq:cvgencoe1-3}\\[12pt]
\text{for the coefficients of the term}\:\: &t^{0},\:\:\:\:c_{14} \left(\frac{4}{2}-1\right) = c_{24}\:.\label{eq:cvgencoe1-4} 
\eq

%============================================================================================================
\subsubsection*{2nd set of equations}

Following that, we now examine the coefficients of the various powers of $t$ in Eq. (\ref{eq:substi2}) which give the following,
for the coefficients of the term $t^{-7/2}$, we have,
\be\label{cvgencoe2-1}
2 \sum_{k=0}^{1} c_{2(1-k)} \left(\frac{1-k}{2}-2\right)c_{1k} = \sum_{k=0}^{1} c_{2(1-k)}c_{2k}-6\sum_{k=0}^{1} \sum_{l=0}^{k} c_{1(1-k)} c_{1(k-l)} c_{2l}, 
\ee
for the coefficients of the term $t^{-3}$, we have,
\be\label{cvgencoe2-2}
2 \sum_{k=0}^{2} c_{2(2-k)} \left(\frac{2-k}{2}-2\right)c_{1k} = \sum_{k=0}^{2} c_{2(2-k)}c_{2k}-6\sum_{k=0}^{2} \sum_{l=0}^{k} c_{1(2-k)} c_{1(k-l)} c_{2l}, 
\ee
for the coefficients of the term $t^{-5/2}$, we have,
\be\label{cvgencoe2-3}
2 \sum_{k=0}^{3} c_{2(3-k)} \left(\frac{3-k}{2}-2\right)c_{1k} = \sum_{k=0}^{3} c_{2(3-k)}c_{2k}
-6\sum_{k=0}^{3} \sum_{l=0}^{k} c_{1(3-k)} c_{1(k-l)} c_{2l},
\ee
for the coefficients of the term $t^{-2}$, we have,
\be\label{cvgencoe2-4}
2 \sum_{k=0}^{4} c_{2(4-k)} \left(\frac{4-k}{2}-2\right)c_{1k} = \sum_{k=0}^{4} c_{2(4-k)}c_{2k}-6\sum_{k=0}^{4} \sum_{l=0}^{k} c_{1(4-k)} c_{1(k-l)} c_{2l}-\frac{1}{6\epsilon} c^{2}_{01}. 
\ee
Consequently, solving the sets of Eqs. (\ref{eq:cvgencoe1-1})-(\ref{eq:cvgencoe1-1}) and (\ref{cvgencoe2-1})-(\ref{cvgencoe2-4}) we find,
\be
c_{11}=c_{21}=0, \label{coef1}
\ee
\be
c_{12}=c_{22}=0, \label{coef2}
\ee
\be
c_{13}=2c_{23}, \label{coef3}
\ee
and
\be
c_{14}=c_{24}=-\frac{1}{36\epsilon}. \label{coef4}
\ee
Thus, the final series representation of the solution has the form:
\begin{eqnarray}
x(t) = \frac{1}{2}\:\:t^{-1} +c_{13}\:\:t^{1/2} -\frac{1}{36\epsilon}\:\: t + \cdots,\label{eq:flatgensol1} \\[15pt] 
y(t) = -\frac{1}{2}\:\:t^{-2} +\frac{c_{13}}{2}\:\:t^{-1/2} -\frac{1}{36\epsilon}\:\: t^{0} + \cdots,\label{eq:flatgensol2}
\end{eqnarray}

and, since $x=H=\dot{a} / a$, we arrive at the asymptotic form of the scale factor around the singularity:
\be \label{eq:flatfinal}
a(t) =\a\:\:t^{1/2}+\frac{2c_{13}\a}{3}\:\:t^{2}-\frac{\a}{72\epsilon}\:\:t^{5/2}+\frac{4\a\:c_{13}^{2}}{9}\:\:t^{7/2}+\cdots, 
\ee
where $\a$ is a constant of integration.

%============================================================================================================
\subsection{Fredholm's alternative}

As a final test for admission of this solution, we use Fredholm's alternative to be satisfied by any admissible solution. This leads to a \emph{compatibility condition} for the positive eigenvalue 3/2 and the associated eigenvector,
\be
v^{\top}\cdot\left(\mathcal{K}-\frac{j}{s}I\right)\mathbf{c}_j=0,
\ee
where $I$ denotes the identity matrix, and we have to satisfy this at the $j=3$ level. This gives the following orthogonality constraint,
\be (2,1,-\frac{8\r}{3}) \cdot \left( \begin{array}{l}
                              -\frac{1}{2}c_{13}+c_{23}  \\
                                 c_{13}-2c_{23}   \\
                               -2\r c_{13}-\frac{3}{2}c_{33}
                                 \end{array}
              \right) = 0.
\label{eq:cc}
\ee
which leads to
\be
c_{13}=2c_{23}. 
\ee
This exactly Eq. (\ref{coef3}), thus leading to the conclusion that Eqs. (\ref{eq:flatgensol1})-(\ref{eq:flatgensol2}) correspond  to a valid asymptotic solution around the singularity.

Our series solution (\ref{eq:flatgensol1})-(\ref{eq:flatgensol2}) has two arbitrary constants, namely, $c_{13}$ and a second one corresponding to the arbitrary position of the singularity (taken here to be zero without loss of generality), and is therefore a local expansion of the \emph{general} solution around the initial singularity. Since the leading order coefficients are real, by a theorem of Goriely and Hyde, cf. \cite{gor-01}, we conclude that there is an open set of initial conditions  for which the general solution blows up at the finite time (initial) singularity at $t=0$.  This proves the stability of our solution in the neighborhood of the singularity.

%--------------------------------------------------------------------------------------------------------------------
%	SECTION 3.3 - Conclusion
%--------------------------------------------------------------------------------------------------------------------
\section{Conclusion}

In this chapter we have considered the possible singular behaviors and asymptotic limits of vacuum, flat isotropic universes in the fully quadratic gravity theory in four dimensions which apart from the Einstein term also contains terms proportional to a linear combination of $R^2, \textrm{Ric}^2$ and $\textrm{Riem}^2$. Using various asymptotic and geometric arguments, we were able to built a solution of the field equations in the form of a Fuchsian formal series expansion compatible with all other constraints, dominated asymptotically to leading order by this solution and having the correct number of arbitrary constants that makes it a general solution of the field equations. In this way, we  conclude that this exact solution is an early time attractor of all homogeneous and isotropic flat vacua of the theory, thus proving stability against such `perturbations'.

In the next chapter we will proceed with the curved cases of the same class of cosmological models. Namely, we will analyze asymptotically the vacuum, curved FRW universes thus completing the profile of the asymptotic uniqueness of the vacuum isotropic universes in the fully quadratic gravity theory and their stability in the neighborhood of the initial singularity.
% Chapter Template

\chapter{Curved vacua} % Main chapter title

\label{Chapter4} % Change X to a consecutive number; for referencing this chapter elsewhere, use \ref{ChapterX}

\lhead{Chapter 4. \emph{Curved vacua}} % Change X to a consecutive number; this is for the header on each page - perhaps a shortened title

%---------------------------------------------------------------
% Introduction
%---------------------------------------------------------------

In this chapter we extend our analysis of the vacuum higher-order cosmological models to the curved cases. In these cases the vector field which describes the evolution of the universe takes a three-dimensional form and a number of extra terms appear that increase the possible ways of asymptotic behavior on approach to the initial singularity. The behavior of the extra terms in our asymptotic analysis will lead us at the end of the chapter to certain conclusions about the way curvature affects the asymptotic of this class of universes.

%--------------------------------------------------------------
%	SECTION 4.1 - Introduction
%--------------------------------------------------------------

\section{Introduction}

As we have seen in the previous chapter when we have a flat vacuum FRW model in the fully quadratic theory of gravity defined by the action (\ref{eq:lagra}), the vector field $\mathbf{f}_{\,0,\textsc{VAC}}$ described by the Eq.(\ref{fvf}) has one admissible asymptotic solution near the initial singularity, namely, the form (\ref{eq:flatgensol1})-(\ref{eq:flatgensol2}). In this family, all flat vacua are asymptotically dominated (or `attracted') at early times by the form $a(t)\sim t^{1/2}$, thus proving the stability of this solution in the flat case.

In order to study the situation of a vacuum but curved family of FRW universes, we will apply the method of asymptotic splittings to the vector field (\ref{vf}) which was obtained in Chapter 2 from the general quadratic action (\ref{eq:lagra}). In this case, there are two extra complications which have to be analyzed separately. Firstly, when $k\neq 0$, the vacuum field $\mathbf{f}_{\textsc{VAC}}$ is 3-dimensional instead of planar as it was in the flat case. Secondly, it has more terms than those present in the flat case, namely, those that contain $k$ in (\ref{vf}). 

Below we shall use the suggestive notation $\mathbf{f}_{\,k,\textsc{VAC}}$ instead of $\mathbf{f}_{\textsc{VAC}}$ to signify that we are dealing with non-flat vacua. The general form of the vector field presently is given by,

\be\label{fkvac}
\mathbf{f}_{\,k,\textsc{VAC}}(x,y,z)=\left( y,\frac{y^{2}}{2x}-3xy+kxz-\frac{k^2 z^2}{2x}- \frac{x}{12\epsilon}-\frac{kz}{12\epsilon x},-2xz \right).
\ee

%================================================================
\section{Vector field decompositions}

As described in Chapter 3, firstly, we are interested in finding the complete list of all possible weight-homogeneous decompositions for the field given by Eq. (\ref{fkvac}) of the general form
\be \label{whdeccurved}
\mathbf{f}_{\,k,\textsc{VAC}}=\mathbf{f}^{(0)}_{\,k,\textsc{VAC}} + \mathbf{f}^{\,(\textrm{sub})}_{\,k,\textsc{VAC}}.
\ee
The vector field $\mathbf{f}_{\,k,\textsc{VAC}}$ (or the basic system (\ref{eq:ds})) can decompose precisely in $2^6-1=63$ different ways presented in the table of the Appendix \ref{AppendixA}. In this table $\mathbf{f}^{(0)}_{\,k,\textsc{VAC}}$ denotes the candidate dominant part of the field, and $\mathbf{f}^{\,(\textrm{sub})}_{\,k,\textsc{VAC}}$ its subdominant one. Each of these 63 different decompositions represents the possible ways the field may dominate the evolution of the system. However, for any one of these ways to be an admissible one, certain conditions have to be satisfied as our subsequent asymptotic analysis will show.

%========================================================================
\section{The all-terms-dominant decomposition}

Having found a complete profile of the dynamical field decompositions, the next step is to look for the admissible dominant feature allowed by each vector field splitting, the so-called dominant balances of the field near the finite-time singularity.

%=====================================================================
\subsection{Dominant exponents analysis}

In order to search for the possible dominant balances, we begin with an analysis of the last decomposition, $\mathbf{f^{63}}_{\,k,\textsc{VAC}}$ (also called the \emph{all-terms-dominant} decomposition), from which we can extract useful qualitative conclusions for the total of the possible dominant decompositions. Substituting the forms, 
\be
\mathbf{x}(t)=\mathbf{a}t^{\mathbf{p}}=(\theta t^{p}, \eta t^{q}, \rho t^{r}). \label{eq:domisol3}
\ee 
into the dominant system $(\dot x,\dot y,\dot z)(t)=\mathbf{f^{63}}^{(0)}_{\,k,\textsc{VAC}}$, we obtain a nonlinear algebraic system for the coefficients and the principal exponents of the dominant balance. Solving this system, we may determine the dominant balance $(\mathbf{a},\mathbf{p})$, as an exact, scale invariant solution, where $\mathbf{a}=(\theta, \eta, \rho)\in\mathbb{C}^3$ are constants and $\mathbf{p}=(p, q, r)\in\mathbb{Q}^3$. The dominant system takes the form,
\begin{eqnarray}
\theta pt^{p-1} &=& \eta t^{q},\label{eq:ATDsys1}\\[15pt]
\eta qt^{q-1} &=& \frac{\eta^2}{2\theta}\:t^{2q-p} -3\theta\eta\:t^{p+q}+k\theta\rho\:t^{p+r}-\frac{k^2\rho^2}{2\theta}\:t^{2r-p}\nonumber\\[10pt]
&-&\frac{\theta}{12\epsilon}\:t^{p}-\frac{k\rho}{12\theta\epsilon}\:t^{r-p},\label{eq:ATDsys2}\\[15pt]
\rho rt^{r-1} &=&-2\theta\rho t^{p+r}.\label{eq:ATDsys3}
\end{eqnarray}
We can make the following general observations for the components of the exponents vector $\mathbf{p}=(p, q, r)$. Since the term $-2xz$ is the third component in the dominant part of every one of the 63 asymptotic decompositions, Eq. (\ref{eq:ATDsys3}) leads to $p=-1$. Consequently this is the only possible value of p common in any dominant balance. 

Following that, and since Eq. (\ref{eq:ATDsys1}) gives $p-1=q$ we conclude that $q=-2$ and this is also the only possible common value of q in any dominant balance, the term $y$ being the first component in the dominant part of every asymptotic decomposition.

Each term of the RHS of Eq. (\ref{eq:ATDsys3}) leads to certain equations for the components of $\mathbf{p}=(p, q, r)$ which have to be satisfied for the specific values $p=-1$ and $q=-2$ found from solving Eqs. (\ref{eq:ATDsys1}) and (\ref{eq:ATDsys3}). Hence, we find that, 
\begin{eqnarray}
q-1 &=& 2q-p,\:\:\:  \text{for the term}\:\:\:\: +\frac{\eta^2}{2\theta}\:t^{2q-p}\label{eq:ATDexp1} \\[10pt]
q-1 &=& p+q,\:\:\:  \text{for the term}\:\:\:\: -3\theta\eta\:t^{p+q}\label{eq:ATDexp2} \\[10pt]
q-1 &=& p+r,\:\:\:  \text{for the term}\:\:\:\: +k\theta\rho\:t^{p+r}\label{eq:ATDexp3} \\[10pt]
q-1 &=& 2r-p,\:\:\:  \text{for the term}\:\:\:\: -\frac{k^2\rho^2}{2\theta}\:t^{2r-p}\label{eq:ATDexp4} \\[10pt]
q-1 &=& p,\:\:\:  \text{for the term}\:\:\:\: -\frac{\theta}{12\epsilon}\:t^{p}\label{eq:ATDexp5} \\[10pt]
q-1 &=& r-p,\:\:\:  \text{for the term}\:\:\:\: -\frac{k\rho}{12\theta\epsilon}\:t^{r-p}\label{eq:ATDexp6}
\end{eqnarray}
Consequently, there is no vector $\mathbf{p}=(p, q, r)$ that satisfies the dominant system (\ref{eq:ATDsys1})-(\ref{eq:ATDsys3}), so we conclude that the all-terms-dominant decomposition $\mathbf{f^{63}}_{\,k,\textsc{VAC}}$ does not admit a dominant balance asymptotically towards the singularity.

In addition, there are a number of very useful observations which can be advanced based on the above analysis:
\begin{itemize}
\item The values $p=-1$ and $q=-2$ are the only acceptable ones for any possible dominant balance.
\item Having said that, we can see that, for these values of p and q, Eq.(\ref{eq:ATDexp5}), which is coming from the only linear term of the system, becomes impossible to satisfy. Consequently, the 32 asymptotic decompositions that contain the linear term $-\frac{x}{12\epsilon}$ in their dominant parts cannot admit a dominant balance and may be ignored in our list in Table (\ref{long}) leaving us with 31 possible asymptotic decompositions.
\item For $p=-1$ and $q=-2$, Eq. (\ref{eq:ATDexp6}) leads to $r=-4$ while Eqs. (\ref{eq:ATDexp3}) and (\ref{eq:ATDexp4}) lead to $r=-2$. Consequently, the term $-\frac{kz}{12\epsilon x}$ cannot coexist with neither one of the terms $+kxz$ and $-\frac{k^2 z^2}{2x}$ in the dominant part of an acceptable asymptotic decomposition. Thus 12 of the remaining 31 possible asymptotic decompositions can be crossed out as well, leaving 19 candidates in all.
\end{itemize}
This last observation leads to an interesting interpretation of our current results. This is related to the fact that Eqs. (\ref{eq:ATDexp3}),(\ref{eq:ATDexp4}) and (\ref{eq:ATDexp6}) are obtained from the terms $-\frac{kz}{12\epsilon x}$, $+kxz$ and $-\frac{k^2 z^2}{2x}$ respectively, namely, the only terms of the vector field (\ref{fkvac}) that contain the curvature term $k$. As shown above the appearance of those terms in the dominant part of a specific possible asymptotic decomposition forces $r$ to take some specific value (i.e. $r=-2$ or $r=-4$), while their simultaneous appearance in the subdominant part leaves $r$ as an arbitrary parameter in this phase of the procedure. As we will see later in this Chapter, it is exactly this fact that causes the appearance of one more arbitrary parameter in one of our final asymptotic solutions which subsequently identifies the obtained solution as a \emph{general} asymptotic solution of the dynamical system corresponding to the vector field $\mathbf{f}_{\,k,\textsc{VAC}}$.

%======================================================================
\subsection{Coefficients analysis}

Before we conclude the analysis of the all-terms-dominant decomposition we study the equations satisfied by the coefficients of the dominant system (\ref{eq:ATDsys1})-(\ref{eq:ATDsys3}), that is the components of the vector $\mathbf{a}=(\theta,\: \eta,\: \rho)$ in the dominant balance. We have
\begin{eqnarray}
\theta p &=& \eta,\label{eq:ATDco1}\\[10pt]
\eta q &=& \frac{\eta^2}{2\theta} -3\theta\eta+k\theta\rho-\frac{k^2\rho^2}{2\theta}\nonumber\\[10pt]
&-&\frac{\theta}{12\epsilon}-\frac{k\rho}{12\theta\epsilon},\label{eq:ATDco2}\\[10pt]
\rho r &=&-2\theta\rho.\label{eq:ATDco3}
\end{eqnarray}
Therefore, substituting the values $-1$ and $-2$ for p and q respectively, we find
\begin{eqnarray}
-\theta &=& \eta,\label{eq:ATDcoe1}\\[10pt]
-2\eta &=& \frac{\eta^2}{2\theta} -3\theta\eta+k\theta\rho-\frac{k^2\rho^2}{2\theta}\nonumber\\[10pt]
&-&\frac{\theta}{12\epsilon}-\frac{k\rho}{12\theta\epsilon},\label{eq:ATDcoe2}\\[10pt]
\rho\left( r+2\theta \right)&=& 0.\label{eq:ATDcoe3}
\end{eqnarray}
Since the first and third components of the vector field (\ref{fkvac}), namely, $y$ and $-2xz$, have only one single term we conclude that Eqs. (\ref{eq:ATDcoe1}) and (\ref{eq:ATDcoe3}) must be satisfied by the components of any possible vector $\mathbf{a}=(\theta,\: \eta,\: \rho)$ in the balance. More specifically:
\begin{itemize}
\item The vector $\mathbf{a}$ of any admissible dominant balance $(\mathbf{a},\mathbf{p})$ will have the form $\mathbf{a}=(\theta,\: -\theta,\: \rho)$.
\item From Eq. (\ref{eq:ATDcoe3}), we have that $\rho=0$ or $\theta=-\frac{r}{2}$.
\end{itemize}

%=======================================================================
\section{The hypothetical all-terms-subdominant decomposition}

Although it is impossible to have an all-terms-subdominant asymptotic decomposition since in that case by definition there would not exist any dominant asymptotic behavior in the first place, it is instructive to study the hypothetical case where the subdominant part of a possible asymptotic decomposition would contain all the terms of the initial vector field (\ref{fkvac}) and the rest would comprise the dominant part $\mathbf{f}^{(0)}_{\,k,\textsc{VAC}}$, so that their sum would be the same, namely that which in this chapter we denote by $\mathbf{f}_{\,k,\textsc{VAC}}$. Therefore, let us suppose that there exists a possible asymptotic decomposition whose subdominant part has the form
\be
\mathbf{f^h}^{(sub)}_{\,k,\textsc{VAC}}=\left( 0,\frac{y^{2}}{2x}-3xy+kxz-\frac{k^2 z^2}{2x}-\frac{x}{12\epsilon}-\frac{kz}{12\epsilon x},0\right),\label{hsub}
\ee
where $h$ stands for hypothetical. Then, as we have already seen in Chapter 3 we need to examine the limit of the following expression as $t\rightarrow 0$, in order to confirm that the subdominant part is indeed subdominant asymptotically.  
\be\label{subcon}
\frac{\mathbf{f}^{\,(\textrm{sub})}_{\,k,\textsc{VAC}}(\mathbf{a}t^{\mathbf{p}})}{t^{\mathbf{p}-1}}.
\ee
Of course, in this case we do not have an admissible dominant balance $(\mathbf{a},\mathbf{p})$ but we may use a dominant balance that satisfies all the conditions that we have found so far. More specifically, we will use a dominant balance of the form $(\mathbf{a_h},\mathbf{p_h})=\left((\theta,\: -\theta,\: \rho),(-1,-2,r)\right)$, having in mind that $r$ can either take the values $-2$, $-4$ or else be unspecified. In this case, we have
\be\begin{split}
\mathbf{f^h}^{(sub)}_{\,k,\textsc{VAC}}\left(\mathbf{a_h}t^{\mathbf{p_h}}\right)=\left(0,\:\:\frac{\eta^2}{2\theta}\:t^{2q-p}-3\theta\eta\:t^{p+q}+k\theta\rho\:t^{p+r}\right.\\[13pt]
\left.-\frac{k^2\rho^2}{2\theta}\:t^{2r-p}-\frac{\theta}{12\epsilon}\:t^{p}-\frac{k\rho}{12\theta\epsilon}\:t^{r-p},\:\:0 \right),\label{hsubconup}
\end{split}\ee
and
\be
t^{\mathbf{p}-1} = \left(t^{p-1},\: t^{q-1},\: t^{r-1}\right).\label{hsubcondown}
\ee
Subsequently, after substituting for $(\mathbf{a_h},\mathbf{p_h})$,  Eqs. (\ref{hsubconup}) and (\ref{hsubcondown}) take the form, 
\be\begin{split}
\mathbf{f^h}^{(sub)}_{\,k,\textsc{VAC}}\left(\mathbf{a_h}t^{\mathbf{p_h}}\right)=\left(0,\:\:\frac{\theta}{2}\:t^{-3}+3\theta^{2}\:t^{-3}+k\theta\rho\:t^{r-1}\right.\\[13pt]
\left.-\frac{k^2\rho^2}{2\theta}\:t^{2r+1}-\frac{\theta}{12\epsilon}\:t^{-1}-\frac{k\rho}{12\theta\epsilon}\:t^{r+1},\:\:0 \right),\label{hsubconup2}
\end{split}\ee
and
\be
t^{\mathbf{p_h}-1} = \left(t^{-2},\: t^{-3},\: t^{r-1}\right),\label{hsubcondown2}
\ee
which gives,
\be\begin{split}
\frac{\mathbf{f^h}^{\,(\textrm{sub})}_{\,k,\textsc{VAC}}(\mathbf{a_h}t^{\mathbf{p_h}})}{t^{\mathbf{p_h}-1}}= \left(0,\:\:-\frac{\theta}{2}\:t^{0} +3\theta^{2}\:t^{0}+k\theta\rho\:t^{r+2}\right.\\[13pt]
\left.-\frac{k^2\rho^2}{2\theta}\:t^{2r+4}-\frac{\theta}{12\epsilon}\:t^{2}-\frac{k\rho}{12\theta\epsilon}\:t^{r+4},\:\:0 \right),\label{hsublim}
\end{split}\ee
or, equivalently,
\begin{eqnarray}
\frac{\mathbf{f^h}^{\,(\textrm{sub})}_{\,k,\textsc{VAC}}(\mathbf{a_h}t^{\mathbf{p_h}})}{t^{\mathbf{p_h}-1}}&=&\underbrace{\left(0,\:\:-\frac{\theta}{2},\:\:0\right)}_{\text{1st term}} +\underbrace{\left(0,3\theta^{2},0\right)}_{\text{2nd term}}\nonumber\\[15pt]
&+&\underbrace{\left(0,\:k\theta\rho,\:\:0\right)t^{r+2}}_{\text{3rd term}}+\underbrace{\left(0,\:\:-\frac{k^2\rho^2}{2\theta},\:\:0\right)t^{2(r+2)}}_{\text{4th term}}\nonumber\\[15pt]
&+&\underbrace{\left(0,\:\:-\frac{\theta}{12\epsilon},\:\:0\right)t^{2}}_{\text{5th term}}+\underbrace{\left(0,\:\:-\frac{k\rho}{12\theta\epsilon},\:\:0\right)t^{r+4}}_{\text{6th term}}.\label{hsublim2}
\end{eqnarray}
Taking now the limit as $t\rightarrow 0$, although we cannot reach to a conclusion about the actual subdominant part of an existing possible asymptotic decomposition, we can gain a valuable insight of the subdominant behavior of each term of the second component of (\ref{fkvac}). Below we examine separately the subdominant behavior of each term of the second component of (\ref{fkvac}) in comparison with their corresponding terms in Eq. (\ref{hsublim2}) and we explain the reasons for which 16 more of the 19 possible asymptotic decompositions left fail to lead to an asymptotic solution leaving us finally with 3 admissible asymptotic decompositions out of the initial 63.

\begin{itemize}
\item The first term, $\frac{y^{2}}{2x}$, which corresponds to the first term of the RHS of Eq. (\ref{hsublim2}) has a strongly dominant character in the sense that when it appears in the subdominant part it is most likely that the expression (\ref{subcon}) will not vanish as $t\rightarrow 0$ except in the cases when $\theta=0$ or when $\theta$ might take such a value that the term $-\frac{\theta}{2}$ will be eliminated by addition to a term with the opposite sign. Solving one by one the dominant systems of the 19 left possible asymptotic decompositions of the initial vector field, (\ref{fkvac}), we find that the appearance of $\frac{y^{2}}{2x}$ in the subdominant part leads indeed to the exclusion of the relevant decompositions, except of one interesting case which we will study in the next section. 
\item The second term, $-3xy$, which corresponds to the second term of the RHS of Eq. (\ref{hsublim2}) has also a strongly dominant character in the same sense as the first term. Again solving one by one the dominant systems of the 19 remaining asymptotic decompositions, we find that its appearance in the subdominant part of a possible asymptotic decomposition leads in every case to the exclusion of that specific decomposition.
\item The third and fourth terms, $+kxz$ and $-\frac{k^2 z^2}{2x}$, corresponding to the third and fourth terms respectively of the RHS of Eq. (\ref{hsublim2}) are, as we have seen previously, the ones responsible for $r$ taking the value $-2$ when they appear in the dominant part of a possible asymptotic decomposition. For that same value of $r$, both terms do not vanish as $t\rightarrow 0$, except in the case when $\rho=0$ or $\theta=0$, for the third term only. Thus, one might expect these two terms are in general `bound' to one another, in the sense that either they both have to appear in the dominant part of a decomposition (causing $r$ to take the value $-2$), or they both have to appear in the subdominant part. If they are split, the one appearing in the dominant part will cause $r$ to take the value $-2$, while the other (appearing in the subdominant part) will cause the expression (\ref{subcon}) not to vanish as $t\rightarrow 0$ for that same value of $r$. Again, we note that this is one more reason causing some of the possible asymptotic decompositions to fail to lead to an acceptable asymptotic solution of the dynamical system in question. Although the above argument makes it highly, below we also treat the case, where these two terms have different dominant behavior but the values of the dominant balance are such that they allow the specific decomposition to lead to an asymptotic solution. 
\item The fifth term, $-\frac{x}{12\epsilon}$, corresponds to the fifth term of the RHS of Eq. (\ref{hsublim2}). As expected, this linear term shows an absolute subdominant asymptotic behavior. It was already clear from the previous subsection that it was impossible to exist in the dominant part of any valid asymptotic decomposition and it was also met presently.
\item Finally, the sixth term $-\frac{kz}{12\epsilon x}$ which corresponds to the sixth term of the RHS of Eq. (\ref{hsublim2}), is one of the terms which include the curvature term $k$. Its subdominant behavior depends on the value of $r$, and as we have already seen it cannot coexist with the third and fourth terms in the dominant part of an admissible asymptotic decomposition.
\end{itemize}

%=====================================================================
\section{The admissible asymptotic decompositions}

For the reasons explained so far, only 3 of the 63 decompositions of Table (\ref{long}) eventually lead to fully acceptable dominant balances, while the rest 60 decompositions fail to do so. Therefore, the only acceptable asymptotic splittings of the vector field  $\mathbf{f}_{\,k,\textsc{VAC}}$ of the general form $\mathbf{f}_{\,k,\textsc{VAC}}=\mathbf{f}^{(0)}_{\,k,\textsc{VAC}} + \mathbf{f}^{\,(\textrm{sub})}_{\,k,\textsc{VAC}}$,
have the following dominant parts
\be \label{eq:curveddom1}
\mathbf{f^{7}}^{(0)}_{\,k,\textsc{VAC}} =\left(y, \frac{y^2}{2x}-3xy,-2xz\right),
\ee

\be \label{eq:curveddom2}
\mathbf{f^{12}}^{(0)}_{\,k,\textsc{VAC}}=\left(y, -3xy+kxz,-2xz\right) ,
\ee

\be \label{eq:curveddom3}
\mathbf{f^{42}}^{(0)}_{\,k,\textsc{VAC}}=\left(y, \frac{y^2}{2x}-3xy+kxz-\frac{k^2 z^2}{2x},-2xz\right) ,
\ee
while their subdominant parts are given respectively by the forms,
\be \label{eq:curvedsub1}
\mathbf{f^{7}}^{(\textrm{sub})}_{\,k,\textsc{VAC}} =\left(0, kxz-\frac{k^2 z^2}{2x}-\frac{x}{12\epsilon}-\frac{kz}{12x\epsilon},0\right) ,
\ee

\be \label{eq:curvedsub2}
\mathbf{f^{12}}^{(\textrm{sub})}_{\,k,\textsc{VAC}}=\left(0, \frac{y^2}{2x}-\frac{k^2 z^2}{2x}-\frac{x}{12\epsilon}-\frac{kz}{12x\epsilon},0\right) ,
\ee

\be \label{eq:curvedsub3}
\mathbf{f^{42}}^{(\textrm{sub})}_{\,k,\textsc{VAC}}=\left(0, -\frac{x}{12\epsilon}-\frac{kz}{12x\epsilon},0\right).
\ee
In the next section, we present the asymptotic analysis of the first of the three finally accepted decompositions and construct the asymptotic solution in the form of a formal series expansion.

%--------------------------------------------------------------------------------------------------------------------
%	SECTION 4.2 - General asymptotic solution
%--------------------------------------------------------------------------------------------------------------------
\section{General asymptotic solutions}

%============================================================================================================
\subsection{The dominant balance of the general solution}

We now proceed with the construction of the asymptotic solution for the first of the three admissible asymptotic decompositions, namely, the one that leads to a \emph{general} solution which will also allow us to make certain conclusions for the stability of that solution. The decomposition that leads to the general solution is $\mathbf{f^{7}}_{\,k,\textsc{VAC}}$ of Table (\ref{long}). 

To obtain its asymptotic balance $\mathcal{B^{7}}_{\,k,\textsc{VAC}} \in\mathbb{C}^3\times\mathbb{Q}^3$, we solve the dominant system $\dot{\mathbf{x}}(t)=\mathbf{f^7}^{(0)}_{\,k,\textsc{VAC}}$ obtained by substituting a scale-invariant solution of the form,
\be
\mathbf{x}(t)=\mathbf{a}t^{\mathbf{p}}=(\theta t^{p}, \eta t^{q}, \rho t^{r}). \label{eq:domsol}
\ee
That is we find,
\begin{eqnarray}
\theta p t^{p-1}&=&\eta t^{q},\label{eq:gdoms1}\\[12pt]
\eta q t^{q-1}&=&\frac{\theta^2}{2\eta}\:t^{2q-p}-3\theta\eta\:t^{p+q},\label{eq:gdoms2}\\[12pt]
\rho r t^{r-1}&=&-2\theta\rho t^{p+r}.\label{eq:gdoms3}
\end{eqnarray}
Consequently, we are led to the dominant balance,

\be \label{curvedbalance1}
\mathcal{B^{7}}_{\,k,\textsc{VAC}}=(\mathbf{a},\mathbf{p})= \left(\left(\frac{1}{2},-\frac{1}{2},\r \right),\: \left(-1,-2,-1\right)\right).
\ee
In particular, this means that the vector field $\mathbf{f^{7}}^{(0)}_{\,k,\textsc{VAC}}$ is a scale-invariant system.

%============================================================================================================
\subsection{Subdominant condition}

Subsequently, we need to show that the higher-order terms (\ref{eq:curvedsub1}) in the basic decomposition of the vacuum field are themselves weight-homogeneous with respect to the balance (\ref{curvedbalance1}) for this to be an acceptable one. To prove this, we first split the subdominant part (\ref{eq:curvedsub1}) by writing
\be
\mathbf{f^{7}}^{\,(\textrm{sub})}_{\,k,\textsc{VAC}}(\mathbf{x}) = \mathbf{f^{7}}^{(1)}_{\,k,\textsc{VAC}}(\mathbf{x}) +
\mathbf{f^{7}}^{(2)}_{\,k,\textsc{VAC}}(\mathbf{x}) + \mathbf{f^{7}}^{(3)}_{\,k,\textsc{VAC}}(\mathbf{x}),
\ee
where
\begin{eqnarray}
\mathbf{f^{7}}^{(1)}_{\,k,\textsc{VAC}}(\mathbf{x})&=&\left(0,kxz,0 \right),\nonumber\\[10pt]
\mathbf{f^{7}}^{(2)}_{\,k,\textsc{VAC}}(\mathbf{x})&=&\left(0,- \frac{k^2 z^2}{2x}-\frac{x}{12\epsilon},0 \right),\nonumber\\[10pt]
\mathbf{f^{7}}^{(3)}_{\,k,\textsc{VAC}}(\mathbf{x})&=&\left(0,-\frac{kz}{12x\epsilon},0 \right),
\end{eqnarray}
and using the balance $\mathcal{B^{7}}_{\,k,\textsc{VAC}} $ defined by Eq. (\ref{curvedbalance1}), we find that

\bq
\frac{\mathbf{f^{7}}^{(1)}_{\,k,\textsc{VAC}}(\mathbf{a}t^{\mathbf{p}})}{t^{\mathbf{p}-1}}&=&
\mathbf{f^{7}}^{(1)}_{\,k,\textsc{VAC}}(\mathbf{a})t =\left(0,\frac{k\:\r}{2}\:\:t,0 \right),\\[10pt] \frac{\mathbf{f^{7}}^{(2)}_{\,k,\textsc{VAC}}(\mathbf{a}t^{\mathbf{p}})}{t^{\mathbf{p}-1}}&=&
\mathbf{f^{7}}^{(2)}_{\,k,\textsc{VAC}}(\mathbf{a})t^{2}=\left(0,\left(- k^2 \r^2-\frac{1}{24\epsilon}\right)\:\:t^2,0 \right),\\[10pt]
\frac{\mathbf{f^{7}}^{(3)}_{\,k,\textsc{VAC}}(\mathbf{a}t^{\mathbf{p}})}{t^{\mathbf{p}-1}}&=&
\mathbf{f^{7}}^{(3)}_{\,k,\textsc{VAC}}(\mathbf{a})t^{3}=\left(0,-\frac{k\r}{6\epsilon}\:\:t^3,0 \right).
\eq
Hence, taking the limit as $t\rightarrow 0$, we see that these forms go to zero asymptotically provided that $\mathbf{f}^{(i)}_{\,k,\textsc{VAC}_1}(\mathbf{a}), i=1,2,3$ are  all different from zero. This happens only when
$\epsilon\neq 0$, that is for all cases except when $3\b +\g =0$.
Since the \emph{subdominant exponents}
\be\label{sub exps}
q^{(0)}=0\ <\ q^{(1)}=1\ <\ q^{(2)}=2\ <\ q^{(3)}=3,
\ee
are ordered, we conclude that the subdominant part (\ref{eq:curvedsub1}) is weight-homogeneous as promised.

%============================================================================================================
\subsection{Construction of the K-matrix}

Further, we calculate the Kovalevskaya matrix given by
\be
\mathcal{K}^{7}_{\,k,\textsc{VAC}}=D\,\mathbf{f^{7}}^{(0)}_{K,\textsc{VAC}}(\mathbf{a})-\textrm{diag}(\mathbf{p}),
\ee
which leads to,
\be
\mathcal{K}^{7}_{\,k,\textsc{VAC}}=\left(
                     \begin{array}{ccc}
                       1 & 1  & 0   \\
                       1 & -1/2 & 0 \\
                       -2\rho & 0 & 0 \\
                     \end{array}
                   \right),
\ee
with spectrum
\be
\textrm{spec}(\mathcal{K}^{7}_{\,k,\textsc{VAC}})=\{0,-1,3/2\}.
\ee

%============================================================================================================
\subsection{Substitution of the Fuchsian series expansion}

In order to find that solution, we substitute the \emph{Fuchsian series expansions} and their derivatives

\begin{equation} \label{eq:cseries}
x(t) = \sum_{i=0}^{\infty} c_{1i} t^{\frac{i}{2}-1}, \:\:\:\:\:
y(t) = \sum_{i=0}^{\infty} c_{2i}  t^{\frac{i}{2}-2},\:\:\:\:\:
z(t) = \sum_{i=0}^{\infty} c_{3i} t^{\frac{i}{2}-1} ,\:\:\:\:\:
\end{equation}
\begin{equation} \label{eq:cseriesder}
\begin{split}
\dot{x}(t) = \sum_{i=0}^{\infty} c_{1i} \left(\frac{i}{2}-1\right) t^{\frac{i}{2}-2},\:\:\:\:\:
\dot{y}(t) = \sum_{i=0}^{\infty} c_{2i} \left(\frac{i}{2}-2\right) t^{\frac{i}{2}-3},\\[10pt]
\dot{z}(t) = \sum_{i=0}^{\infty} c_{3i} \left(\frac{i}{2}-\rho\right) t^{\frac{i}{2}-2} ,\:\:\:\:\:
\end{split}
\end{equation}
where  $c_{10}=1/2$ and $\:c_{20}=-1/2$, $\:c_{30}=\r$, in the following equivalent form of the original system (\ref{eq:ds}), namely
\begin{eqnarray}
\dot{x} &=& y,\label{cds1}\\[10pt]
2x\dot{y} &=& y^{2}-6x^{2}y +2kx^{2}z-k^2z^{2}-\frac{1}{6\epsilon}x^{2}-\frac{k}{6\epsilon}z,\label{cds2}\\[10pt]
\dot{z} &=& -2xz.\label{cds3}
\end{eqnarray}
from which we will be led to various recursion relations that determine the unknowns $c_{1i}, c_{2i}, c_{3i}$ term by term. More specifically from Eq. (\ref{cds1}) after substitution we have

\be
\sum_{i=0}^{\infty} c_{1i} \left(\frac{i}{2}-1\right) t^{\frac{i}{2}-2}=\sum_{i=0}^{\infty} c_{2i}  t^{\frac{i}{2}-2},
\ee
which leads to
\be
c_{1i} \left(\frac{i}{2}-2\right) = c_{2i}\:.\label{eq:substi1c}
\ee
From Eq. (\ref{cds2}) we calculate separately each term after substitution:

\begin{eqnarray}
2x\dot{y}&=& 2\left(\sum_{i=0}^{\infty}c_{1i}\:\:t^{\frac{i}{2}-1}\right)\left(\sum_{i=0}^{\infty}c_{2i} \left(\frac{i}{2}-2\right)\:\: t^{\frac{i}{2}-3} \right)\nonumber\\[13pt]
&=& 2\:t^{-4} \sum_{i=0}^{\infty} \sum_{l=0}^{i} c_{2(i-l)} \left(\frac{i-l}{2}-2\right)c_{1l}\:\: t^{\frac{i}{2}},
\end{eqnarray}
\begin{eqnarray}
y^{2}&=&  \left( \sum_{i=0}^{\infty} c_{2i}\:\: t^{\frac{i}{2}-2} \right)\left( \sum_{i=0}^{\infty} c_{2i}\:\:  t^{\frac{i}{2}-2} \right) \nonumber\\[13pt]
&=& t^{-4} \sum_{i=0}^{\infty} \sum_{l=0}^{i} c_{2(i-l)} c_{2l}\:\: t^{\frac{i}{2}},\\[15pt]
-6x^{2}y &=& -6  \left(\sum_{i=0}^{\infty} c_{1i}\:\: t^{\frac{i}{2}-1}\right) \left(\sum_{i=0}^{\infty} c_{1i}\:\: t^{\frac{i}{2}-1}\right) \left( \sum_{i=0}^{\infty} c_{2i} \:\: t^{\frac{i}{2}-2} \right) \nonumber\\[13pt]
&=& -6\:t^{-4} \sum_{i=0}^{\infty} \sum_{l=0}^{i} \sum_{m=0}^{k} c_{1(i-l)} c_{1(l-l)} c_{2m} \:\:t^{\frac{i}{2}}, \\[15pt]
2kx^{2}z &=& 2k  \left(\sum_{i=0}^{\infty} c_{1i}\:\: t^{\frac{i}{2}-1}\right) \left(\sum_{i=0}^{\infty} c_{1i} \:\:t^{\frac{i}{2}-1}\right) \left( \sum_{i=0}^{\infty} c_{3i} \:\: t^{\frac{i}{2}-1} \right) \nonumber\\[13pt]
&=& 2k \:t^{-3} \sum_{i=0}^{\infty} \sum_{l=0}^{i} \sum_{m=0}^{k} c_{1(i-l)} c_{1(m-l)} c_{3l}\:\: t^{\frac{i}{2}},\\[15pt] 
-k^{2}z^{2}&=& -k^{2} \left( \sum_{i=0}^{\infty} c_{3i} \:\: t^{\frac{i}{2}-1} \right)\left( \sum_{i=0}^{\infty} c_{3i}\:\:  t^{\frac{i}{2}-1} \right) \nonumber\\[13pt]
&=& -k^{2}\:t^{-2} \sum_{i=0}^{\infty} \sum_{l=0}^{i} c_{3(i-l)} c_{3l}\:\: t^{\frac{i}{2}},\\[15pt]
-\frac{1}{6\epsilon}x^{2}&=& -\frac{1}{6\epsilon}\left(\sum_{i=0}^{\infty} c_{1i}\:\: t^{\frac{i}{2}-1}\right) \left(\sum_{i=0}^{\infty} c_{1i}\:\: t^{\frac{i}{2}-1}\right)  \nonumber \\[13pt]
&=& -\frac{1}{6\epsilon}\:t^{-2} \sum_{i=0}^{\infty} \sum_{k=0}^{i} c_{1(i-k)} c_{1k} \:\:t^{\frac{i}{2}},\\[15pt]
-\frac{k}{6\epsilon}z&=& -\frac{k}{6\epsilon}\:t^{-1} \sum_{i=0}^{\infty} c_{3i} \:\: t^{\frac{i}{2}}.
\end{eqnarray}
Subsequently, we are led to the following form of (\ref{cds2}),
\begin{eqnarray}
2\:\:t^{-4} \sum_{i=0}^{\infty} \sum_{l=0}^{i} c_{2(i-l)} \left(\frac{i-l}{2}-2\right)c_{1l}\:\: t^{\frac{i}{2}}=t^{-4} \sum_{i=0}^{\infty} \sum_{l=0}^{i} c_{2(i-l)} c_{2l}\:\: t^{\frac{i}{2}}\nonumber\\[13pt]
-6\:t^{-4} \sum_{i=0}^{\infty} \sum_{l=0}^{i} \sum_{m=0}^{k} c_{1(i-l)} c_{1(l-m)} c_{2m}\:\: t^{\frac{i}{2}}+2k \:t^{-3} \sum_{i=0}^{\infty} \sum_{l=0}^{i} \sum_{m=0}^{k} c_{1(i-l)} c_{1(m-l)} c_{3l} t^{\frac{i}{2}}\nonumber\\[13pt] 
-k^{2}\:t^{-2} \sum_{i=0}^{\infty} \sum_{l=0}^{i} c_{3(i-l)} c_{3l}\: t^{\frac{i}{2}}
-\frac{1}{6\epsilon}\:t^{-2} \sum_{i=0}^{\infty} \sum_{k=0}^{i} c_{1(i-k)} c_{1k}\: t^{\frac{i}{2}}-\frac{k}{6\epsilon}\:t^{-1} \sum_{i=0}^{\infty} c_{3i}\:t^{\frac{i}{2}}
\label{eq:substi2c}
\end{eqnarray}
Finally, from Eq. (\ref{cds3}) after substitution we have
\be
\sum_{i=0}^{\infty} c_{3i} \left(\frac{i}{2}-1\right)\: t^{\frac{i}{2}-2}=-2\left(\sum_{i=0}^{\infty} c_{1i}  t^{\frac{i}{2}-1}\right)\left(\sum_{i=0}^{\infty} c_{3i}\:\:t^{\frac{i}{2}-1}\right),
\ee
which leads to
\be
t^{-2}\sum_{i=0}^{\infty} c_{3i} \left(\frac{i}{2}-1\right)\: t^{\frac{i}{2}}=-2\:t^{-2}\sum_{i=0}^{\infty}\sum_{l=0}^{i} c_{1(i-l)}c_{3l}\:\:t^{\frac{i}{2}}.\label{eq:substi3c}
\ee

%============================================================================================================
\subsection{Calculation of the final series coefficients}

Eqs. (\ref{eq:substi1c}),(\ref{eq:substi2c}) and (\ref{eq:substi3c}) constitute the system from which we calculate term by term the coefficients $c_{1i}$, $c_{2i}$ and $c_{3i}$ of the asymptotic solution of the initial dynamical system (\ref{cds1})-(\ref{cds3}) in the form of the Fuchsian series expansions (\ref{eq:cseries}), that is
\begin{eqnarray}
x(t)&=&\frac{1}{2}t^{-1} +c_{11}\:\:t^{-1/2} +c_{12}\:\:t^{0}+c_{13}\:\:t^{1/2} +c_{14}\:\:t^{1} + \cdots,\nonumber\\[10pt]
y(t)&=&-\frac{1}{2}t^{-2} +c_{21}\:\:t^{-3/2} +c_{22}\:\:t^{-1}+c_{23}\:\:t^{-1/2} +c_{24}\:\:t^{0} + \cdots,\nonumber\\[10pt]
z(t)&=&\rho t^{-1} +c_{31}\:\:t^{-1/2} +c_{32}\:\:t^{0}+c_{33}\:\:t^{1/2} +c_{34}\:\:t^{1} + \cdots,
\end{eqnarray}
We determine for each of the Eqs. (\ref{eq:substi1c}), (\ref{eq:substi2c}) and (\ref{eq:substi3c}) a different set of equations for the coefficients of the various powers of $t$ which will eventually give us the values of $c_{1i}$, $c_{2i}$ and $c_{3i}$. 

%============================================================================================================
\subsubsection*{1st set of equations}

For the coefficients of the different powers of $t$, Eq. (\ref{eq:substi1c}) leads to the following equations,
\bq
\text{for the coefficients of the term}\:\: &t^{-3/2},\:\:\:\:c_{11} \left(\frac{1}{2}-1\right) = c_{21}\:,\\[12pt]
\text{for the coefficients of the term}\:\: &t^{-1},\:\:\:\:c_{12} \left(\frac{2}{2}-1\right) = c_{22}\:,\\[12pt]
\text{for the coefficients of the term}\:\: &t^{-1/2},\:\:\:\:c_{13} \left(\frac{3}{2}-1\right) = c_{23}\:,\\[12pt]
\text{for the coefficients of the term}\:\: &t^{0},\:\:\:\:c_{14} \left(\frac{4}{2}-1\right) = c_{24}\:. 
\eq

%============================================================================================================
\subsubsection*{2nd set of equations}

Subsequently, for the coefficients of the different powers of $t$, Eq. (\ref{eq:substi2c}) leads to the following equations:\\
For the coefficients of the term $t^{-7/2}$, we have
\be
2\sum_{l=0}^{1} c_{2(i-l)} \left(\frac{i-l}{2}-2\right)c_{1l}= \sum_{l=0}^{1} c_{2(i-l)} c_{2l}-6\sum_{l=0}^{1} \sum_{m=0}^{k} c_{1(i-l)} c_{1(l-m)} c_{2m}\:,
\ee
which leads to 
\be
c_{11}=c_{21}\label{coef2-1c}.
\ee
For the coefficients of the term $t^{-3}$, we have
\be
\begin{split}
2\sum_{l=0}^{2} c_{2(i-l)} \left(\frac{i-l}{2}-2\right)c_{1l}= \sum_{l=0}^{2} c_{2(i-l)} c_{2l}-6\sum_{l=0}^{2} \sum_{m=0}^{l} c_{1(i-l)} c_{1(l-m)} c_{2m}\\[12pt]
+2k \sum_{l=0}^{0} \sum_{m=0}^{l} c_{1(i-l)} c_{1(m-l)} c_{3l}\:,
\end{split}
\ee
which leads to
\be
c_{12}=\frac{k\rho}{2}\label{coef2-2c}.
\ee
For the coefficients of the term $t^{-5/2}$, we have
\be
\begin{split}
2\sum_{l=0}^{3} c_{2(i-l)} \left(\frac{i-l}{2}-2\right)c_{1l}= \sum_{l=0}^{3} c_{2(i-l)} c_{2l}-6\sum_{l=0}^{3} \sum_{m=0}^{l} c_{1(i-l)} c_{1(l-m)} c_{2m}\\[12pt]
+2k \sum_{l=0}^{1} \sum_{m=0}^{l} c_{1(i-l)} c_{1(m-l)} c_{3l}\:,
\end{split}
\ee
which leads to 
\be
c_{23}=\frac{c_{13}}{2} \label{coef2-3c}.
\ee
For the coefficients of the term $t^{-2}$, we have
\be
\begin{split}
2\sum_{l=0}^{4} c_{2(i-l)} \left(\frac{i-l}{2}-2\right)c_{1l}= \sum_{l=0}^{4} c_{2(i-l)} c_{2l}-6\sum_{l=0}^{4} \sum_{m=0}^{k} c_{1(i-l)} c_{1(l-m)} c_{2m} \\[12pt]
+2k \sum_{l=0}^{2} \sum_{m=0}^{k} c_{1(i-l)} c_{1(m-l)} c_{3l}-k^{2}\sum_{l=0}^{0} c_{3(i-l)} c_{3l}-\frac{1}{6\epsilon} \sum_{l=0}^{0} c_{1(i-l)} c_{1l}
\end{split}
\ee
which leads to 
\be
5c_{24}-2c_{14}=-\frac{3\rho^{2}}{2}-\frac{1}{12\epsilon}.\label{coef2-4c}
\ee

%========================================================================
\subsubsection*{3rd set of equations}

Finally, for the coefficients of the different powers of $t$, Eq. (\ref{eq:substi3c}) leads to the following equations:
\\For the coefficients of the term $t^{-3/2}$,
\be
c_{13} \left(\frac{1}{2}-1\right) = -2\sum_{l=0}^{1} c_{1(1-l)}c_{3l}\:,\ee 

leading to 
\be
c_{13}=-4\rho c_{11}. \label{coef3-1c}
\ee
For the coefficients of the term $t^{-1}$,

\be
c_{32} \left(\frac{2}{2}-1\right) = -2\sum_{l=0}^{2} c_{1(2-l)}c_{3l}, 
\ee
which leads to
\be
2\rho c_{12}+c_{32}+2c_{11}c_{31}=0. \label{coef3-2c}
\ee
for the coefficients of the term $t^{-1/2}$,

\be
c_{33} \left(\frac{3}{2}-1\right)=-2\sum_{l=0}^{3} c_{1(3-l)}c_{3l}, 
\ee
which lead to 

\be
3c_{33}+4\rho c_{13}+4c_{12}c_{31}+4c_{11}c_{32}=0  \label{coef3-3c}, 
\ee
for the coefficients of the term $t^{0}$,

\be
c_{14} \left(\frac{4}{2}-1\right)=-2\sum_{l=0}^{4} c_{1(4-l)}c_{3l}, 
\ee
which lead to 

\be
c_{34}+\rho c_{14}+c_{13}c_{31}+c_{12}c_{32}+c_{11}c_{33}=0. \label{coef3-4c}
\ee

%=======================================================================
\subsection{Final form of the general solution}

Consequently, we arrive at the following asymptotic series representation for the decomposition (\ref{eq:curveddom1}):
\be
x(t) = \frac{1}{2}\:\:t^{-1} -\frac{k\r}{2} + c_{13}\:\:t^{1/2} - \left(\frac{k^2 \r^2}{4}+\frac{1}{36\epsilon} \right)\:\: t + \cdots,
\label{eq:gensol}
\ee
while the corresponding series expansion for $y(t)$ is given by the first time derivative of the above expression, while that for $z(t)$ is given by
\be
z(t) = \r\:\:t^{-1} -k\r^2 -\frac{4\r \: c_{13}}{3} \:\:t^{1/2} + \left(\frac{k^2 \r^2 (1+2\r)}{4}+\frac{1}{36\epsilon} \right)\:\: t + \cdots.
\label{eq:gensolz}
\ee
For the scale factor, we find
\be \label{eq:curvedfinalgen}
a(t) =\a\:\:t^{1/2}-\frac{k\r \a}{2}\:\:t^{3/2}+\frac{2c_{13}\a}{3}\:\:t^{2}-\left( \frac{k^2 \r^2 \a}{8}+\frac{\a}{72\epsilon} \right)\:\:t^{5/2}+\cdots, 
\ee
where $\a$ is a constant of integration and $\a^{-2}=\r$. 

This series (\ref{eq:gensol}) has three arbitrary constants, $\r$, $c_{13}$ with the third one corresponding to the arbitrary position of the singularity. Therefore, this represents a local expansion of a \emph{general} solution around the initial singularity. The transformation $c_{13}={3c^{'}_{13}/2\a}$ and $\epsilon = k/6 $ in the series expansion (\ref{eq:curvedfinalgen}), leads to the form which is obtained by setting $\z=0$ in the series expansion found for the curved, radiation case, cf. Eq. (4.13) of \cite{cot-kol-tso-13}. In addition, by setting $k=0$ we are lead to the form (\ref{eq:flatgensol1})-(\ref{eq:flatgensol2}) found for the flat vacuum.

We note that because of the square root, limits can only be taken in the backward direction, $t\downarrow 0$, in the solution (\ref{eq:curvedfinalgen}), another way of expressing the curious fact that this solution (along with Eq. (\ref{eq:flatfinal}) found in the previous Section) is \emph{only} valid at early times and corresponds to a past singularity.

%-------------------------------------------------------------------
%	SECTION 4.3  - Milne states
%-------------------------------------------------------------------
\section{Milne states}

%===================================================================
\subsection{Milne - Dominant balance}

We now move on to the analysis of the last two decompositions, namely, those with dominant parts (\ref{eq:curveddom2}) and (\ref{eq:curveddom3}). We show below that these lead to particular solutions (meaning having less number of arbitrary constants than in a general solution) for $k=-1$ and $k=+1$. 

In the case of open universes,  $k=-1$, and the dominant parts take the forms
\be \label{eq:curveddom2-1}
\mathbf{f^{12}}^{(0)}_{\,-1,\textsc{VAC}}=\left(y, -3xy-xz, -2xz\right) , \ee \be \label{eq:curveddom3 -1} \mathbf{f^{42}}^{(0)}_{\,-1,\textsc{VAC}}=\left(y, \frac{y^2}{2x}-3xy-xz-\frac{z^2}{2x}, -2xz\right) ,
\ee
with subdominant parts given by
\be \label{eq:curvedsub2-1} \mathbf{f^{12}}^{(sub)}_{\,-1,\textsc{VAC}}=\left(0, \frac{y^2}{2x}-\frac{z^2}{2x}-\frac{x}{12\epsilon}+\frac{z}{12x\epsilon},0\right) ,
\ee
\be \label{eq:curvedsub3-1} \mathbf{f^{42}}^{(sub)}_{\,-1,\textsc{VAC}}=\left(0, -\frac{x}{12\epsilon}+\frac{z}{12x\epsilon},0\right) ,
\ee
These two forms  lead, however, to the same acceptable asymptotic balance, namely,
\be \label{curvedbalance2-1} \mathcal{B}^{12,42}_{\,-1,\textsc{VAC}}=(\mathbf{a},\mathbf{p})= \left(\left(1,-1,1 \right),\: \left(-1,-2,-2\right)\right),
\ee

%=====================================================================
\subsection{Milne - subdominant condition}

As we have already seen in Chapter 3, we have to examine the limit of the following expression 
\be\label{subconM}
\frac{\mathbf{f}^{\,(\textrm{sub})}_{\,-1,\textsc{VAC}}(\mathbf{a}t^{\mathbf{p}})}{t^{\mathbf{p}-1}}.
\ee
as $t\rightarrow 0$, in order to confirm that the subdominant part is indeed subdominant asymptotically. Accordingly, we need to show that the higher-order terms (\ref{eq:curvedsub2-1}) and (\ref{eq:curvedsub3-1}) in the basic decompositions $\mathbf{f^{12}}_{\,-1,\textsc{VAC}}$ and $\mathbf{f^{42}}_{\,-1,\textsc{VAC}}$ respectively, are themselves weight-homogeneous with respect to the balance (\ref{curvedbalance2-1}) for this to be an acceptable one.  To prove this, we first split the subdominant parts (\ref{eq:curvedsub2-1}) and (\ref{eq:curvedsub3-1}) by writing  
\be
\mathbf{f^{12}}^{\,(\textrm{sub})}_{\,-1,\textsc{VAC}}(\mathbf{x}) = \mathbf{f^{12}}^{(1)}_{\,-1,\textsc{VAC}}(\mathbf{x}) +
\mathbf{f^{12}}^{(2)}_{\,-1,\textsc{VAC}}(\mathbf{x}) + \mathbf{f^{12}}^{(3)}_{\,-1,\textsc{VAC}}(\mathbf{x}) + \mathbf{f^{12}}^{(4)}_{\,-1,\textsc{VAC}}(\mathbf{x}),
\ee
\be
\mathbf{f^{42}}^{\,(\textrm{sub})}_{\,-1,\textsc{VAC}}(\mathbf{x}) = \mathbf{f^{42}}^{(1)}_{\,-1,\textsc{VAC}}(\mathbf{x}) +
\mathbf{f^{42}}^{(2)}_{\,-1,\textsc{VAC}}(\mathbf{x}).
\ee
Thus, we are led to the forms
\be\label{subcon12-}
\mathbf{f^{12}}^{\,(\textrm{sub})}_{\,k,\textsc{VAC}}(\mathbf{x})=\left(0,\frac{y^2}{2x},0 \right)+\left(0,-\frac{z^2}{2x},0 \right)+\left(0,-\frac{x}{12\epsilon},0 \right)+\left(0,\frac{z}{12\epsilon},0 \right),
\ee
\be\label{subcon42-}
\mathbf{f^{42}}^{\,(\textrm{sub})}_{\,k,\textsc{VAC}}(\mathbf{x})=\left(0,-\frac{x}{12\epsilon},0 \right)+\left(0,\frac{z}{12\epsilon},0 \right).
\ee
And, subsequenlty,
\be\label{subcon12--}
\begin{split}
\mathbf{f^{12}}^{\,(\textrm{sub})}_{\,k,\textsc{VAC}}(\mathbf{a}t^{\mathbf{p}})=\left(0,\frac{\eta^2}{2\theta}t^{2q-p},0 \right)+\left(0,-\frac{\rho^2}{2\theta}t^{2r-p},0 \right)+\\[10pt]
+\left(0,-\frac{\theta}{12\epsilon}t^{p},0 \right)+\left(0,\frac{\rho}{12\epsilon\theta}t^{r-p},0 \right),
\end{split}
\ee
\be\label{subcon42--}
\mathbf{f^{42}}^{\,(\textrm{sub})}_{\,k,\textsc{VAC}}(\mathbf{a}t^{\mathbf{p}})=\left(0,-\frac{\theta}{12\epsilon}t^{p},0 \right)+\left(0,\frac{\rho}{12\epsilon\theta}t^{r-p},0 \right).
\ee
Now, using the $\mathcal{B}^{12,42}_{\,-1,\textsc{VAC}} $ defined by Eq. (\ref{curvedbalance2-1}), we find that
\be\label{subcon12---}
\begin{split}
\mathbf{f^{12}}^{\,(\textrm{sub})}_{\,k,\textsc{VAC}}(\mathbf{a}t^{\mathbf{p}})=\left(0,-\frac{1}{2}t^{-3},0 \right)+\left(0,\frac{1}{2}t^{-3},0 \right)\\[10pt]
+\left(0,-\frac{1}{12\epsilon}t^{-1},0  \right)+\left(0,\frac{1}{12\epsilon}t^{-1},0 \right)=(0,0,0)
\end{split}
\ee
\be\label{subcon42---}
\mathbf{f^{42}}^{\,(\textrm{sub})}_{\,k,\textsc{VAC}}(\mathbf{a}t^{\mathbf{p}})=\left(0,-\frac{1}{12\epsilon}t^{-1},0  \right)+\left(0,\frac{1}{12\epsilon}t^{-1},0  \right)=(0,0,0)
\ee
As we can see, in the case of open universes, the subdominant parts of the asymptotic decompositions $\mathbf{f^{12}}_{\,-1,\textsc{VAC}}$ and $\mathbf{f^{42}}_{\,-1,\textsc{VAC}}$ deviate, by suitable values of the asymptotic balance, the conditions described in our analysis of the subdominant character of each term and vanish for every value of $t$. Thus, there is no need to examine the limit of the expression (\ref{subconM}) for any of those two decompositions. Looking at the above equations more carefully (as well as the ones below for the closed case) we observe the key role of the curvature term $kxz$ underlying the fact, that in contradiction with the flat case studied in Chapter 3, curvature is a key feature of the universe for the existence of those Milne states in the neighborhood of the spacetime initial singularity.

%============================================================================================================
\subsection{Milne - K-matrices}

Following that, we find that the structure of the $\mathcal{K}$-matrices is
\be
\mathcal{K}^{12}_{\,-1,\textsc{VAC}}=\left(
                     \begin{array}{ccc}
                       1 & 1   & 0\\
                       2 & -1 & -1 \\
                       -2 & 0  & 0
                     \end{array}
                   \right),\quad\textrm{spec}(\mathcal{K}^{12}_{\,-1,\textsc{VAC}})=\{-1,-1,2\},
\ee
and
\be
\mathcal{K}^{42}_{\,-1,\textsc{VAC}_3}=\left(
                     \begin{array}{ccc}
                       1 & 1   & 0\\
                       2 & -2 & -2 \\
                       -2 &0   & 0
                     \end{array}
                   \right),\quad\textrm{spec}(\mathcal{K}^{42}_{\,-1,\textsc{VAC}})=\{-1,-2,2\}.
\ee

%============================================================================================================
\subsection{Milne - Solutions}

Since we are interested in the behavior of solutions near finite-time singularities (as opposed to singularities at infinity), we may set the arbitrary constants corresponding to the negative eigenvalues equal to zero, and led to a form for $x(t)$ common for both decompositions, namely,
\be \label{eq:parsol1}
x(t) =t^{-1}+c_{12}\:\:t-\left(\frac{c_{12}-18\: \epsilon \: c_{12}^2}{60\epsilon}\right) \:\: t^3+\cdots . \ee
The corresponding series expansion for $y(t)$ is given by the first time derivative of the above expression, while the corresponding series expansion for $z(t)$ is given by \be
z(t) = t^{-2} -c_{12} + \left(\frac{c_{12} \:(\:42\epsilon \:c_{12}+1\:)}{120 \epsilon}\right)\:\: t^2 + \cdots . \label{eq:parsol1z} \ee
Finally, we arrive at the following asymptotic form for the scale factor $\alpha (t)$ around the singularity: \be
a(t)=\alpha\:t+\frac{\alpha\: c_{12}}{2}\:\:t^3-\frac{\alpha \left(c_{12}-18\:\epsilon \: c_{12}^2 \right)}{240\epsilon}\:\:t^5 + \cdots , \ee where  $\alpha =\pm1$ as dictated by the definition $z(t)=1/a(t)^2$.

This solution  has therefore two arbitrary constants, $c_{12}$ and a second one corresponding to the arbitrary position of the singularity (taken here to be zero without loss of generality), and is therefore a local expansion of a \emph{particular} solution around the  singularity. Since the time singularity can be approached here from either the past or the future direction, we conclude  that it represents  a 2-parameter family of  past, or  future Milne states for these open vacua. This is also reminiscent of the Frenkel-Brecher horizonless solutions, cf. \cite{fre-bre-82}, with the important difference that their solutions are matter-filled an possibly valid only in the past direction.

%============================================================================================================
\subsection{Milne - Closed universe}

On the other hand, when $k=+1$,  the decomposition  (\ref{eq:curveddom2}) does not lead to an acceptable dominant balance, but (\ref{eq:curveddom3}) does, namely,
\be \label{eq:curveddom3+1} \mathbf{f^{42}}^{(0)}_{\,+1,\textsc{VAC}}=\left(y, \frac{y^2}{2x}-3xy+xz-\frac{z^2}{2x}, -2xz\right) , \ee with subdominant part \be \label{eq:curvedsub3 +1} \mathbf{f}^{(sub)}_{\,+1,\textsc{VAC}_3}=\left(0, -\frac{x}{12\epsilon}-\frac{z}{12x\epsilon}, 0\right) ,
\ee
 and we obtain \be \label{curvedbalance2+1} \mathcal{B}^{42}_{\,+1,\textsc{VAC}}=(\mathbf{a},\mathbf{p})= \left(\left(1,-1,3 \right),\: \left(-1,-2,-2\right)\right).
\ee
The corresponding $\mathcal{K}$-matrix is

\be
\mathcal{K}^{42}_{\,+1,\textsc{VAC}}=\left(
                     \begin{array}{ccc}
                       1 & 1   & 0 \\
                       10 & -2 & -2 \\
                       -6 &0   & 0
                      \end{array}
                    \right),\quad\textrm{spec}(\mathcal{K}_{\,+1,\textsc{VAC}_3})=\{-1,-2\sqrt{3},2\sqrt{3}\},
\ee
and we expect  particular solutions in this case with the given leading order, however,  due to the irrational Kowalevskaya exponents the resulting series will contain logarithmic terms.

%--------------------------------------------------------------------------------------------------------------------
%	SECTION  - Conclusion
%--------------------------------------------------------------------------------------------------------------------

\section{Conclusion} 

It turns out that a prominent role in the \emph{early} asymptotic evolution of both flat and curved vacua in this theory is played by a scaling form that behaves as $t^{1/2}$ near the initial singularity. Using various asymptotic and geometric arguments, we were able to built a solution of the field equations in the form of a Fuchsian formal series expansion compatible with all other constraints, dominated asymptotically to leading order by this solution and having the correct number of arbitrary constants that makes it a general solution of the field equations. In this way, we  conclude that this exact solution is an early time attractor of all homogeneous and isotropic vacua of the theory, thus proving stability against such `perturbations'.

For open vacua, there is a 2-parameter family of Fuchsian solutions that is dominated asymptotically by the Milne form both for past and future singularities. In the case of closed models, we have logarithmic solutions coming from a manifold of initial conditions with smaller dimension than the full phase space but dominated asymptotically by the same $a(t)\sim t$ form.

% Chapter Template

\chapter{Radiation-filled universes} % Main chapter title

\label{Chapter5} % Change X to a consecutive number; for referencing this chapter elsewhere, use \ref{ChapterX}

\lhead{Chapter 5. \emph{Radiation-filled universes}} % Change X to a consecutive number; this is for the header on each page - perhaps a shortened title

In this chapter we will start from the basic dynamical system and the equivalent vector field which describe the dynamical evolution of any radiation-filled FRW universe in higher order gravity and, assuming the existence of a finite time singularity (taken here to lie at t=0 without loss of generality) in its set of solutions, we will study its asymptotic behavior in the neighborhood of the singularity.

%--------------------------------------------------------------------------------------------------------------------
%	SECTION  - The basic curvature-radiation vector field and associated dynamical system
%--------------------------------------------------------------------------------------------------------------------
\section{Introduction}

In this section we shall derive the basic vector field and the equivalent dynamical system which completely describe the dynamical evolution of any radiation-filled FRW universe in higher order gravity. 

These cosmological models are determined by the Robertson-Walker metric, which is derived by assuming homogeneity and isotropy of a universe with constant curvature and has the general form (\ref{rwmetrics}).

We assume that these spaces are filled with a radiation fluid and the energy-momentum tensor has the form
$T_{\mu\nu}=(p+\rho)u_\mu u_\nu +pg_{\mu\nu}$,
where the fluid velocity 4-vector is $u^\mu=\delta^\mu_0$ an equation of state of the form, $p=\rho/3$.

The general higher-order action is the same as the one we have used so far, namely, Eq. (\ref{eq:lagra}) which as it has already been discussed in Chapter 2 leads to the field equations,
\be
R^{\mu\nu}-\frac{1}{2}g^{\mu\nu}R+
      \frac{\xi}{6} \left[2RR^{\mu\nu}-\frac{1}{2}R^2g^{\mu\nu}-2(g^{\mu\rho}g^{\nu\s}-g^{\mu\nu}g^{\rho\s})\nabla_{\rho}\nabla_{\s}R \right]=T^{\mu\nu},
\label{rcfe}
\ee

Using  the metric (\ref{rwmetrics}), the field equation (\ref{rcfe}) leads to our basic cosmological equation  in the form
\be \label{rcbasiceq}
\frac{k+\dot{a}^2}{a^2}+\xi\left[2\: \frac{\dddot{a}\:\dot{a}}{a^2} + 2\:\frac{\ddot{a}\dot{a}^2}{a^3}-\frac{\ddot{a}^2}{a^2} - 3\:
\frac{\dot{a}^4}{a^4} -2k\frac{\dot{a}^2}{a^4} + \frac{k^2}{a^4}\right] = \frac{\zeta^2}{a^4} ,
\ee
where $\zeta$ is a constant defined by the constraint
\be
\frac{\rho}{3}=\frac{\z^2}{a^4},\quad (\textrm{from}\,\,\nabla_{\mu}T^{\mu 0}=0).
\ee
This is exactly like Eq. (\ref{eq:fe2}) except for the RHS where the `radiation' term, $\frac{\zeta^2}{a^4}$, appears instead of zero.

Setting in this case
\be \label{rcvar}
x=a,\:\:\:\: y=\dot{a}\:\:\:\: \text{and}\:\:\:\: z=\ddot{a},
\ee
Eq. (\ref{rcbasiceq}) can be written as an autonomous dynamical
system of the form
\be \label{rcdsform}
\mathbf{\dot{x}}=\mathbf{f}_{\,k,\textsc{RAD}}(\mathbf{x}),\quad \mathbf{x}=(x,y,z),
\ee
where
\begin{eqnarray}
\dot{x} &=& y,\label{eq:rcds1}\\[11pt] 
\dot{y} &=& z,\label{eq:rcds2}\\[13pt]
\dot{z} &=& \frac{\z^2}{2\xi x^2y} -\frac{k^2}{2 x^2y} + \frac{3y^3}{2x^2} + \frac{z^2}{2y} -\frac{yz}{x} - \frac{y}{2\xi}
-\frac{k}{2\xi y} + \frac{ky}{x^2},\label{eq:rcds3}
\end{eqnarray}
 equivalent to the \emph{curvature-radiation} vector field $\mathbf{f}_{\,k,\textsc{RAD}}:\mathbb{R}^3\rightarrow\mathbb{R}^3:(x,y,z)\mapsto\mathbf{f}_{\,k,\textsc{RAD}}(x,y,z)$ with
\be\label{rcvf}
\mathbf{f}_{\,k,\textsc{RAD}}(x,y,z)=\left( y,z,\frac{\z^2}{2\xi x^2y} -\frac{k^2}{2 x^2y} + \frac{3y^3}{2x^2} + \frac{z^2}{2y} -\frac{yz}{x} -\frac{y}{2\xi}
-\frac{k}{2\xi y} + \frac{ky}{x^2}\right).
\ee
The curvature-radiation field $\mathbf{f}_{\,k,\textsc{RAD}}$, or equivalently the dynamical system (\ref{eq:rcds1})-(\ref{eq:rcds3}), combines the effects of curvature and radiation, and completely describes the dynamical evolution of any radiation-filled FRW universe in higher-order gravity. In the following Section we shall see how this field can split asymptotically and determine all dominant asymptotic modes developing on approach to the initial singularity in higher-order gravity.

%--------------------------------------------------------------------------------------------------------------------
%	SECTION  - Asymptotic splittings of curvature - radiation field
%--------------------------------------------------------------------------------------------------------------------
\section{Asymptotic splittings of the curvature-radiation vector field}

How many admissible asymptotic decompositions does the vector field $\mathbf{f}_{\,k,\textsc{RAD}}$ given by Eq. (\ref{rcvf}) possess on approach to the initial state at $t=0$? We recall that when $k=0$, as shown in \cite{cot-tso-07}, the vector field $\mathbf{f}_{\,0,\textsc{RAD}}$ has two admissible asymptotic solutions near the initial singularity. In the first family, all flat, radiation solutions are dominated (or attracted) at early times by the form $a(t)\sim t^{1/2}$, thus proving the stability of this solution in the flat case. There is also a second possible asymptotic form near the singularity in the flat case, $a(t)\sim t$, but this contains only two arbitrary constants and hence it corresponds to a \emph{particular} solution of the theory (cf. \cite{cot-tso-07}).

When $k\neq 0$, and we have the present situation of a radiation-filled, curved family of FRW universes to follow asymptotically near the past singularity, the field $\mathbf{f}_{\,k,\textsc{RAD}}$ has more terms - those that contain $k$ in (\ref{rcvf}) - than in the flat case. Since we already have a precise picture of the asymptotic forms of the flat case, we can now study the combined effects of curvature and radiation alone asymptotically. A simple combinatorial calculation shows that $\mathbf{f}_{\,k,\textsc{RAD}}$ (or the basic system (\ref{eq:rcds1})-(\ref{eq:rcds3})) can decompose precisely in
\be
(\begin{array}{c}
8\\[0.1pt]
1
\end{array})+(\begin{array}{c}
8\\[0.1pt]
2
\end{array})+\cdots +(\begin{array}{c}
8\\[0.1pt]
8
\end{array})=255
\ee
different ways. Each one of these 255 different modes leads to an asymptotic splitting of the form (\ref{general split3}), which may contain many possible dominant balances and so needs to be checked for admissibility. Any candidate asymptotic splitting of the form (\ref{general split3}) will accordingly be acceptable in principle, provided the candidate subdominant part tends to zero asymptotically, that is it indeed behaves as a subdominant contribution to the dominant asymptotic form the field splits into (\ref{general split3}). The complete list of the 255 asympototic splittings of the vector field (\ref{rcvf}) is given in the Appendix \ref{AppendixB}. In that table $\mathbf{f}^{(0)}_{\,k,\textsc{RAD}}$ denotes the candidate dominant part of the field, and $\mathbf{f}^{\,(\textrm{sub})}_{\,k,\textsc{RAD}}$ its subdominant one. Each one of these 255 decompositions represents the possible ways the vector field may dominate the evolution of the system. However, as discussed, for any one of these asymptotic splittings to be an admissible one, certain conditions that will be examined below, have to be satisfied.

%============================================================================================================
\section{The all-terms-dominant decomposition}

%============================================================================================================
\subsection{Dominant exponents analysis}

In order to take a quick glance at the general picture and make some preliminary qualitative observations, we start our asymptotic decomposition analysis with the case of the \textit{all-terms-dominant} decomposition, $\mathbf{f}^{255}_{\,k,\textsc{RAD}}$, that is the asymptotic splitting where,
\be
\mathbf{f^{255}}^{\,(\textrm{sub})}_{\,k,\textsc{RAD}}=\mathbf{0},
\ee
and by consequence,
\be
\mathbf{f}^{\,(\textrm{0})}_{\,k,\textsc{RAD}}=\left( y,z,\frac{\z^2}{2\xi x^2y} -\frac{k^2}{2 x^2y} + \frac{3y^3}{2x^2} + \frac{z^2}{2y} - \frac{yz}{x} - \frac{y}{2\xi}-\frac{k}{2\xi y} + \frac{ky}{x^2}\right).
\ee
Substituting the dominant solution 
\be
\mathbf{x}(t)=\mathbf{a}t^{\mathbf{p}}=(\theta t^{p}, \eta t^{q}, \rho t^{r}). \label{domisol3}
\ee 
into the corresponding dominant system, (\ref{eq:rcds1})-(\ref{eq:rcds3}), we find,
\begin{eqnarray}
\theta p\:t^{p-1}&=& \eta\: t^{q},\label{eq:rcatd-domsys1}\\[13pt] 
\eta q\:t^{q-1} &=& \rho\: t^{r},\label{eq:rcatd-domsys2}\\[13pt] 
\rho r\:t^{r-1} &=& \frac{\zeta^2}{2\xi\theta^{2}\eta}\:\:t^{-2p-q} -\frac{k^2}{2\theta^{2}\eta}\:\:t^{-2p-q} +\frac{3\eta^{3}}{2\theta^{2}}\:\:t^{-2p+3q} + \frac{\rho^{2}}{2\h}\:\:t^{2r-q}- \nonumber\\[11pt]
&-&\frac{\eta\rho}{\theta}\:\: t^{q+r-p} -\frac{\eta}{2\xi}\:\:t^{q} -\frac{k}{2\xi \eta}\:t^{-q} +\frac{k\eta}{\theta^{2}}\:\: t^{-2p+q}.\label{eq:rcatd-domsys3}
\end{eqnarray}
Firstly, we need to solve the above system for the dominant exponents, that is for the components $\mathbf{p}=(p,q,r)\in\mathbb{Q}^3$. Eqns. (\ref{eq:rcatd-domsys1}) and (\ref{eq:rcatd-domsys2}) lead to,
\be
p-1=q\:\:\:\:\text{and}\:\:\:\:q-1=r.\label{rcatd-domexp}
\ee
We note, that, these two equations must be satisfied by all possible dominant balances since they are derived from Eqns. (\ref{eq:rcatd-domsys1}) and (\ref{eq:rcatd-domsys2}) as are part of every possible dominant system of the 255 possible asymptotic decompositions. Thus, any possible vector $\mathbf{p}=(p,q,r)$ will have the form, 
\be
\mathbf{p}=(q-1,\:\:q,\:\:q+1).\label{crexp-genform}
\ee
Eq. (\ref{eq:rcatd-domsys3}) then leads to a set of equations for each one of the eight terms in the right-hand side. The first and second terms, namely, $\frac{\zeta^2}{2\xi\theta^{2}\eta}\:\:t^{-2p-q}$ and $-\frac{k^2}{2\theta^{2}\eta}\:\:t^{-2p-q}$, become,
\be
r-1=2p-q.\label{cratd-exp12}
\ee
Similarly, the third term, namely, $\frac{3\eta^{3}}{2\theta^{2}}\:\:t^{-2p+3q}$, gives
\be 
r-1=-2p+3q,\label{cratd-exp3}
\ee
the fourth term, namely, $\frac{\rho^{2}}{2\h}\:\:t^{2r-q}$, gives
\be 
r-1=2r-q,\label{cratd-exp4}
\ee
the fifth term, namely, $-\frac{\eta\rho}{\theta}\:\: t^{q+r-p}$, gives
\be
r-1=q+r-p,\label{cratd-exp5}
\ee
the sixth term, namely, $-\frac{\eta}{2\xi}\:\:t^{q}$, gives
\be
r-1=q,\label{cratd-exp6}
\ee
the seventh term, namely, $-\frac{k}{2\xi \eta}\:t^{-q}$, gives
\be
r-1=-q,\label{cratd-exp7}
\ee
and finaly, the eighth term, namely, $\frac{k\eta}{\theta^{2}}\:\: t^{-2p+q}$, leads to
\be
r-1=-2p+q.\label{cratd-exp8}
\ee
Since this is the \textit{all-terms-dominant} decomposition, it is clear that the dominant exponents in any other asymptotic splitting will satisfy a subset of the Eqs. (\ref{cratd-exp12})-(\ref{cratd-exp8}) depending on the way that the terms of the RHS of Eq. (\ref{eq:rcatd-domsys3}) are distributed between the dominant and the subdominant part.

By substituting the general form (\ref{crexp-genform}) in Eqs. (\ref{cratd-exp12})-(\ref{cratd-exp8}), we are led to the following observations:
\begin{enumerate}
\item Eq. (\ref{cratd-exp6}) is impossible and the linear term $-\frac{y}{2\xi}$ cannot be included in the dominant part of any asymptotic splitting as expected. Since there are 127 asymptotic splittings that contain the linear term $-\frac{y}{2\xi}$ in their dominant part, we may cross them out of the list of Table (\ref{long2}) and left with 128 possible asymptotic decompositions.
\item Eq. (\ref{cratd-exp7}) leads to the value $q=1$, and thus when the term $-\frac{k}{2\xi \eta}\:t^{-q}$ is included in the dominant part of an asymptotic splitting the only possible value for the vector $\mathbf{p}$ is 
\be
\mathbf{p}=(p,\:q,\:r)=(2,\:1,\:0).\label{crexp-form1}
\ee 
\item Each one of Eqs. (\ref{cratd-exp12}) and (\ref{cratd-exp8}) lead to the value $q=0$, and thus when any of the terms $\:\:\frac{\zeta^2}{2\xi\theta^{2}\eta}\:\:t^{-2p-q}\:$, $\:\:\:-\frac{k^2}{2\theta^{2}\eta}\:\:t^{-2p-q}\:$, or $\:\:\frac{k\eta}{\theta^{2}}\:\: t^{-2p+q}\:\:$, is included in the dominant part of an asymptotic splitting, we find 
\be
\mathbf{p}=(p,\:q,\:r)=(1,\:0,\:-1).\label{crexp-form2}
\ee 
\item Taking into consideration the last two observations we can safely conclude that it is impossible for an asymptotic splitting to lead to an acceptable dominant balance when it contains in its dominant part in addition to the seventh term, $-\frac{k}{2\xi \eta}\:t^{-q}$, any one of the first, second or eighth term, namely, the terms $\:\:\frac{\zeta^2}{2\xi\theta^{2}\eta}\:\:t^{-2p-q}\:$, $\:\:-\frac{k^2}{2\theta^{2}\eta}\:\:t^{-2p-q}\:\:$ or $\:\:\frac{k\eta}{\theta^{2}}\:\: t^{-2p+q}\:\:$ respectively. Of the 128 remaining asymptotic splittings of the Table (\ref{long2}), there are precisely 57 that satisfy this condition and can be safely deleted from our list, thus leaving 71 possible asymptotic decompositions whose behavior remains to be examined. One of these 57 rejected decompositions is $\mathbf{f}^{255}_{\,k,\textsc{RAD}}$.
\end{enumerate}

%============================================================================================================
\subsection{Coefficients analysis}

Since the dominant system, (\ref{eq:rcatd-domsys1})-(\ref{eq:rcatd-domsys3}) corresponding to the all-terms-dominant possible asymptotic decomposition, $\mathbf{f^{255}}_{\,k,\textsc{RAD}}$ cannot be solved for the dominant exponents $\mathbf{p}=(p,q,r)$, Eq. (\ref{eq:rcatd-domsys3}) cannot lead to a general qualitative conclusion for the conditions needed to be satisfied by the coefficients $\mathbf{a}=(\theta, \eta, \rho)$ in order to have an acceptable dominant balance from any other asymptotic splitting. Nevertheless, Eqs. (\ref{eq:rcatd-domsys1}) and (\ref{eq:rcatd-domsys2}) are valid in the dominant system of any possible asymptotic decomposition and thus, we may proceed and examine below the possible forms of the coefficient vector $\mathbf{a}$ for each one of the special cases (\ref{crexp-form1}) and (\ref{crexp-form2}) for the values of the exponent vector $\mathbf{p}$.

%===========================================================================================================
\subsubsection*{First case p=(2,1,0)}

When $\mathbf{p}=(2, 1, 0)$, that is, when the seventh term of the RHS of (\ref{eq:rcds3}) is considered as the dominant one. Eqs. (\ref{eq:rcatd-domsys1})-(\ref{eq:rcatd-domsys2}) take the form,
\begin{eqnarray}
2\theta \:t &=& \eta\: t,\label{eq:rcatd-domsys11}\\
\eta \:t^{0} &=& \rho\: t^{0},\label{eq:rcatd-domsys21}
\end{eqnarray}
so that,
\begin{eqnarray}
2\theta &=& \eta,\label{eq:rcatd-coe11}\\
\eta    &=& \rho.\label{eq:rcatd-coe21}
\end{eqnarray}
Concluding, in the cases where the term $-\frac{k}{2\xi \eta}\:t^{-q}$ appears in the dominant part of an asymptotic splitting, the only possible form of a dominant balance is 
\be\label{rcdombalf1}
\mathcal{B}_{\,k,\textsc{RAD}}=(\mathbf{a},\mathbf{p})= \left(\left(\theta,\:\:2\theta,\:\:2\theta\right),\:\left(2,1,0\right)\right).
\ee

%============================================================================================================
\subsubsection*{Second case p=(-1,0,1)}

When $\mathbf{p}=(-1, 0, 1)$, that is, when at least one of the first, second or eighth term of the RHS of (\ref{eq:rcds3}) is considered as the dominant one. Eqs. (\ref{eq:rcatd-domsys1})-(\ref{eq:rcatd-domsys2}) take the form,
\begin{eqnarray}
\theta \:t^{0} &=& \eta\: t^{0},\label{eq:rcatd-domsys12}\\
0\:t^{-1} &=& \rho\: t^{-1},\label{eq:rcatd-domsys22}
\end{eqnarray}
so that, 
\begin{eqnarray}
\theta &=& \eta,\label{eq:rcatd-coe12}\\
\rho   &=& 0 .\label{eq:rcatd-coe22}
\end{eqnarray}
Concluding, in the cases where at least one of the terms $\:\:\frac{\zeta^2}{2\xi\theta^{2}\eta}\:\:t^{-2p-q}\:$, $\:\:\:-\frac{k^2}{2\theta^{2}\eta}\:\:t^{-2p-q}\:$ or $\:\:\frac{k\eta}{\theta^{2}}\:\: t^{-2p+q}\:\:$ appear in the dominant part of an asymptotic splitting the only possible form of a dominant balance is 
\be\label{rcdombalf2}
\mathcal{B}_{\,k,\textsc{RAD}}=(\mathbf{a},\mathbf{p})= \left(\left(\theta,\:\:\theta,\:\:0\right),\:\left(1,0,-1\right)\right).
\ee
In any other case, the dominant balance takes the form,
\be\label{rcdombalf3}
\mathcal{B}_{\,k,\textsc{RAD}}=(\mathbf{a},\mathbf{p})= \left(\left(\theta,\:\:\eta,\:\:\rho\right),\:\left(q+1,\:\:q,\:\:q-1\right)\right).
\ee

%========================================================================================================
\subsection{The hypothetical all-terms-subdominant decomposition}

Let us now move on to examine the possible subdominant behavior of each of the terms in (\ref{eq:rcds1})-(\ref{eq:rcds3}), $\mathbf{f}_{\,k,\textsc{RAD}}$, with respect to the specific forms of possible dominant balances that we have found in the previous subsection, namely, (\ref{rcdombalf1}), (\ref{rcdombalf2}) and (\ref{rcdombalf3}).

In order to do that we suppose the existence of a \emph{hypothetical} possible asymptotic decomposition which contains in its subdominant part all the terms in the initial radiation-curved vector field (\ref{rcvf}). In that case the subdominant part would have the form,
\be\label{rchsubvf}
\mathbf{f^{h}}^{(sub)}_{\,k,\textsc{RAD}}=\left( 0, 0, \frac{\z^2}{2\xi x^2y} -\frac{k^2}{2 x^2y} +\frac{3y^3}{2x^2} +\frac{z^2}{2y} -\frac{yz}{x} -\frac{y}{2\xi} -\frac{k}{2\xi y} +\frac{ky}{x^2}\right),
\ee
where $h$ stands for `hypothetical'. As discussed in the previous Chapter, it is impossible to have such a possible asymptotic decomposition since, in that case, we would have no terms left to justify the existence of a dominant asymptotic behavior.

Nevertheless, our current discussion aims at shedding light on the subdominant character of each term by finding the conditions under which it remains weight-homogeneous with respect to the dominant balances (\ref{rcdombalf1}), (\ref{rcdombalf2}) and (\ref{rcdombalf3}). To examine this, we first split the subdominant part (\ref{rchsubvf}) in the following way, 
\be
\begin{split}
\mathbf{f^{h}}^{\,(\textrm{sub})}_{\,k,\textsc{RAD}}(\mathbf{x}) = \mathbf{f^{h}}^{(1)}_{\,k,\textsc{RAD}}(\mathbf{x}) +
\mathbf{f^{h}}^{(2)}_{\,k,\textsc{RAD}}(\mathbf{x}) + \mathbf{f^{h}}^{(3)}_{\,k,\textsc{RAD}}(\mathbf{x}) + \mathbf{f^{h}}^{(4)}_{\,k,\textsc{RAD}}(\mathbf{x}) +\\[10pt]
+ \mathbf{f^{h}}^{(5)}_{\,k,\textsc{RAD}}(\mathbf{x}) +
\mathbf{f^{h}}^{(6)}_{\,k,\textsc{RAD}}(\mathbf{x}) + \mathbf{f^{h}}^{(7)}_{\,k,\textsc{RAD}}(\mathbf{x}) + \mathbf{f^{h}}^{(8)}_{\,k,\textsc{RAD}}(\mathbf{x}),
\end{split}
\ee
where
\begin{eqnarray}
\mathbf{f^{h}}^{(1)}_{\,k,\textsc{RAD}}(\mathbf{x})&=&\left(\:0,\:0,\:\frac{\z^2}{2\xi x^2y}\right),\label{eq:rchsub1}\\[12pt]
\mathbf{f^{h}}^{(2)}_{\,k,\textsc{RAD}}(\mathbf{x})&=&\left(\:0,\:0,\:-\frac{k^2}{2 x^2y}\right),\label{eq:rchsub2}\\[12pt]
\mathbf{f^{h}}^{(3)}_{\,k,\textsc{RAD}}(\mathbf{x})&=&\left(\:0,\:0,\:\frac{3y^3}{2x^2}\right),\label{eq:rchsub3}\\[12pt]
\mathbf{f^{h}}^{(4)}_{\,k,\textsc{RAD}}(\mathbf{x})&=&\left(\:0,\:0,\:\frac{z^2}{2y}\right),\label{eq:rchsub4}\\[12pt]
\mathbf{f^{h}}^{(5)}_{\,k,\textsc{RAD}}(\mathbf{x})&=&\left(\:0,\:0,\:-\frac{yz}{x}\right),\label{eq:rchsub5}\\[12pt]
\mathbf{f^{h}}^{(6)}_{\,k,\textsc{RAD}}(\mathbf{x})&=&\left(\:0,\:0,\:-\frac{y}{2\xi}\right),\label{eq:rchsub6}\\[12pt]
\mathbf{f^{h}}^{(7)}_{\,k,\textsc{RAD}}(\mathbf{x})&=&\left(\:0,\:0,\:-\frac{k}{2\xi y}\right),\label{eq:rchsub7}\\[12pt]
\mathbf{f^{h}}^{(8)}_{\,k,\textsc{RAD}}(\mathbf{x})&=&\left(\:0,\:0,\:\frac{ky}{x^2}\right),\label{eq:rchsub8}.
\end{eqnarray}
Subsequently, we are going to describe the contribution of each of the forms (\ref{eq:rchsub1})-(\ref{eq:rchsub8}) to the behavior of the expression, 
\be\label{rcsubcon}
\frac{\mathbf{f}^{\,(\textrm{sub})}_{\,k,\textsc{RAD}}(\mathbf{a}t^{\mathbf{p}})}{t^{\mathbf{p}-1}},
\ee
as $\:t\rightarrow0\:$ for each one of the possible dominant balances (\ref{rcdombalf1}), (\ref{rcdombalf2}) and (\ref{rcdombalf3}) separately.

%========================================================================================================
\subsubsection*{First case p=(2,1,0)}

The first case is when the dominant balance is of the form (\ref{rcdombalf1}), that is, 
\be\label{rcdombalf11}
\mathcal{B}_{\,k,\textsc{RAD}}=(\mathbf{a},\mathbf{p})= \left(\left(\theta,\:\:2\theta,\:\:2\theta\right),\:\left(2,1,0\right)\right).
\ee
As discussed earlier, this happens when the seventh term of the RHS of (\ref{eq:rcds3}) is included in the dominant part of a possible asymptotic splitting.
We also note that in this case, $\theta\neq0$ because $\textbf{a}$ cannot vanish. In order to deal with the expression (\ref{rcsubcon}), we substitute the dominant balance (\ref{rcdombalf11}) in Eqs. (\ref{eq:rchsub1})-(\ref{eq:rchsub8}) and we divide by $t^{\mathbf{p}-1}$. Thus, we have,
\begin{eqnarray}
\frac{\mathbf{f^{h}}^{(1)}_{\,k,\textsc{RAD}}(\mathbf{a}t^{\mathbf{p}})}{t^{\mathbf{p}-1}}&=& \mathbf{f^{h}}^{(1)}_{\,k,\textsc{RAD}}(\mathbf{a})\:t^{-4}=\left(\:0,\:0,\:\frac{\zeta^2}{4\xi\theta^{3}}\right)\:t^{-4},\label{eq:rchsub1B1}\\[12pt]
\frac{\mathbf{f^{h}}^{(2)}_{\,k,\textsc{RAD}}(\mathbf{a}t^{\mathbf{p}})}{t^{\mathbf{p}-1}}&=& \mathbf{f^{h}}^{(2)}_{\,k,\textsc{RAD}}(\mathbf{a})\:t^{-4}=\left(\:0,\:0,\:-\frac{k^2}{4 \theta^{3}}\right)\:t^{-4},\label{eq:rchsub2B1}\\[12pt]
\frac{\mathbf{f^{h}}^{(3)}_{\,k,\textsc{RAD}}(\mathbf{a}t^{\mathbf{p}})}{t^{\mathbf{p}-1}}&=& \mathbf{f^{h}}^{(3)}_{\,k,\textsc{RAD}}(\mathbf{a})\:t^{0}=\left(\:0,\:0,\:12\theta\right)\:t^{0},\label{eq:rchsub3B1}\\[12pt]
\frac{\mathbf{f^{h}}^{(4)}_{\,k,\textsc{RAD}}(\mathbf{a}t^{\mathbf{p}})}{t^{\mathbf{p}-1}}&=& \mathbf{f^{h}}^{(4)}_{\,k,\textsc{RAD}}(\mathbf{a})\:t^{0}=\left(\:0,\:0,\:2\theta\right)\:t^{0},\label{eq:rchsub4B1}\\[12pt]
\frac{\mathbf{f^{h}}^{(5)}_{\,k,\textsc{RAD}}(\mathbf{a}t^{\mathbf{p}})}{t^{\mathbf{p}-1}}&=& \mathbf{f^{h}}^{(5)}_{\,k,\textsc{RAD}}(\mathbf{a})\:t^{0}=\left(\:0,\:0,\:-4\theta\right)\:t^{0},\label{eq:rchsub5B1}\\[12pt]
\frac{\mathbf{f^{h}}^{(6)}_{\,k,\textsc{RAD}}(\mathbf{a}t^{\mathbf{p}})}{t^{\mathbf{p}-1}}&=& \mathbf{f^{h}}^{(6)}_{\,k,\textsc{RAD}}(\mathbf{a})\:t^{2}=\left(\:0,\:0,\:-\frac{\theta}{\xi}\right)\:t^{2},\label{eq:rchsub6B1}\\[12pt]
\frac{\mathbf{f^{h}}^{(7)}_{\,k,\textsc{RAD}}(\mathbf{a}t^{\mathbf{p}})}{t^{\mathbf{p}-1}}&=& \mathbf{f^{h}}^{(7)}_{\,k,\textsc{RAD}}(\mathbf{a})\:t^{0}=\left(\:0,\:0,\:-\frac{k}{2\xi \theta}\right)\:t^{0},\label{eq:rchsub7B1}\\[12pt]
\frac{\mathbf{f^{h}}^{(8)}_{\,k,\textsc{RAD}}(\mathbf{a}t^{\mathbf{p}})}{t^{\mathbf{p}-1}}&=& \mathbf{f^{h}}^{(8)}_{\,k,\textsc{RAD}}(\mathbf{a})\:t^{-2}=\left(\:0,\:0,\:\frac{2k}{\theta}\right)\:t^{-2},\label{eq:rchsub8B1}.
\end{eqnarray}
Since, the balance (\ref{rcdombalf11}) occurs only when the term $-\frac{k}{2\xi y}$ is in the dominant part of a decomposition, Eq. (\ref{eq:rchsub7B1}) does not have a `real' meaning in this context and we will not use it to make any conclusions. From the rest of the above expressions we can make the following general observations:
\begin{enumerate}
\item Firstly, we recall that the expressions (\ref{eq:rchsub1B1}), (\ref{eq:rchsub2B1}) and (\ref{eq:rchsub8B1}) will always be included in the subdominant part of any asymptotic splitting leading to (\ref{rcdombalf11}), since their corresponding terms lead to a different balance, namely, (\ref{rcdombalf2}). 
\item Taking the limit of the expression (\ref{eq:rchsub8B1}) as $t\rightarrow0$, we see that it always goes to infinity. Hence, it is impossible for the eighth term, at least, to exist in the subdominant part of any splitting of this case, in direct contradiction with our first observation above. Consequently, the specific balance (\ref{rcdombalf11}) is not a valid one and the term $-\frac{k}{2\xi y}$ can never be part of the dominant part of any possible asymptotic decomposition. Of the 71 remaining asymptotic splittings of the Table (\ref{long2}), there are 8 that still contain that term in their dominant part, and so we are left with 63 possible asymptotic decompositions whose behavior remains to be examined.
\item Taking the limit of the expressions (\ref{eq:rchsub1B1}) and (\ref{eq:rchsub2B1}) as $t\rightarrow0$ they go to infinity because of the negative exponent $-4$, revealing the strong dominant character of the corresponding terms.  
\item Expressions (\ref{eq:rchsub3B1}), (\ref{eq:rchsub4B1}) and (\ref{eq:rchsub5B1}) go to $12\theta$, $2\theta$ and $-4\theta$ respectively as  $t\rightarrow0$ showing as well a strongly dominant behavior in this case while exrpession (\ref{eq:rchsub6B1}) corresponding to the only linear term is the only one which shows a strongly subdominant character.
\end{enumerate} 
We will now continue this `subdominant analysis' with the second possible form of dominant balance.

%============================================================================================================
\subsubsection*{Second case p=(-1,0,1)}

The only other case is when the dominant balance is of the form (\ref{rcdombalf2}), that is,
\be\label{rcdombalf22}
\mathcal{B}_{\,k,\textsc{RAD}}=(\mathbf{a},\mathbf{p})= \left(\left(\theta,\:\:\theta,\:\:0\right),\:\left(1,0,-1\right)\right).
\ee
This dominant balance occurs when at least one of the first, second or eighth term of the RHS of (\ref{eq:rcds3}) is present in the dominant part of an asymptotic splitting. In order to construct the expression (\ref{rcsubcon}) we substitute the dominant balance (\ref{rcdombalf22}) in Eqs. (\ref{eq:rchsub1})-(\ref{eq:rchsub8}) and we divide by $t^{\mathbf{p}-1}$. Thus, we have,
\begin{eqnarray}
\frac{\mathbf{f^{h}}^{(1)}_{\,k,\textsc{RAD}}(\mathbf{a}t^{\mathbf{p}})}{t^{\mathbf{p}-1}}&=& \mathbf{f^{h}}^{(1)}_{\,k,\textsc{RAD}}(\mathbf{a})\:t^{0}=\left(\:0,\:0,\:\frac{\zeta^2}{2\xi\theta^{3}}\right)\:t^{0},\label{eq:rchsub1B2}\\[12pt]
\frac{\mathbf{f^{h}}^{(2)}_{\,k,\textsc{RAD}}(\mathbf{a}t^{\mathbf{p}})}{t^{\mathbf{p}-1}}&=& \mathbf{f^{h}}^{(2)}_{\,k,\textsc{RAD}}(\mathbf{a})\:t^{0}=\left(\:0,\:0,\:-\frac{k^2}{2 \theta^{3}}\right)\:t^{0},\label{eq:rchsub2B2}\\[12pt]
\frac{\mathbf{f^{h}}^{(3)}_{\,k,\textsc{RAD}}(\mathbf{a}t^{\mathbf{p}})}{t^{\mathbf{p}-1}}&=& \mathbf{f^{h}}^{(3)}_{\,k,\textsc{RAD}}(\mathbf{a})\:t^{0}=\left(\:0,\:0,\:\frac{3\theta}{2}\right)\:t^{0},\label{eq:rchsub3B2}
\end{eqnarray}
\begin{eqnarray}
\frac{\mathbf{f^{h}}^{(4)}_{\,k,\textsc{RAD}}(\mathbf{a}t^{\mathbf{p}})}{t^{\mathbf{p}-1}}&=& \mathbf{f^{h}}^{(4)}_{\,k,\textsc{RAD}}(\mathbf{a})\:t^{0}=\left(\:0,\:0,\:0\right)\:t^{0}=\mathbf{0},\label{eq:rchsub4B2}\\[12pt]
\frac{\mathbf{f^{h}}^{(5)}_{\,k,\textsc{RAD}}(\mathbf{a}t^{\mathbf{p}})}{t^{\mathbf{p}-1}}&=& \mathbf{f^{h}}^{(5)}_{\,k,\textsc{RAD}}(\mathbf{a})\:t^{0}=\left(\:0,\:0,\:0\right)\:t^{0}=\mathbf{0},\label{eq:rchsub5B2}\\[12pt]
\frac{\mathbf{f^{h}}^{(6)}_{\,k,\textsc{RAD}}(\mathbf{a}t^{\mathbf{p}})}{t^{\mathbf{p}-1}}&=& \mathbf{f^{h}}^{(6)}_{\,k,\textsc{RAD}}(\mathbf{a})\:t^{2}=\left(\:0,\:0,\:-\frac{\theta}{2\xi}\right)\:t^{2},\label{eq:rchsub6B2}\\[12pt]
\frac{\mathbf{f^{h}}^{(7)}_{\,k,\textsc{RAD}}(\mathbf{a}t^{\mathbf{p}})}{t^{\mathbf{p}-1}}&=& \mathbf{f^{h}}^{(7)}_{\,k,\textsc{RAD}}(\mathbf{a})\:t^{2}=\left(\:0,\:0,\:-\frac{k}{2\xi \theta}\right)\:t^{2},\label{eq:rchsub7B2}\\[12pt]
\frac{\mathbf{f^{h}}^{(8)}_{\,k,\textsc{RAD}}(\mathbf{a}t^{\mathbf{p}})}{t^{\mathbf{p}-1}}&=& \mathbf{f^{h}}^{(8)}_{\,k,\textsc{RAD}}(\mathbf{a})\:t^{0}=\left(\:0,\:0,\:\frac{k}{\theta}\right)\:t^{0},\label{eq:rchsub8B2}.
\end{eqnarray}
From the above expressions (\ref{eq:rchsub1B2})-(\ref{eq:rchsub8B2}) we can make the following general observations:
\begin{enumerate}
\item As $t\rightarrow0$, the expressions (\ref{eq:rchsub1B2}) and (\ref{eq:rchsub2B2}) go to $\frac{\zeta^2}{2\xi\theta^{3}}$ and  $-\frac{k^2}{2 \theta^{3}}$ respectively revealing a strong dominant character of the corresponding terms. Consequently, when one of the first or the second term of the RHS of (\ref{eq:rcds3}) shows a dominant behavior in an asymptotic splitting, and thus leads to the balance (\ref{rcdombalf22}), the other one cannot be part of the subdominant part of that same asymptotic splitting. In other words these two terms, namely $\frac{\z^2}{2\xi x^2y}$ and $-\frac{k^2\xi}{2\xi x^2y}$, cannot be separated between the dominant and the subdominant parts of a possible asymptotic decomposition. This happens in 32 cases of the 63 possible asymptotic decompositions we have left. Hence, we are now left with 31 candidates which can still lead to an asymptotic solution of our initial radiation-curved dynamical system.
\item The third expression, (\ref{eq:rchsub3B2}), goes to $\frac{3\theta}{2}$ as $t\rightarrow0$. Consequently, the corresponding term $\frac{3y^3}{2x^2}$ shows a very strong dominant character in this case and it is impossible to occur in the subdominant part in any possible asymptotic decomposition that admits the dominant balance (\ref{rcdombalf22}). There are 11 asymptotic splittings that have this characteristic out of the 31 we have left from the previous observation and accordingly we have now 20 possible asymptotic splittings whose asymptotic behavior needs further analysis.
\item The fourth, fifth, sixth and seventh term of the RHS of (\ref{eq:rcds3}) all reveal a very strong subdominant character for different reasons. More specifically, in Eqs. (\ref{eq:rchsub4B2}) and (\ref{eq:rchsub5B2}) there is no need to take the limit as $t\rightarrow0$ since the value $r=0$ causes  
$\mathbf{f^{h}}^{(4)}_{\,k,\textsc{RAD}}(\mathbf{a})$  and $\mathbf{f^{h}}^{(5)}_{\,k,\textsc{RAD}}(\mathbf{a})$ to vanish independently of t. Expression (\ref{eq:rchsub6B1}) corresponds to the linear term and as expected it vanishes as $t\rightarrow0$. The same happens for the expression (\ref{eq:rchsub7B2}) which was also expected since that term is the one that leads to a different dominant balance, namely (\ref{rcdombalf11}) which was studied in the previous case.
\item The last expression, (\ref{eq:rchsub4B2}), corresponds to the eighth term of (\ref{eq:rcds3}) and one of the terms that leads to the dominant balance of the present case. Consequently, that term was expected to show a dominant character as it does. Taking the limit of (\ref{eq:rchsub8B2}) as $t\rightarrow0$, it goes to $\frac{k}{\theta}$. Hence, it is impossible for an asymptotic splitting which admits the dominant balance (\ref{rcdombalf22}), i.e. has the first and second terms in its dominant part, to have the term $\frac{ky}{x^2}$ in its subdominant part. There 4 asymptotic spllitings left with this characteristic out of the 20 we have in total, thus leaving 16 asymptotic splittings to be examined.
\end{enumerate}
There are 9 more asymptotic splittings, out of the 16 we have left, that admit the dominant balance (\ref{rcdombalf22}). Five of them, namely $\mathbf{f^{8}}_{\,k,\textsc{RAD}}$, $\mathbf{f^{26}}_{\,k,\textsc{RAD}}$, $\mathbf{f^{76}}_{\,k,\textsc{RAD}}$, $\mathbf{f^{79}}_{\,k,\textsc{RAD}}$ and $\mathbf{f^{150}}_{\,k,\textsc{RAD}}$  admit the present dominant balance, (\ref{rcdombalf22}), because the eighth term is included in their dominant part while the first and second terms of the RHS of (\ref{eq:rcds3}) are included in their subdominant part. Combining the first and fourth observations above it is clear that for these five asymptotic splittings the expression (\ref{rcsubcon}) does not vanish as $t\rightarrow0$. The other four asymptotic splittings, namely $\mathbf{f^{97}}_{\,k,\textsc{RAD}}$, $\mathbf{f^{166}}_{\,k,\textsc{RAD}}$, $\mathbf{f^{169}}_{\,k,\textsc{RAD}}$ and $\mathbf{f^{221}}_{\,k,\textsc{RAD}}$ contain in their dominant part all of the necessary terms, i.e. the first, the second and the eighth, and lead to an admissible dominant balance of the form (\ref{rcdombalf22}) but the spectrum of their Kovalevskaya matrices does not contain the eigenvalue -1, which as we have discussed in earlier chapters, corresponds to the position of the singularity in the final asymptotic solution. Hence, they fail to show consistency with the overall scheme of the method of asymptotic splittings.

%==========================================================================
\subsubsection*{ The general form of dominant balance p=(q+1,q,q-1)}

After having discussed extensively the two special cases of the forms of possible dominant balances, namely (\ref{rcdombalf11}) and (\ref{rcdombalf22}), we  
have explained the reasons for which 248 asympotic splittings out of the 255 which are listed in Table (\ref{long2}) at the end of this chapter, fail to lead to an asymptotic solution of the radiation-curved vector field (\ref{rcvf}) or the associated dynamical system (\ref{eq:rcds1})-(\ref{eq:rcds3}),
we are left with 7 asympotic splittings that do not fall in any of these two categories, namely $\mathbf{f^{3}}_{\,k,\textsc{RAD}}$, $\mathbf{f^{4}}_{\,k,\textsc{RAD}}$, $\mathbf{f^{5}}_{\,k,\textsc{RAD}}$, $\mathbf{f^{22}}_{\,k,\textsc{RAD}}$, $\mathbf{f^{23}}_{\,k,\textsc{RAD}}$, $\mathbf{f^{27}}_{\,k,\textsc{RAD}}$ and $\mathbf{f^{73}}_{\,k,\textsc{RAD}}$. If any of these 8 asymptotic splittings admits a dominant balance, it must be of the general form (\ref{rcdombalf3}), that is 
\be\label{rcdombalf33}
\mathcal{B}_{\,k,\textsc{RAD}}=(\mathbf{a},\mathbf{p})= \left(\left(\theta,\:\:\eta,\:\:\rho\right),\:\left(q+1,\:\:q,\:\:q-1\right)\right).
\ee

%============================================================================================================
\subsection{The unique asymptotic decomposition}

After solving one by one the dominant systems of the final 7 asympotic splittings we conclude that the only acceptable asymptotic splitting of the vector field  $\mathbf{f}_{\,k,\textsc{RAD}}$ is 
\be
\mathbf{f^{73}}_{\,k,\textsc{RAD}}=\mathbf{f^{73}}^{(0)}_{\,k,\textsc{RAD}} + \mathbf{f^{73}}^{\,(\textrm{sub})}_{\,k,\textsc{RAD}},
\ee
with dominant part
\be \label{rcfinaldom}
\mathbf{f^{73}}^{(0)}_{\,k,\textsc{RAD}}(\mathbf{x})=\left(y,z,\frac{3y^3}{2x^2} + \frac{z^2}{2y} -\frac{yz}{x}
\right), 
\ee
and subdominant part
\be \label{rcfinalsub}
\mathbf{f^{73}}^{\,(\textrm{sub})}_{\,k,\textsc{RAD}}(\mathbf{x})=
\left(0,0, \frac{\z^2}{2\xi x^2y} -\frac{k^2}{2 x^2y} -\frac{y}{2\xi} -\frac{k}{2\xi y} + \frac{ky}{x^2}\right).
\ee
A final comment about the asymptotic splittings of the field equations of the present Section is in order. It is interesting that the \emph{dominant} part of the vector field given by Eq. (\ref{rcfinaldom}) is precisely the same as that of the field in the flat, radiation dominated case treated in \cite{cot-tso-07} (see Eq. (16) in that paper). Their \emph{difference} lies in the subdominant parts of the two cases, the curved one treated here and the flat case in \cite{cot-tso-07}: Here the subdominant part given by (\ref{rcfinalsub}) contains precisely the terms of the vector field  $\mathbf{f}^{\,(\textrm{sub})}_{\,0,\textsc{RAD}}$ (radiation and linear terms),  plus the three curvature terms (those with a $k$ in Eq. (\ref{rcvf}).

In this Section we have found that the dynamical system describing radiation universes with curvature admits one and only one possible asymptotic behavior as we approach the finite-time singularity. This unique asymptotic splitting allows us to construct in the following sections an asymptotic solution of the radiation-curved vector field (\ref{rcvf}) or the associated dynamical system (\ref{eq:rcds1})-(\ref{eq:rcds3}) in the neighborhood of the initial singularity. We described some basic qualitative characteristics of the behavior of the basic vector field by examining the exhaustive list of all possible dominant asymptotic systems without solving them. Thus, we completed the first part of our asymptotic analysis through the method of asymptotic splittings. In the next Section we will examine further the consistency of the found asymptotic decomposition and we will construct the final asymptotic solution in the form of a Fuchsian series.

%--------------------------------------------------------------------------------------------------------------------
%	SECTION 5.5  - Stability of the curvature-radiation asymptotic solutions 
%--------------------------------------------------------------------------------------------------------------------
\section{Stability of the curvature-radiation asymptotic solution}

%============================================================================================================
\subsection{Dominant balance}

In this Section we will look for the possible asymptotic solutions, asymptotic forms of integral curves  of the curvature-radiation field $\mathbf{f}_{\,k,\textsc{RAD}}$. In other words, we search for the dominant balances determined by the dominant part $\mathbf{f^{73}}^{(0)}_{\,k,\textsc{RAD}}$ given by Eq. (\ref{rcfinaldom}). In order to do that, we substitute the forms (\ref{eq:domisol3}) in the dominant system $(\dot x,\dot y,\dot z)(t)=\mathbf{f}^{(0)}_{\,k,\textsc{RAD}}$ and solve the resulting nonlinear algebraic system  aiming to determine the dominant balance $(\mathbf{a},\mathbf{p})$ in the form of an exact, scale invariant solution. Hence, we have
\begin{eqnarray}
\theta p\:t^{p-1}&=& \eta\: t^{q},\label{eq:rcgensol-domsys1}\\[13pt] 
\eta q\:t^{q-1} &=& \rho\: t^{r},\label{eq:rcgensol-domsys2}\\[13pt] 
\rho r\:t^{r-1} &=& \frac{3\eta^{3}}{2\theta^{2}}\:\:t^{-2p+3q} +\frac{\rho^{2}}{2\eta}\:\:t^{2r-q} -\frac{\eta\rho}{\theta}\:\:t^{q+r-p}.\label{eq:rcgensol-domsys3}
\end{eqnarray}
This leads to the unique \emph{curvature-radiation} balance $\mathcal{B}^{73}_{\,k,\textsc{RAD}}\in\mathbb{C}^3\times\mathbb{Q}^3$,  with
\be\label{rcfinaldb}
\mathcal{B}^{73}_{\,k,\textsc{RAD}}=(\mathbf{a},\mathbf{p})= \left(\left(\theta,\frac{\theta}{2},-\frac{\theta}{4}\right),\:
\left(\frac{1}{2},-\frac{1}{2},-\frac{3}{2}\right)\right),
\ee
where $\theta$ is a real, arbitrary constant. Consequently, for reasons that we have discussed in previous chapters we are lead to the fact that the vector field $\mathbf{f}^{(0)}_{\,k,\textsc{RAD}}$ is \emph{a scale invariant system}.

%============================================================================================================
\subsection{Subdominant condition}

Accordingly, we must also check that, in the basic decomposition of the curvature-radiation field, (\ref{general split3}), the higher order terms (\ref{rcfinalsub})are weight-homogeneous with respect to the curvature-radiation balance (\ref{rcfinaldb}). We begin by splitting the subdominant part (\ref{rcfinaldom}) in the following way,
\be
\mathbf{f^{73}}^{\,(\textrm{sub})}_{\,k,\textsc{RAD}}(\mathbf{x}) = \mathbf{f^{73}}^{(1)}_{\,k,\textsc{RAD}}(\mathbf{x}) +
\mathbf{f^{73}}^{(2)}_{\,k,\textsc{RAD}}(\mathbf{x}) + \mathbf{f^{73}}^{(3)}_{\,k,\textsc{RAD}}(\mathbf{x}),
\ee
where
\begin{eqnarray}
\mathbf{f^{73}}^{(1)}_{\,k,\textsc{RAD}}(\mathbf{x})&=&\left(0,0,\frac{ky}{x^2}\right),\\[10pt]
\mathbf{f^{73}}^{(2)}_{\,k,\textsc{RAD}}(\mathbf{x})&=&\left(0,0,\frac{\z^2-k^2\xi}{2\xi x^2 y}-\frac{y}{2\xi}\right),\\[10pt]
\mathbf{f^{73}}^{(3)}_{\,k,\textsc{RAD}}(\mathbf{x})&=&\left(0,0,-\frac{k}{2\xi y}\right).
\end{eqnarray}
and we can now look for the required condition by examining the subdominant character of these expressions. Using the balance $\mathcal{B}^{73}_{\,k,\textsc{RAD}} $ defined by Eq. (\ref{rcfinaldb}), and taking into consideration that $t^{r-1}=t^{-5/2}$, we construct the fractions,
\bq
\frac{\mathbf{f^{73}}^{(1)}_{\,k,\textsc{RAD}}(\mathbf{a}t^{\mathbf{p}})}{t^{\mathbf{p}-1}}&=& \mathbf{f^{73}}^{(1)}_{\,k,\textsc{RAD}}(\mathbf{a})t,\\
\frac{\mathbf{f^{73}}^{(2)}_{\,k,\textsc{RAD}}(\mathbf{a}t^{\mathbf{p}})}{t^{\mathbf{p}-1}}&=& \mathbf{f^{73}}^{(2)}_{\,k,\textsc{RAD}}(\mathbf{a})t^{2},\\
\frac{\mathbf{f^{73}}^{(3)}_{\,k,\textsc{RAD}}(\mathbf{a}t^{\mathbf{p}})}{t^{\mathbf{p}-1}}&=& \mathbf{f^{73}}^{(3)}_{\,k,\textsc{RAD}}(\mathbf{a})t^{3}.
\eq
Provided that the forms $\mathbf{f}^{(i)}_{\,k,\textsc{RAD}}(\mathbf{a}), i=1,2,3,$ do not vanish, these fractions go to zero asymptotically as $t\rightarrow 0$. This happens for all cases except when $3\b +\g=0$, that is when $\xi\neq 0$. We conclude that this result is true in all higher order gravity theories \emph{except} perhaps the so-called conformally invariant Bach-Weyl gravity\footnote{Apparently, this case needs a separate treatment altogether.}, cf. \cite{wey-19a, bac-21}.
We conclude that the subdominant part (\ref{rcfinalsub}) is weight-homogeneous because the \emph{subdominant exponents} are ordered,
\be\label{sub exps3}
q^{(0)}=0\ <\ q^{(1)}=1\ <\ q^{(2)}=2\ <\ q^{(3)}=3.
\ee

%============================================================================================================
\subsection{Construction of the K-matrix}

In this last phase of the asymptotic analysis, our aim is to construct a final series representation of the asymptotic solutions of the curved, radiation-filled vector field valid in the neighborhood of the singularity in such a way that will allow the dominant balance solutions we have construct until now to dominate it. The number of arbitrary constants in that final series expansion will determine whether we have a \emph{general} or a \emph{particular} solution and it is expected that such constants will appear in certain places of our final formal developments. Subsequently, we move on to check the \emph{consistency} of our asymptotic solutions with the broader mathematical context we are using and proceed to the calculation of the precise positions of the arbitrary constants in the final series representation.

We recall that the arbitrary constants of any solution first appear in those terms in the asymptotic series expansion whose coefficients $\mathbf{c}_{i}$ have indices $i=\varrho s$, where $\varrho$ is a non-negative $\mathcal{K}$-exponent. The least common multiple of the denominators of the set of all subdominant exponents (\ref{sub exps3}) and those of all the $\mathcal{K}$-exponents with positive real parts is denoted by $s$  (in our case, $s=2$). In order to calculate those exponents we need to find the spectrum of the Kovalevskaya matrix of the dominant part of our asymptotic decomposition with respect to the dominant balance,
\be
\mathcal{K}^{73}_{\,k,\textsc{RAD}}=D\,\mathbf{f^{73}}^{(0)}_{\,k,\textsc{RAD}}(\mathbf{a})-\textrm{diag}(\mathbf{p}).
\ee
Hence, the $\mathcal{K}$-exponents depend of the dominant part of the vector field as well as the dominant balance. In our case, the Kovalevskaya matrix is
\be
\mathcal{K}^{73}_{\,k,\textsc{RAD}}=\left(
                     \begin{array}{ccc}
                       -1/2& 1  & 0\\
                       0 & 1/2&1 \\
                       -1/2&5/4&1/2
                     \end{array}
                   \right),
\ee
with spectrum
\be
\textrm{spec}(\mathcal{K}^{73}_{\,k,\textsc{RAD}})=\{-1,0,3/2\},
\ee
and corresponding eigenvectors
\be \{(4,-2,3),(4,2,-1),(1,2,2)\}.  \ee
The number of non-negative $\mathcal{K}$-exponents equals the number of arbitrary constants that appear in the series expansions. There is always the $-1$ exponent that corresponds to an arbitrary constant, the position of the singularity, and because the $\textrm{spec}(\mathcal{K}_{\,k,\textsc{RAD}})$ in our case possesses two non-negative eigenvalues, the balance $\mathcal{B}^{73}_{\,k,\textsc{RAD}}$ indeed corresponds to the dominant behavior of a \emph{general} solution having the form of a formal series and valid locally around the initial singularity.

%============================================================================================================
\subsection{Construction of the formal expansion series}

To find it, we substitute the \emph{Puiseux series expansions}
\begin{equation} \label{eq:series3}
x(t) = \sum_{i=0}^{\infty} c_{1i} t^{\frac{i}{2}+\frac{1}{2}}, \:\:\:\:\:
y(t) = \sum_{i=0}^{\infty} c_{2i}  t^{\frac{i}{2}-\frac{1}{2}},\:\:\:\:\:
z(t) = \sum_{i=0}^{\infty} c_{3i} t^{\frac{i}{2}-\frac{3}{2}} ,\:\:\:\:\:
\end{equation}
\begin{equation} \label{eq:rcseriesder}
\begin{split}
\dot{x}(t) = \sum_{i=0}^{\infty} c_{1i} \left(\frac{i}{2}+\frac{1}{2}\right) t^{\frac{i}{2}-\frac{1}{2}},\:\:\:\:\:
\dot{y}(t) = \sum_{i=0}^{\infty} c_{2i} \left(\frac{i}{2}-\frac{1}{2}\right) t^{\frac{i}{2}-\frac{3}{2}},\\[9pt]
\dot{z}(t) = \sum_{i=0}^{\infty} c_{3i} \left(\frac{i}{2}-\frac{3}{2}\right) t^{\frac{i}{2}-\frac{5}{2}} ,\:\:\:\:\:
\end{split}
\end{equation}
where  $c_{10}=\theta$, $c_{20}=\theta /2$, $c_{30}=-\theta/4$, in the following equivalent form of the original system(\ref{eq:rcds1})-(\ref{eq:rcds3}),
\begin{eqnarray}
\dot{x} &=& y,\label{eq:rcds11}\\[09pt] 
\dot{y} &=& z,\label{eq:rcds22}\\[12pt]
x^2y\dot{z} &=& -\frac{x^2y^2}{2\xi} +\frac{x^2z^2}{2} -xy^2z +\frac{3y^4}{2} +\frac{\z^2\xi}{2\xi} -\frac{k^2}{2}    
-\frac{kx^2}{2\xi} + ky^2,\label{eq:rcds33}
\end{eqnarray}
and we are led to various recursion relations that determine the unknowns $c_{1i}$, $c_{2i}$, $c_{3i}$ term by term. More specifically from Eq. (\ref{eq:rcds11}) after substitution we have
\be
\sum_{i=0}^{\infty} c_{1i} \left(\frac{i+1}{2}\right)\:\:t^{\frac{i-1}{2}}=\sum_{i=0}^{\infty} c_{2i}\:\:t^{\frac{i-1}{2}},
\ee
which leads to
\be
\left(\frac{i+1}{2}\right)c_{1i} = c_{2i}\:.\label{eq:rcsubsti1}
\ee
From Eq. (\ref{eq:rcds22}) after substitution we have
\be
\sum_{i=0}^{\infty} c_{2i} \left(\frac{i-1}{2}\right)\: t^{\frac{i-3}{2}}=\left(\sum_{i=0}^{\infty} c_{3i}  t^{\frac{i-3}{2}}\right),
\ee
which leads to
\be
c_{3i} = \left(\frac{i-1}{2}\right)c_{2i}\:.\label{eq:rcsubsti2}
\ee
From Eq. (\ref{eq:rcds33}) we calculate separately each term after substitution:

\begin{eqnarray}
x^2y\dot{z} &=& \left(\sum_{i=0}^{\infty} c_{1i}\:\: t^{\frac{i+1}{2}}\right) \left(\sum_{i=0}^{\infty} c_{1i}\:\: t^{\frac{i+1}{2}}\right)\left( \sum_{i=0}^{\infty} c_{2i}\:\: t^{\frac{i-1}{2}} \right)\left( \sum_{i=0}^{\infty} c_{3i} \left(\frac{i}{2}-\frac{3}{2}\right) t^{\frac{i}{2}-\frac{5}{2}} \right)\nonumber\\[13pt]
&=& t^{-2} \sum_{i=0}^{\infty} \sum_{l=0}^{i} \sum_{m=0}^{l} \sum_{n=0}^{m} c_{3(i-l)}c_{2(l-m)}c_{1(m-n)} c_{1n}\:\left(\frac{i-l}{2}-\frac{3}{2}\right)\: 
t^{\frac{i}{2}},\\[15pt]
-\frac{x^2y^2}{2\xi} &=& -\frac{1}{2\xi}\left(\sum_{i=0}^{\infty} c_{1i}\:\: t^{\frac{i+1}{2}}\right) \left(\sum_{i=0}^{\infty} c_{1i}\:\: 
t^{\frac{i+1}{2}}\right)\left( \sum_{i=0}^{\infty} c_{2i}\:\: t^{\frac{i-1}{2}} \right)\left( \sum_{i=0}^{\infty} c_{2i}\:\:  t^{\frac{i-1}{2}} \right) \nonumber\\[13pt]
&=& -\frac{1}{2\xi}\:\: t^{0} \sum_{i=0}^{\infty} \sum_{l=0}^{i} \sum_{m=0}^{l} \sum_{n=0}^{m} c_{2(i-l)}c_{2(l-m)}c_{1(m-n)} c_{1n}\:\: t^{\frac{i}{2}},\\[15pt]
\frac{x^2z^2}{2} &=& \frac{1}{2}\left(\sum_{i=0}^{\infty} c_{1i}\:\: t^{\frac{i+1}{2}}\right) \left(\sum_{i=0}^{\infty} c_{1i}\:\: t^{\frac{i+1}{2}}\right)\left( \sum_{i=0}^{\infty} c_{3i}\:\: t^{\frac{i-3}{2}} \right)\left( \sum_{i=0}^{\infty} c_{3i}\:\:  t^{\frac{i-3}{2}} \right) \nonumber\\[13pt]
&=& \frac{1}{2}\:\: t^{-2} \sum_{i=0}^{\infty} \sum_{l=0}^{i} \sum_{m=0}^{l} \sum_{n=0}^{m} c_{3(i-l)}c_{3(l-m)}c_{1(m-n)} c_{1n}\:\: t^{\frac{i}{2}},\\[15pt]
-xy^2z &=& - \left(\sum_{i=0}^{\infty} c_{1i}\:\: t^{\frac{i+1}{2}}\right)\left( \sum_{i=0}^{\infty} c_{2i}\:\: t^{\frac{i-1}{2}} \right)\left( \sum_{i=0}^{\infty} c_{2i}\:\:  t^{\frac{i-1}{2}} \right)\left( \sum_{i=0}^{\infty} c_{3i}\:\: t^{\frac{i-3}{2}} \right)\nonumber\\[13pt]
&=& - t^{-2} \sum_{i=0}^{\infty} \sum_{l=0}^{i} \sum_{m=0}^{l} \sum_{n=0}^{m} c_{3(i-l)}c_{1(l-m)}c_{2(m-n)} c_{2n}\:\: t^{\frac{i}{2}},\\[15pt]
\frac{3y^{4}}{2}&=& \frac{3}{2} \left( \sum_{i=0}^{\infty} c_{2i}\:\: t^{\frac{i-1}{2}} \right)\left( \sum_{i=0}^{\infty} c_{2i}\:\:  t^{\frac{i-1}{2}} \right)\left( \sum_{i=0}^{\infty} c_{2i}\:\: t^{\frac{i-1}{2}} \right)\left( \sum_{i=0}^{\infty} c_{2i}\:\: t^{\frac{i-1}{2}} \right) \nonumber\\[13pt]
&=&\frac{3}{2}\:\: t^{-2} \sum_{i=0}^{\infty} \sum_{l=0}^{i} \sum_{m=0}^{l} \sum_{n=0}^{m} c_{2(i-l)}c_{2(l-m)}c_{2(m-n)} c_{2n}\:\: t^{\frac{i}{2}},
\end{eqnarray}
\begin{eqnarray}
-\frac{kx^{2}}{2\xi}&=& -\frac{k}{2\xi}\left(\sum_{i=0}^{\infty} c_{1i}\:\: t^{\frac{i+1}{2}}\right) \left(\sum_{i=0}^{\infty} c_{1i}\:\: t^{\frac{i+1}{2}}\right)  \nonumber \\[13pt]
&=&-\frac{k}{2\xi}\:\: t \sum_{i=0}^{\infty} \sum_{k=0}^{i} c_{1(i-k)} c_{1k} \:\:t^{\frac{i}{2}},\\[15pt]
ky^{2}&=& k \left( \sum_{i=0}^{\infty} c_{2i}\:\: t^{\frac{i-1}{2}} \right)\left( \sum_{i=0}^{\infty} c_{2i}\:\:  t^{\frac{i-1}{2}} \right) \nonumber\\[13pt]
&=& k\:\: t^{-1} \sum_{i=0}^{\infty} \sum_{l=0}^{i} c_{2(i-l)} c_{2l}\:\: t^{\frac{i}{2}},
\end{eqnarray}
Subsequently, we are led to the following form of (\ref{eq:rcds33}),
\be
\begin{split}
\sum_{i=0}^{\infty} \sum_{l=0}^{i} \sum_{m=0}^{l} \sum_{n=0}^{m}\left( c_{3(i-l)}c_{2(l-m)}c_{1(m-n)} c_{1n}\:\left(\frac{i-l}{2}-\frac{3}{2}\right)-\frac{1}{2}\:\:c_{3(i-l)}c_{3(l-m)}c_{1(m-n)} c_{1n}+\right.\\[13pt] 
\left.  +c_{3(i-l)}c_{1(l-m)}c_{2(m-n)} c_{2n} -\frac{3}{2}\:\:c_{2(i-l)}c_{2(l-m)}c_{2(m-n)} c_{2n}\: \right)t^{\frac{i}{2}-2} = \\[13pt] 
=\frac{\zeta^2}{2\xi}-\frac{k^2}{2} + \sum_{i=0}^{\infty}\sum_{l=0}^{i}\left( k\:\: c_{2(i-l)} c_{2l} \right)t^{\frac{i}{2}-1}- \sum_{i=0}^{\infty} \sum_{l=0}^{i} \left( \frac{k}{2\xi}\:\:c_{1(i-l)} c_{1l}\right)t^{\frac{i}{2}+1} -\\[13pt]
-\sum_{i=0}^{\infty} \sum_{l=0}^{i} \sum_{m=0}^{l} \sum_{n=0}^{m}\left( \frac{1}{2\xi}\:\: c_{2(i-l)}c_{2(l-m)}c_{1(m-n)}c_{1n} \right)t^{\frac{i}{2}}.
\label{eq:rcsubsti3}
\end{split}
\ee

%============================================================================================================
\subsection{Calculation of the final series coefficients}

Eqs. (\ref{eq:rcsubsti1}), (\ref{eq:rcsubsti2}) and (\ref{eq:rcsubsti3}) constitute the system from which we will now calculate term by term the coefficients $c_{1i}$, $c_{2i}$ and $c_{3i}$ of the asymptotic solution of the initial dynamical system (\ref{cds1})-(\ref{cds3}) in the form of the Fuchsian series expansions (\ref{eq:cseries}), that is
\begin{eqnarray}
x(t)&=& \theta t^{1/2} +c_{11}\:\:t^{1} +c_{12}\:\:t^{3/2}+c_{13}\:\:t^{2} +c_{14}\:\:t^{5/2} + \cdots,\nonumber\\[10pt]
y(t)&=& \frac{\theta}{2}t^{-1/2} +c_{21}\:\:t^{0} +c_{22}\:\:t^{1/2}+c_{23}\:\:t^{1} +c_{24}\:\:t^{3/2} + \cdots,\nonumber\\[10pt]
z(t)&=& -\frac{\theta}{4} t^{-3/2} +c_{31}\:\:t^{-1} +c_{32}\:\:t^{-1/2}+c_{33}\:\:t^{0} +c_{34}\:\:t^{1/2} + \cdots.
\end{eqnarray}
In what follows we will calculate the values of $c_{1i}$, $c_{2i}$ and $c_{3i}$ using Eqs. (\ref{eq:rcsubsti1}), (\ref{eq:rcsubsti2}) and (\ref{eq:rcsubsti3})
For each one of these three, we will determine a different set of equations for the coefficients of the various powers of $t$. 

%============================================================================================================
\subsubsection*{1st set of equations}

Eq. (\ref{eq:rcsubsti1}) will lead to the following equations, for the coefficients of the different powers of $t$, 
\bq
\text{for the coefficients of the term} &t^{0},\:\:\:\:c_{21} = c_{11}\:,\label{coef1-1cr}\\[12pt]
\text{for the coefficients of the term} &t^{1/2},\:\:\:\:c_{22} = \frac{3}{2} c_{12}\:,\label{coef1-2cr}\\[12pt]
\text{for the coefficients of the term} &t^{1},\:\:\:\:c_{23} = 2c_{13}\:,\label{coef1-3cr}\\[12pt]
\text{for the coefficients of the term} &t^{3/2},\:\:\:\:c_{24} = \frac{5}{2}c_{14}\:.\label{coef1-4cr} 
\eq

%============================================================================================================
\subsubsection*{2nd set of equations}

Calculating the coefficients of the different powers of $t$ in Eq. (\ref{eq:rcsubsti2}), we are led to the following equations,
for the coefficients of the term $t^{-1}$ ($i=1$), we have
\be
c_{31} = 0\:,\label{coef2-1cr}
\ee
for the coefficients of the term $t^{-1/2}$ ($i=2$), we have
\be
c_{32} = \frac{1}{2}c_{22}\:,
\ee
which, taking into consideration (\ref{coef1-2cr}) leads to
\be
c_{32} = \frac{3}{4}c_{12}\:,\label{coef2-2cr}
\ee
for the coefficients of the term $t^{0}$ ($i=3$), we have
\be
c_{33} = c_{23}\:,
\ee
which, taking into consideration (\ref{coef1-3cr}) leads to
\be
c_{33} = 2c_{13}\:, \label{coef2-3cr}
\ee
for the coefficients of the term $t^{1/2}$ ($i=4$), we have
\be
c_{34} = \frac{3}{2}c_{24}\:, 
\ee
which, taking into consideration (\ref{coef1-4cr}) leads to
\be
c_{34} = \frac{15}{4}c_{14}\:.\label{coef2-4cr}
\ee

%============================================================================================================
\subsubsection*{3rd set of equations}

Subsequently, Eq. (\ref{eq:rcsubsti3}) will lead to the following equations by writing the conditions for the coefficients of the different powers of $t$, \\
for the coefficients of the term $t^{-3/2}$, we have
\be
\begin{split}
\sum_{l=0}^{1} \sum_{m=0}^{l} \sum_{n=0}^{m} c_{3(1-l)}c_{2(l-m)}c_{1(m-n)} c_{1n}\:\left(\frac{1-l}{2}-\frac{3}{2}\right) -\frac{1}{2}c_{3(1-l)}c_{3(l-m)}c_{1(m-n)} c_{1n} + \\[13pt] 
+c_{3(1-l)}c_{1(l-m)}c_{2(m-n)} c_{2n}-\frac{3}{2}\:\:c_{2(1-l)}c_{2(l-m)}c_{2(m-n)} c_{2n}  \: = 0\:,
\end{split}
\ee
which leads to 
\be
2c_{11} =5c_{21}\label{coef3-1cr},
\ee
for the coefficients of the term $t^{-1}$, we have
\be
\begin{split}
\sum_{l=0}^{2} \sum_{m=0}^{l} \sum_{n=0}^{m} c_{3(2-l)}c_{2(l-m)}c_{1(m-n)} c_{1n}\:\left(\frac{2-l}{2}-\frac{3}{2}\right) -\frac{1}{2}c_{3(2-l)}c_{3(l-m)}c_{1(m-n)} c_{1n} +\\[13pt]   
+c_{3(2-l)}c_{1(l-m)}c_{2(m-n)} c_{2n}-\frac{3}{2}\:\:c_{2(2-l)}c_{2(l-m)}c_{2(m-n)} c_{2n} \: =  k\:\:\sum_{l=0}^{0} c_{2(0-l)} c_{2l} \:,
\end{split}
\ee
which leads to
\be
c_{12}=-\frac{k}{2\theta}\label{coef3-2cr},
\ee
for the coefficients of the term $t^{-1/2}$, we have
\be
\begin{split}
\sum_{l=0}^{3} \sum_{m=0}^{l} \sum_{n=0}^{m} c_{3(3-l)}c_{2(l-m)}c_{1(m-n)} c_{1n}\:\left(\frac{3-l}{2}-\frac{3}{2}\right) -\frac{1}{2}c_{3(3-l)}c_{3(l-m)}c_{1(m-n)} c_{1n} +\\[13pt]   
+c_{3(3-l)}c_{1(l-m)}c_{2(m-n)} c_{2n}-\frac{3}{2}\:\:c_{2(3-l)}c_{2(l-m)}c_{2(m-n)} c_{2n} \: =  k\:\:\sum_{l=0}^{1} c_{2(1-l)} c_{2l} \:,
\end{split}
\ee
which leads to 
\be
2c_{13}-5c_{23}+4c_{33}=0\label{coef3-3cr},
\ee
for the coefficients of the term $t^{0}$, we have
\be
\begin{split}
\sum_{l=0}^{4} \sum_{m=0}^{l} \sum_{n=0}^{m} c_{3(4-l)}c_{2(l-m)}c_{1(m-n)} c_{1n}\:\left(\frac{4-l}{2}-\frac{3}{2}\right)  -\frac{1}{2}c_{3(4-l)}c_{3(l-m)}c_{1(m-n)} c_{1n}+ \\[13pt]
+c_{3(4-l)}c_{1(l-m)}c_{2(m-n)} c_{2n} -\frac{3}{2}\:\:c_{2(4-l)}c_{2(l-m)}c_{2(m-n)} c_{2n} \: = \\[13pt]  
=\frac{\zeta^2}{2\xi}-\frac{k^2}{2} + k\:\:\sum_{l=0}^{2} c_{2(2-l)} c_{2l} - \frac{1}{2\xi}\:\:\sum_{l=0}^{0} \sum_{m=0}^{l} \sum_{n=0}^{m} c_{2(0-l)}c_{2(l-m)}c_{1(m-n)}c_{1n},
\end{split}
\ee
which leads to 
\be
c_{14}=\frac{4\zeta^2-\theta^4}{12\xi\theta^3}-\frac{k^2}{8\theta^3}\:\:.\label{coef3-4cr}
\ee

%============================================================================================================
\subsection{Final form of the general solution}

In the end, we are lead to the final series representation of the solution in the form:
\be
x(t) = \theta \:\:t^{1/2} - \frac{k}{2\theta}\:\:t^{3/2} + c_{13} \:\: t^{2} + \displaystyle \left(\frac{4\z^2-\theta^4}{12\xi\t^3}-\frac{k^2}{8\theta^3}\right)\:\:t^{5/2} + \cdots .
\label{eq:gensol3}
\ee
The corresponding series expansions for $y(t)$ and $z(t)$ are
given by  the first and second time derivatives of the above
expression respectively.

%
%%============================================================================================================
\subsubsection*{Fredholm's alternative}

As a final test for admission of this solution, we use Fredholm's
alternative to be satisfied by any admissible solution. This leads to
a \emph{compatibility condition} for the positive
eigenvalue 3/2 and the associated eigenvector: This condition has the form
\be
v^{\top}\cdot\left(\mathcal{K}-\frac{j}{s}I\right)\mathbf{c}_j=0,
\ee
where $I$ denotes the identity matrix, and we have to satisfy
this at the $j=3$ level. This gives the following orthogonality constraint,
\be (1,2,2) \cdot \left( \begin{array}{l}
                       -2c_{13}+c_{23}  \\
                       -c_{23}+c_{33}  \\
                       -\frac{1}{2}c_{13}+\frac{5}{4}c_{23}-c_{33}
                     \end{array}
              \right) = 0\:\:.
\label{eq:cc3}
\ee
Since this is indeed  satisfied, we are led to the conclusion that (\ref{eq:gensol3}) corresponds indeed to a valid asymptotic solution around the singularity. We note that in comparison with the series expansion we are led form found for the flat, radiation case in \cite{cot-tso-07} we arrive at the exact same form by setting $k=0$, cf. Eq. (21)\footnote{In that reference, $12\xi\t^3$ was mistakenly written as $24\xi\t^3$.}.

Eq. (\ref{eq:gensol3}) is  a local expansion of a \emph{general} solution around the initial singularity since it has exactly three arbitrary constants, $\theta, c_{13}$ and a third one corresponding to the arbitrary position of the singularity, taken here to be zero without loss of generality. Additionaly, using a theorem of Goriely and Hyde, cf. \cite{gor-01}, we can safely conclude that there is an open set of initial conditions for which the general solution blows up at the finite time (initial) singularity at $t=0$, since the leading order coefficients are real. Thus, the stability of our solutions in the neighborhood of the singularity is proved.

%--------------------------------------------------------------------------------------------------------------------
%	SECTION   - Conclusion
%--------------------------------------------------------------------------------------------------------------------

\section{Conclusion}

In this Chapter we have analyzed the asympotic behavior of the curved, radiation-filled FRW universes in the general quadratic gravity theory on approach to the initial singularity. We concluded that all curved radiation solutions tend asymptotically to the flat, vacuum $t^{1/2}$ solution of these theories, with the possible exception of the solutions in the conformally invariant Bach-Weyl theory.

We found a set of 255 different ways that the basic curvature-radiation vector field of this problem can decompose asymptotically and we were able to formulate certain qualitative remarks.  Each one of these asymptotic splittings could in principle contain various solutions on approach to the initial singularity but through performing a number of tests on necessary conditions that have to hold in order to exist an admissible asymptotic solution, we are left with only one exact solution of the scale invariant system of a unique possible asymptotic decomposition. This form is precisely the one that, after the use of various asymptotic and geometric techniques leads to the construction of a solution of the associated dynamical system in the form of a Puiseux formal series expansion dominated by the $t^{1/2}$ solution near the singularity. Our formal series expansion possesses the exact number of arbitrary constants in order to be a general solution, concluding that this solution is stable under such `perturbations' since it acts as an attractor of all homogeneous and isotropic radiation solutions of the theory.

Compared with the results of the previous chapters where the stability of the same solution was proved in the context of vacuum, flat universes and taking into consideration the impressive restrictions placed by the higher order field equations on the structure of the possible initial cosmological states of the theory we are lead to believe that the initial state of radiation-filled universes of this class possibly resembles the one of the vacuum, flat models. This remark is supported not only by the fact that the unique possible mode of approach to the singularity was found to be the one in which the curvature as well as the radiation parameters enter \emph{only} in the subdominant part of the vector field asymptotically but also by the general behavior of these parameters through out our asymptotic analysis. Meaning that in the great majority of the possible asymptotic decompositions, the existence of these parameters was the main reason of their failure to lead to an asymptotic balance.

% Chapter Template

\chapter{Discussion} % Main chapter title

\label{Chapter6} % Change X to a consecutive number; for referencing this chapter elsewhere, use \ref{ChapterX}

\lhead{Chapter 6. \emph{Discussion}} % Change X to a consecutive number; this is for the header on each page - perhaps a shortened title

%--------------------------------------------------------------------------------------------------------------------
%	SECTION 6.1  - Introduction
%--------------------------------------------------------------------------------------------------------------------

%\section{Discussion}

%======================================================================================================
%======================================================================================================
%VVVVVVVVVVVVVVVVVVV=======ATTENTION PLAGIARISM CHECK FAIL !!!!!! =========VVVVVVVVVVVVVVVVVVVVVVVVVVVV
%%=================================================
%\subsection{Ekpyrosis}

%It is instructive to also comment on this Chapter's results in connection with the results of Chapter 5 on the stability of radiation, curved universes for the same class of theories.
In this thesis, we studied the asymptotic behavior of the vacuum flat and curved, isotropic and homogeneous universes with a general action of the form,
\be\label{genact}
\mathcal{S}=\int_{\mathcal{M}}\mathcal{L}(R)d\mu_{g},
\ee
where
\be
\label{genlag}
\mathcal{L}(R)=\mathcal{L}(0)+ aR + bR^2 + cR^{\mu\nu}R_{\mu\nu} + dR^{\mu\nu\kappa\lambda}R_{\mu\nu\kappa\lambda},
\ee
as well as the ones filled with radiation in the curved case. For each one of these classes of cosmological models we constructed an autonomous dynamical system and an associated vector field which fully describe their evolution. Subsequently, we applied the method of asymptotic splittings presented in \cite{cot-bar-07} through which we decomposed the relevant vector fields in all the possible ways they could show a dominant behavior on approach to the initial singularity. Through a series of qualitative and analytical arguments we discovered the decompositions which dominate asympoticaly the dynamical systems in question and constructed their asymptotic solutions in the form of convergent formal series. The specific way we performed our asymptotic analysis allows us to observe in great detail how the different features of curvature and radiation affect the asymptotic behavior of the universes we studied. Additionaly, the final form of the asymptotic solutions we obtained proves the stability of those solutions with respect to such `pertubations'. That is, to the addition of higher-order curvature invariants to the Einstein-Hilbert action.

Our treatment showed that both radiation and vacuum, flat or curved cosmological models are attracted by the `universal' $t^{1/2}$ asymptote near the initial singularity and, in all these universes, this is the most dominant feature . However, open vacua show a more complex behavior in these models because they admit particular asymptotic solutions, that is universes that emerge from initial data sets of smaller dimension and valid for both early and late times. These universes asymptote to the $\a t$ Milne form during their early and late evolution toward finite-time singularities. On the other hand, closed vacua advance in time as more complicated solutions that are identified by logarithmic formal series, but on approach to the singularity their leading order is described again by singularities similar to the open case studied here.

The solutions found for the radiation-curved universes indeed correspond to the vacuum ones by letting the constant $\zeta$ tend to zero, meaning that these forms are indeed possible in the general vacuum evolution. In the curved case this comes from the asymptotic splitting $\mathbf{f^{7}}_{\,k,\textsc{VAC}}$. However, the vacuum field has more decompositions, namely $\mathbf{f^{12}}_{\,k,\textsc{VAC}}$ and $\mathbf{f^{42}}_{\,k,\textsc{VAC}}$, that now include the effects of vacuum and curvature appearing in their dominant part asymptotically (cf. beginning of Chapter 4) impossible in the radiation problem. Using these forms, we were able to find new asymptotic vacua not having any relation to those obtained from the radiation ones by letting the constant $\zeta$ tend to zero. These in turn lead to Milne type attractors monitoring precisely the dominant effects of vacuum and curvature in the asymptotic evolution.

As explained in Chapter 2, the Lagrangian of these theories is equivalent to the form $R+\xi R^2$. Nevertheless, this case is not the same as the general $R+\alpha R^2$, where $\alpha$ is a completely arbitrary parameter, since in the case studied here the parameter $\xi$ is a function of the coefficients of the quadratic corrections in the general Lagrangian (\ref{genlag}). Thus, it is an interesting problem to explore in what ways the various results concerning the $R+\xi R^2$ Lagrangian are affected by the different `weights' of the higher curvature invariants in the initial general Lagrangian. In our case specifically,  we can easily see that by substituting the parameter $\epsilon = \xi/6$ with the equivalent expression $b+\frac{1}{3}c+\frac{4}{3}d-1)$ we will obtain an equivalent system whose terms would include the parameters $a$, $b$, $c$ and $d$. Treating, this new vector field, with the method of asymptotic splittings, could possibly reach to very interesting conclusions about the way each higher-order curvature invariant affects the asymptotic behavior of these universes.

The attractor properties of our solutions and the existence of the Milne singularity bear a potential significance for the ekpyrotic scenario and its cyclic extension, wherein the passage through the singularity in these models, `the linchpin of the cyclic picture', depends on the stability of a Milne-type state under various kinds of perturbations \cite{kho-ovr-ste-tur-01,kho-ovr-sei-ste-tur-02,ste-tur-02a,ste-tur-02b}. In particular, during the brane collision it is found that the spacetime asymptotes to Milne and so it is expected that higher derivative corrections will be small during such a phase, cf. \cite{tol-tur-ste-04,eri-wes-ste-tur-04,leh-08}. Our work indicates that such Milne states may indeed dynamically  emerge as stable asymptotes during the evolution in any theory with higher order corrections in vacuum or with a radiation content. What remains is an interesting issue (that can be fully addressed with our asymptotic methods), that is to find whether the `compactified Milne mod $\mathbb{Z}_2$'$\times\mathbb{R}_3$ space  monitoring the reversal phase in the ekpyrotic and cyclic scenarios also emerges asymptotically as a stable attractor in the dynamics of higher order gravity when the matter content is a fluid with a general equation of state.

As far as it regards the general problem of having pure radiation or vacuum substituted by a fluid with a general equation of state $p=w\rho$ who would like to point out that new terms would appear in place of simple radiation terms, in this case,  for instance of the form
\be
\left(y,z,\frac{\z^2}{2\xi x^{3w+1}y}\right).
\ee
In such an approach, we would need to consider all different ranges of values of the fluid parameter $w$ to see if new forms of asymptotic evolution are possible even though in the limits of radiation and vacuum, it reduces to the known forms.

%%=================================================
%\subsection{Gravitational waves}

%\input{Chapters/Chapter7} 
%\input{Chapters/Chapter8} 
%\input{Chapters/Chapter9} 

%\input{Chapters/Chapter10}  % LaTeX GUIDE

%---------------------------------------------------------------------
%	THESIS CONTENT - APPENDICES
%---------------------------------------------------------------------

\addtocontents{toc}{\vspace{2em}} % Add a gap in the Contents, for aesthetics

\appendix % Cue to tell LaTeX that the following 'chapters' are Appendices

% Include the appendices of the thesis as separate files from the Appendices folder
% Uncomment the lines as you write the Appendices

% Appendix A

\chapter{Asymptotic splittings of the vacuum-curved vector field} % Main appendix title

\label{AppendixA} % For referencing this appendix elsewhere, use \ref{AppendixA}

\lhead{Appendix A. \emph{Asymptotic splittings of the vacuum-curved vector field}} % This is for the header on each page - perhaps a shortened title

%============================================================================================
%       Begining  of  TABLE OF VACUUM CURVED ASYMPOTIC SPLITTINGS
%============================================================================================
% \begin{center}
% \begin{tabular}{||c c c c||} 
% \hline
% Col1 & Col2 & Col2 & Col3 \\ [0.5ex] 
% \hline\hline
% $\mathbf{f^1}_{\,k,\textsc{VAC}}$ & 6 & 87837 & 787 \\ 
% \hline
% $\mathbf{f^2}_{\,k,\textsc{VAC}}$ & 7 & 78 & 5415 \\
% \hline
% $\mathbf{f^3}_{\,k,\textsc{VAC}}$ & 545 & 778 & 7507 \\
% \hline
% $\mathbf{f^4}_{\,k,\textsc{VAC}}$ & 545 & 18744 & 7560 \\
% \hline
% $\mathbf{f^5}_{\,k,\textsc{VAC}}$ & 88 & 788 & 6344 \\ [1ex] 
% \hline
%\end{tabular}
%\end{center}

%\hskip-4.0cm\
%\hspace*{-2cm}
%\makebox[\textwidth][c]{\includegraphics[width=1.2\textwidth]{image}}

%\begin{adjustbox}{center}
%\centerline{} 

%\begingroup
%\setlength{\LTleft}{-20cm plus -1fill}
%\setlength{\LTright}{\LTleft}
\begin{landscape}
\renewcommand*{\arraystretch}{1.6}  
%\changetext{0pt}{}{-1.2cm}{-1.2cm}{}%
\begin{longtable}[c]{||p{0.16\textwidth}||p{0.65\textwidth}|p{0.65\textwidth}||}
%\begin{longtable}[c]{||p{0.095\textwidth}||p{0.555\textwidth}|p{0.46\textwidth}||}%\begin{adjustwidth}{-8em}{}
%\includegraphics{wide} 
% \begin{widepage}
\caption{List of $\mathbf{f}_{\,k,\textsc{VAC}}$ possible asymptotic decompositions.\label{long}}\\
 
\hline
\multicolumn{3}{c}{Asymptotic splittings of the vacuum-curved vector field}\\

\hline
$\mathbf{f^n}_{\,k,\textsc{VAC}}$ & $\mathbf{f^n}^{(0)}_{\,k,\textsc{VAC}}$ & $\mathbf{f^n}^{(sub)}_{\,k,\textsc{VAC}}$\\
\hline
\endfirsthead
 
\hline
\multicolumn{3}{c}{Asymptotic splittings of the vacuum-curved vector field}\\
\hline
$\mathbf{f^n}_{\,k,\textsc{VAC}}$ & $\mathbf{f^n}^{(0)}_{\,k,\textsc{VAC}}$ & $\mathbf{f^n}^{(sub)}_{\,k,\textsc{VAC}}$\\
\hline
\endhead
 
\hline
\endfoot
  
 $\mathbf{f^1}_{\,k,\textsc{VAC}}$ & $\left( y,\frac{y^{2}}{2x},-2xz \right)$ & $\left(0,-3xy+kxz-\frac{k^2 z^2}{2x}-\frac{x}{12\epsilon}-\frac{kz}{12\epsilon x},0 \right)$\\
 $\mathbf{f^2}_{\,k,\textsc{VAC}}$ & $\left( y,-3xy,-2xz \right)$ & $\left(0,\frac{y^{2}}{2x}+kxz-\frac{k^2 z^2}{2x}-\frac{x}{12\epsilon}-\frac{kz}{12\epsilon x},0 \right)$\\
 $\mathbf{f^3}_{\,k,\textsc{VAC}}$ & $\left( y,+kxz,-2xz \right)$ & $\left(0,\frac{y^{2}}{2x}-3xy-\frac{k^2 z^2}{2x}-\frac{x}{12\epsilon}-\frac{kz}{12\epsilon x},0 \right)$\\
 $\mathbf{f^4}_{\,k,\textsc{VAC}}$ & $\left( y,-\frac{k^2 z^2}{2x},-2xz \right)$ & $\left(0,\frac{y^{2}}{2x}-3xy+kxz-\frac{x}{12\epsilon}-\frac{kz}{12\epsilon x},0 \right)$\\
 $\mathbf{f^5}_{\,k,\textsc{VAC}}$ & $\left( y,-\frac{x}{12\epsilon},-2xz \right)$ & $\left(0,\frac{y^{2}}{2x}-3xy+kxz-\frac{k^2 z^2}{2x}-\frac{kz}{12\epsilon x},0 \right)$\\
 $\mathbf{f^6}_{\,k,\textsc{VAC}}$ & $\left( y,-\frac{kz}{12\epsilon x},-2xz \right)$ & $\left(0,\frac{y^{2}}{2x}-3xy+kxz-\frac{k^2 z^2}{2x}-\frac{x}{12\epsilon},0 \right)$\\
 \hline
 $\mathbf{f^7}_{\,k,\textsc{VAC}}$ & $\left( y,\frac{y^{2}}{2x}-3xy,-2xz \right)$ & $\left(0,+kxz-\frac{k^2 z^2}{2x}-\frac{x}{12\epsilon}-\frac{kz}{12\epsilon x},0 \right)$\\
 $\mathbf{f^8}_{\,k,\textsc{VAC}}$ & $\left( y,\frac{y^{2}}{2x}+kxz,-2xz \right)$ & $\left(0,-3xy-\frac{k^2 z^2}{2x}-\frac{x}{12\epsilon}-\frac{kz}{12\epsilon x},0 \right)$\\
 $\mathbf{f^9}_{\,k,\textsc{VAC}}$ & $\left( y,\frac{y^{2}}{2x}-\frac{k^2 z^2}{2x},-2xz \right)$ & $\left(0,-3xy+kxz-\frac{x}{12\epsilon}-\frac{kz}{12\epsilon x},0 \right)$\\
 $\mathbf{f^{10}}_{\,k,\textsc{VAC}}$ & $\left( y,\frac{y^{2}}{2x}-\frac{x}{12\epsilon},-2xz \right)$ & $\left(0,-3xy+kxz-\frac{k^2 z^2}{2x}-\frac{kz}{12\epsilon x},0 \right)$\\
 $\mathbf{f^{11}}_{\,k,\textsc{VAC}}$ & $\left( y,\frac{y^{2}}{2x}-\frac{kz}{12\epsilon x},-2xz \right)$ & $\left(0,-3xy+kxz-\frac{k^2 z^2}{2x}-\frac{x}{12\epsilon},0 \right)$\\
 $\mathbf{f^{12}}_{\,k,\textsc{VAC}}$ & $\left( y,-3xy+kxz,-2xz \right)$ & $\left(0,\frac{y^{2}}{2x}-\frac{k^2 z^2}{2x}-\frac{x}{12\epsilon}-\frac{kz}{12\epsilon x},0 \right)$\\
 $\mathbf{f^{13}}_{\,k,\textsc{VAC}}$ & $\left( y,-3xy-\frac{k^2 z^2}{2x},-2xz \right)$ & $\left(0,\frac{y^{2}}{2x}+kxz-\frac{x}{12\epsilon}-\frac{kz}{12\epsilon x},0 \right)$\\
 $\mathbf{f^{14}}_{\,k,\textsc{VAC}}$ & $\left( y,-3xy-\frac{x}{12\epsilon},-2xz \right)$ & $\left(0,\frac{y^{2}}{2x}+kxz-\frac{k^2 z^2}{2x}-\frac{kz}{12\epsilon x},0 \right)$\\
 $\mathbf{f^{15}}_{\,k,\textsc{VAC}}$ & $\left( y,-3xy-\frac{kz}{12\epsilon x},-2xz \right)$ & $\left(0,\frac{y^{2}}{2x}+kxz-\frac{k^2 z^2}{2x}-\frac{x}{12\epsilon},0 \right)$\\
 $\mathbf{f^{16}}_{\,k,\textsc{VAC}}$ & $\left( y,+kxz-\frac{k^2 z^2}{2x},-2xz \right)$ & $\left(0,\frac{y^{2}}{2x}-3xy-\frac{x}{12\epsilon}-\frac{kz}{12\epsilon x},0 \right)$\\
 $\mathbf{f^{17}}_{\,k,\textsc{VAC}}$ & $\left( y,+kxz-\frac{x}{12\epsilon},-2xz \right)$ & $\left(0,\frac{y^{2}}{2x}-3xy-\frac{k^2 z^2}{2x}-\frac{kz}{12\epsilon x},0 \right)$\\
 $\mathbf{f^{18}}_{\,k,\textsc{VAC}}$ & $\left( y,+kxz-\frac{kz}{12\epsilon x},-2xz \right)$ & $\left(0,\frac{y^{2}}{2x}-3xy-\frac{k^2z^2}{2x}-\frac{x}{12\epsilon},0 \right)$\\
 $\mathbf{f^{19}}_{\,k,\textsc{VAC}}$ & $\left( y,-\frac{k^2 z^2}{2x}-\frac{x}{12\epsilon},-2xz \right)$ & $\left(0,\frac{y^{2}}{2x}-3xy+kxz-\frac{kz}{12\epsilon x},0 \right)$\\
 $\mathbf{f^{20}}_{\,k,\textsc{VAC}}$ & $\left( y,-\frac{k^2 z^2}{2x}-\frac{kz}{12\epsilon x},-2xz \right)$ & $\left(0,\frac{y^{2}}{2x}-3xy+kxz-\frac{x}{12\epsilon},0 \right)$\\
 $\mathbf{f^{21}}_{\,k,\textsc{VAC}}$ & $\left( y,-\frac{x}{12\epsilon}-\frac{kz}{12\epsilon x},-2xz \right)$ & $\left(0,\frac{y^{2}}{2x}-3xy+kxz-\frac{k^2 z^2}{2x},0 \right)$\\
 \hline
 $\mathbf{f^{22}}_{\,k,\textsc{VAC}}$ & $\left( y,\frac{y^{2}}{2x}-3xy+kxz,-2xz \right)$ & $\left(0,-\frac{k^2 z^2}{2x}-\frac{x}{12\epsilon}-\frac{kz}{12\epsilon x},0 \right)$\\
 $\mathbf{f^{23}}_{\,k,\textsc{VAC}}$ & $\left( y,\frac{y^{2}}{2x}-3xy-\frac{k^2 z^2}{2x},-2xz \right)$ & $\left(0,+kxz-\frac{x}{12\epsilon}-\frac{kz}{12\epsilon x},0 \right)$\\
 $\mathbf{f^{24}}_{\,k,\textsc{VAC}}$ & $\left( y,\frac{y^{2}}{2x}-3xy-\frac{x}{12\epsilon},-2xz \right)$ & $\left(0,+kxz-\frac{k^2 z^2}{2x}-\frac{kz}{12\epsilon x},0 \right)$\\
 $\mathbf{f^{25}}_{\,k,\textsc{VAC}}$ & $\left( y,\frac{y^{2}}{2x}-3xy-\frac{kz}{12\epsilon x},-2xz \right)$ & $\left(0,+kxz-\frac{k^2 z^2}{2x}-\frac{x}{12\epsilon},0 \right)$\\
 $\mathbf{f^{26}}_{\,k,\textsc{VAC}}$ & $\left( y,\frac{y^{2}}{2x}+kxz-\frac{k^2 z^2}{2x},-2xz \right)$ & $\left(0,-3xy-\frac{x}{12\epsilon}-\frac{kz}{12\epsilon x},0 \right)$\\
 $\mathbf{f^{27}}_{\,k,\textsc{VAC}}$ & $\left( y,\frac{y^{2}}{2x}+kxz-\frac{x}{12\epsilon},-2xz \right)$ & $\left(0,-3xy-\frac{k^2 z^2}{2x}-\frac{kz}{12\epsilon x},0 \right)$\\
 $\mathbf{f^{28}}_{\,k,\textsc{VAC}}$ & $\left( y,\frac{y^{2}}{2x}+kxz-\frac{kz}{12\epsilon x},-2xz \right)$ & $\left(0,-3xy-\frac{k^2 z^2}{2x}-\frac{x}{12\epsilon},0 \right)$\\
 $\mathbf{f^{29}}_{\,k,\textsc{VAC}}$ & $\left( y,\frac{y^{2}}{2x}-\frac{k^2 z^2}{2x}-\frac{x}{12\epsilon},-2xz \right)$ & $\left(0,-3xy+kxz-\frac{kz}{12\epsilon x},0 \right)$\\
 $\mathbf{f^{30}}_{\,k,\textsc{VAC}}$ & $\left( y,\frac{y^{2}}{2x}-\frac{k^2 z^2}{2x}-\frac{kz}{12\epsilon x},-2xz \right)$ & $\left(0,-3xy+kxz-\frac{x}{12\epsilon},0 \right)$\\
 $\mathbf{f^{31}}_{\,k,\textsc{VAC}}$ & $\left( y,\frac{y^{2}}{2x}-\frac{x}{12\epsilon}-\frac{kz}{12\epsilon x},-2xz \right)$ & $\left(0,-3xy+kxz-\frac{k^2 z^2}{2x},0 \right)$\\
 $\mathbf{f^{32}}_{\,k,\textsc{VAC}}$ & $\left( y,-3xy+kxz-\frac{k^2 z^2}{2x},-2xz \right)$ & $\left(0,\frac{y^{2}}{2x}-\frac{x}{12\epsilon}-\frac{kz}{12\epsilon x},0 \right)$\\
 $\mathbf{f^{33}}_{\,k,\textsc{VAC}}$ & $\left( y,-3xy+kxz-\frac{x}{12\epsilon},-2xz \right)$ & $\left(0,\frac{y^{2}}{2x}-\frac{k^2 z^2}{2x}-\frac{kz}{12\epsilon x},0 \right)$\\
 $\mathbf{f^{34}}_{\,k,\textsc{VAC}}$ & $\left( y,-3xy+kxz-\frac{kz}{12\epsilon x},-2xz \right)$ & $\left(0,\frac{y^{2}}{2x}-\frac{k^2 z^2}{2x}-\frac{x}{12\epsilon},0 \right)$\\
 $\mathbf{f^{35}}_{\,k,\textsc{VAC}}$ & $\left( y,-3xy-\frac{k^2 z^2}{2x}-\frac{x}{12\epsilon},-2xz \right)$ & $\left(0,\frac{y^{2}}{2x}+kxz-\frac{kz}{12\epsilon x},0 \right)$\\
 $\mathbf{f^{36}}_{\,k,\textsc{VAC}}$ & $\left( y,-3xy-\frac{k^2 z^2}{2x}-\frac{kz}{12\epsilon x},-2xz \right)$ & $\left(0,\frac{y^{2}}{2x}+kxz-\frac{x}{12\epsilon},0 \right)$\\
 $\mathbf{f^{37}}_{\,k,\textsc{VAC}}$ & $\left( y,-3xy-\frac{x}{12\epsilon}-\frac{kz}{12\epsilon x},-2xz \right)$ & $\left(0,\frac{y^{2}}{2x}+kxz-\frac{k^2 z^2}{2x},0 \right)$\\
 $\mathbf{f^{38}}_{\,k,\textsc{VAC}}$ & $\left( y,+kxz-\frac{k^2 z^2}{2x}-\frac{x}{12\epsilon},-2xz \right)$ & $\left(0,\frac{y^{2}}{2x}-3xy-\frac{kz}{12\epsilon x},0 \right)$\\
 $\mathbf{f^{39}}_{\,k,\textsc{VAC}}$ & $\left( y,+kxz-\frac{k^2 z^2}{2x}-\frac{kz}{12\epsilon x},-2xz \right)$ & $\left(0,\frac{y^{2}}{2x}-3xy-\frac{x}{12\epsilon},0 \right)$\\
 $\mathbf{f^{40}}_{\,k,\textsc{VAC}}$ & $\left( y,+kxz-\frac{x}{12\epsilon}-\frac{kz}{12\epsilon x},-2xz \right)$ & $\left(0,\frac{y^{2}}{2x}-3xy-\frac{k^2 z^2}{2x},0 \right)$\\
 $\mathbf{f^{41}}_{\,k,\textsc{VAC}}$ & $\left( y,-\frac{k^2 z^2}{2x}-\frac{x}{12\epsilon}-\frac{kz}{12\epsilon x},-2xz \right)$ & $\left(0,\frac{y^{2}}{2x}-3xy+kxz,0 \right)$\\
 \hline
 $\mathbf{f^{42}}_{\,k,\textsc{VAC}}$ & $\left( y,\frac{y^{2}}{2x}-3xy+kxz-\frac{k^2 z^2}{2x},-2xz \right)$ & $\left(0,-\frac{x}{12\epsilon}-\frac{kz}{12\epsilon x},0 \right)$\\
 $\mathbf{f^{43}}_{\,k,\textsc{VAC}}$ & $\left( y,\frac{y^{2}}{2x}-3xy+kxz-\frac{x}{12\epsilon},-2xz \right)$ & $\left(0,-\frac{k^2 z^2}{2x}-\frac{kz}{12\epsilon x},0 \right)$\\
 $\mathbf{f^{44}}_{\,k,\textsc{VAC}}$ & $\left( y,\frac{y^{2}}{2x}-3xy+kxz-\frac{kz}{12\epsilon x},-2xz \right)$ & $\left(0,-\frac{k^2 z^2}{2x}-\frac{x}{12\epsilon},0 \right)$\\
 $\mathbf{f^{45}}_{\,k,\textsc{VAC}}$ & $\left( y,\frac{y^{2}}{2x}-3xy-\frac{k^2 z^2}{2x}-\frac{x}{12\epsilon},-2xz \right)$ & $\left(0,+kxz-\frac{kz}{12\epsilon x},0 \right)$\\
 $\mathbf{f^{46}}_{\,k,\textsc{VAC}}$ & $\left( y,\frac{y^{2}}{2x}-3xy-\frac{k^2 z^2}{2x}-\frac{kz}{12\epsilon x},-2xz \right)$ & $\left(0,+kxz-\frac{x}{12\epsilon},0 \right)$\\
 $\mathbf{f^{47}}_{\,k,\textsc{VAC}}$ & $\left( y,\frac{y^{2}}{2x}-3xy-\frac{x}{12\epsilon}-\frac{kz}{12\epsilon x},-2xz \right)$ & $\left(0,+kxz-\frac{k^2 z^2}{2x},0 \right)$\\
 $\mathbf{f^{48}}_{\,k,\textsc{VAC}}$ & $\left( y,\frac{y^{2}}{2x}+kxz-\frac{k^2 z^2}{2x}-\frac{x}{12\epsilon},-2xz \right)$ & $\left(0,-3xy-\frac{kz}{12\epsilon x},0 \right)$\\
 $\mathbf{f^{49}}_{\,k,\textsc{VAC}}$ & $\left( y,\frac{y^{2}}{2x}+kxz-\frac{k^2 z^2}{2x}-\frac{kz}{12\epsilon x},-2xz \right)$ & $\left(0,-3xy-\frac{x}{12\epsilon},0 \right)$\\
 $\mathbf{f^{50}}_{\,k,\textsc{VAC}}$ & $\left( y,\frac{y^{2}}{2x}+kxz-\frac{x}{12\epsilon}-\frac{kz}{12\epsilon x},-2xz \right)$ & $\left(0,-3xy-\frac{k^2 z^2}{2x},0 \right)$\\
 $\mathbf{f^{51}}_{\,k,\textsc{VAC}}$ & $\left( y,\frac{y^{2}}{2x}-\frac{k^2 z^2}{2x}-\frac{x}{12\epsilon}-\frac{kz}{12\epsilon x},-2xz \right)$ & $\left(0,-3xy+kxz,0 \right)$\\
 $\mathbf{f^{52}}_{\,k,\textsc{VAC}}$ & $\left( y,-3xy+kxz-\frac{k^2 z^2}{2x}-\frac{x}{12\epsilon},-2xz \right)$ & $\left(0,\frac{y^{2}}{2x}-\frac{kz}{12\epsilon x},0 \right)$\\
 $\mathbf{f^{53}}_{\,k,\textsc{VAC}}$ & $\left( y,-3xy+kxz-\frac{k^2 z^2}{2x}-\frac{kz}{12\epsilon x},-2xz \right)$ & $\left(0,\frac{y^{2}}{2x}-\frac{x}{12\epsilon},0 \right)$\\
 $\mathbf{f^{54}}_{\,k,\textsc{VAC}}$ & $\left( y,-3xy+kxz-\frac{x}{12\epsilon}-\frac{kz}{12\epsilon x},-2xz \right)$ & $\left(0,\frac{y^{2}}{2x}-\frac{k^2 z^2}{2x},0 \right)$\\
 $\mathbf{f^{55}}_{\,k,\textsc{VAC}}$ & $\left( y,-3xy-\frac{k^2 z^2}{2x}-\frac{x}{12\epsilon}-\frac{kz}{12\epsilon x},-2xz \right)$ & $\left(0,\frac{y^{2}}{2x}+kxz,0 \right)$\\
 $\mathbf{f^{56}}_{\,k,\textsc{VAC}}$ & $\left( y,+kxz-\frac{k^2 z^2}{2x}-\frac{x}{12\epsilon}-\frac{kz}{12\epsilon x},-2xz \right)$ & $\left(0,\frac{y^{2}}{2x}-3xy,0 \right)$\\
 \hline
 $\mathbf{f^{57}}_{\,k,\textsc{VAC}}$ & $\left( y,\frac{y^{2}}{2x}-3xy+kxz-\frac{k^2 z^2}{2x}-\frac{x}{12\epsilon},-2xz \right)$ & $\left(0,-\frac{kz}{12\epsilon x},0 \right)$\\
 $\mathbf{f^{58}}_{\,k,\textsc{VAC}}$ & $\left( y,\frac{y^{2}}{2x}-3xy+kxz-\frac{k^2 z^2}{2x}-\frac{kz}{12\epsilon x},-2xz \right)$ & $\left(0,-\frac{x}{12\epsilon},0 \right)$\\
 $\mathbf{f^{59}}_{\,k,\textsc{VAC}}$ & $\left( y,\frac{y^{2}}{2x}-3xy+kxz-\frac{x}{12\epsilon}-\frac{kz}{12\epsilon x},-2xz \right)$ & $\left(0,-\frac{k^2 z^2}{2x},0 \right)$\\
 $\mathbf{f^{60}}_{\,k,\textsc{VAC}}$ & $\left( y,\frac{y^{2}}{2x}-3xy-\frac{k^2 z^2}{2x}-\frac{x}{12\epsilon}-\frac{kz}{12\epsilon x},-2xz \right)$ & $\left(0,+kxz,0 \right)$\\
 $\mathbf{f^{61}}_{\,k,\textsc{VAC}}$ & $\left( y,\frac{y^{2}}{2x}+kxz-\frac{k^2 z^2}{2x}-\frac{x}{12\epsilon}-\frac{kz}{12\epsilon x},-2xz \right)$ & $\left(0,-3xy,0 \right)$\\
 $\mathbf{f^{62}}_{\,k,\textsc{VAC}}$ & $\left( y,-3xy+kxz-\frac{k^2 z^2}{2x}-\frac{x}{12\epsilon}-\frac{kz}{12\epsilon x},-2xz \right)$ & $\left(0,\frac{y^{2}}{2x},0 \right)$\\
 \hline
 $\mathbf{f^{63}}_{\,k,\textsc{VAC}}$ & $\left( y,\frac{y^{2}}{2x}-3xy+kxz-\frac{k^2 z^2}{2x}-\frac{x}{12\epsilon}-\frac{kz}{12\epsilon x},-2xz \right)$ & $\left(0,0,0 \right)$\\
%\end{adjustwidth}
%\end{widepage}
\end{longtable}
%\endgroup
%\end{adjustbox}
\end{landscape}
%============================================================================================
%       End  of  TABLE OF VACUMM CURVED ASYMPOTIC SPLITTINGS
%============================================================================================

% Appendix B

\chapter{Asymptotic splittings of the radiation-curved vector field} % Main appendix title

\label{AppendixB} % For referencing this appendix elsewhere, use \ref{AppendixA}

\lhead{Appendix B. \emph{Asymptotic splittings of the radiation-curved vector field}} % This is for the header on each page - perhaps a shortened title

%============================================================================================
%       Begining of  TABLE OF RADIATION CURVED ASYMPOTIC SPLITTINGS
%============================================================================================
\begin{landscape}
\renewcommand*{\arraystretch}{1.6}  
%\changetext{0pt}{}{-1.2cm}{-1.2cm}{}%
\begin{longtable}[c]{||p{0.16\textwidth}||p{0.65\textwidth}|p{0.65\textwidth}||}
%\begin{adjustwidth}{-8em}{}
%\includegraphics{wide} 
% \begin{widepage}

\caption{List of $\mathbf{f}_{\,k,\textsc{RAD}}$ possible asymptotic decompositions.\label{long2}}\\

\hline
\multicolumn{3}{c}{Asymptotic splittings of the radiation-curved vector field}\\

\hline
$\mathbf{f^n}_{\,k,\textsc{RAD}}$ & $\mathbf{f^n}^{(0)}_{\,k,\textsc{RAD}}$ & $\mathbf{f^n}^{(sub)}_{\,k,\textsc{RAD}}$\\
\hline
\endfirsthead

\hline
\multicolumn{3}{c}{Asymptotic splittings of the radiation-curved vector field}\\
\hline
$\mathbf{f^n}_{\,k,\textsc{RAD}}$ & $\mathbf{f^n}^{(0)}_{\,k,\textsc{RAD}}$ & $\mathbf{f^n}^{(sub)}_{\,k,\textsc{RAD}}$\\
\hline
\endhead

\hline
\endfoot

% ONE TERM
% X
  $\mathbf{f^{1}}_{\,k,\textsc{RAD}}$ & $\left( y, z, \frac{\zeta^{2}}{2\xi x^2y}  \right)$ & $\left(0, 0, -\frac{k^2}{2 x^2y} +\frac{3y^3}{2x^2} +\frac{z^2}{2y} -\frac{yz}{x} - \frac{y}{2\xi} -\frac{k}{2\xi y} +\frac{ky}{x^2} \right)$\\
  $\mathbf{f^{2}}_{\,k,\textsc{RAD}}$ & $\left( y, z, -\frac{k^2\xi}{2\xi x^2y}  \right)$ & $\left(0, 0, \frac{\zeta^{2}}{2\xi x^2y} +\frac{3y^3}{2x^2} +\frac{z^2}{2y} -\frac{yz}{x} - \frac{y}{2\xi} -\frac{k}{2\xi y} +\frac{ky}{x^2} \right)$\\  
  $\mathbf{f^{3}}_{\,k,\textsc{RAD}}$ & $\left( y, z, \frac{3y^3}{2x^2}  \right)$ & $\left(0, 0, \frac{\z^2}{2\xi x^2y} -\frac{k^2}{2 x^2y} +\frac{z^2}{2y} -\frac{yz}{x} - \frac{y}{2\xi} -\frac{k}{2\xi y} +\frac{ky}{x^2} \right)$\\ 
  $\mathbf{f^{4}}_{\,k,\textsc{RAD}}$ & $\left( y, z, \frac{z^2}{2y} \right)$ & $\left(0, 0, \frac{\z^2}{2\xi x^2y} -\frac{k^2}{2 x^2y} +\frac{3y^3}{2x^2} -\frac{yz}{x} - \frac{y}{2\xi }-\frac{k}{2\xi y} +\frac{ky}{x^2} \right)$\\   
  $\mathbf{f^{5}}_{\,k,\textsc{RAD}}$ & $\left( y, z, -\frac{yz}{x} \right)$ & $\left( 0, 0, \frac{\z^2}{2\xi x^2y} -\frac{k^2}{2 x^2y} +\frac{3y^3}{2x^2} +\frac{z^2}{2y} - \frac{y}{2\xi} -\frac{k}{2\xi y} +\frac{ky}{x^2}  \right)$\\ 
  $\mathbf{f^{6}}_{\,k,\textsc{RAD}}$ & $\left( y, z, -\frac{y}{2\xi} \right)$ & $\left( 0, 0, \frac{\z^2}{2\xi x^2y} -\frac{k^2}{2 x^2y} + \frac{3y^3}{2x^2} + \frac{z^2}{2y} -\frac{yz}{x} -\frac{k}{2\xi y} + \frac{ky}{x^2}  \right)$\\ 
  $\mathbf{f^{7}}_{\,k,\textsc{RAD}}$ & $\left( y, z, -\frac{k}{2\xi y}  \right)$ & $\left(0,0, \frac{\z^2}{2\xi x^2y} -\frac{k^2}{2 x^2y} + \frac{3y^3}{2x^2} + \frac{z^2}{2y} -\frac{yz}{x} - \frac{y}{2\xi}+ \frac{ky}{x^2} \right)$\\ 
  $\mathbf{f^{8}}_{\,k,\textsc{RAD}}$ & $\left( y, z, \frac{ky}{x^2}\right)$ & $\left( 0, 0, \frac{\z^2}{2\xi x^2y} -\frac{k^2}{2 x^2y} +\frac{3y^3}{2x^2} +\frac{z^2}{2y} -\frac{yz}{x} -\frac{y}{2\xi}-\frac{k}{2\xi y}  \right)$\\
\end{longtable}
%\endgroup
%\end{adjustbox}
\end{landscape}
%=======================================================================
%       End of  TABLE OF RADIATION CURVED ASYMPOTIC SPLITTINGS
%=======================================================================
%\input{Appendices/AppendixC}

\addtocontents{toc}{\vspace{2em}} %Add a gap in the Contents, for aesthetics

\backmatter

%---------------------------------------------------------------------
%	BIBLIOGRAPHY
%---------------------------------------------------------------------

\label{Bibliography}

\lhead{\emph{Bibliography}} % Change the page header to say "Bibliography"

\bibliographystyle{unsrtnat} % Use the "unsrtnat" BibTeX style for formatting the Bibliography

\bibliography{Bibliography} % The references (bibliography) information are stored in the file named "Bibliography.bib"

\end{document}